%% file: thesis.tex
\documentclass[11pt,oneside]{report}
\usepackage[a4paper, bindingoffset=10mm]{geometry}                
\usepackage{graphicx}
\usepackage{amssymb}
\usepackage{amsmath}
\usepackage{slashed}
\usepackage{epstopdf}
\usepackage{subfig}
\usepackage{cite}
\usepackage{hyperref}

\DeclareMathOperator{\Tr}{Tr}
\begin{document}

\input{Chapters/Title}

\begin{abstract}
\input{Chapters/abstract}

\end{abstract}

\chapter*{Declaration}
\input{Chapters/Declaration}
\newpage

\input{Chapters/Appendix_A}

\tableofcontents
\chapter{Introduction}
\input{Chapters/Introduction}

\chapter{Inflationary Cosmology}
\input{Chapters/Inflation}

\chapter{Supersymmetry \& supergravity}

\input{Chapters/Supersymmetry}

\chapter{Gravitino condensation I}

\input{Chapters/Condensation}

\chapter{Gravitino condensation II}

\input{Chapters/Condensation_II}

\chapter{One-loop analysis}

\input{Chapters/Analysis}

\chapter{Conclusions}

\input{Chapters/Conclusions}

\newpage
\input{Chapters/Appendix_B}

\input{Chapters/Appendix_C}

\input{Chapters/Appendix_D}

\input{Chapters/Appendix_E}

\chapter*{Acknowledgements}
\input{Chapters/Acknowledgements}

\end{document}

%% file: Chapters/Title.tex
 \thispagestyle{empty}

  \begin{center}
{\large \vspace*{10mm} {{\includegraphics[width=85mm]{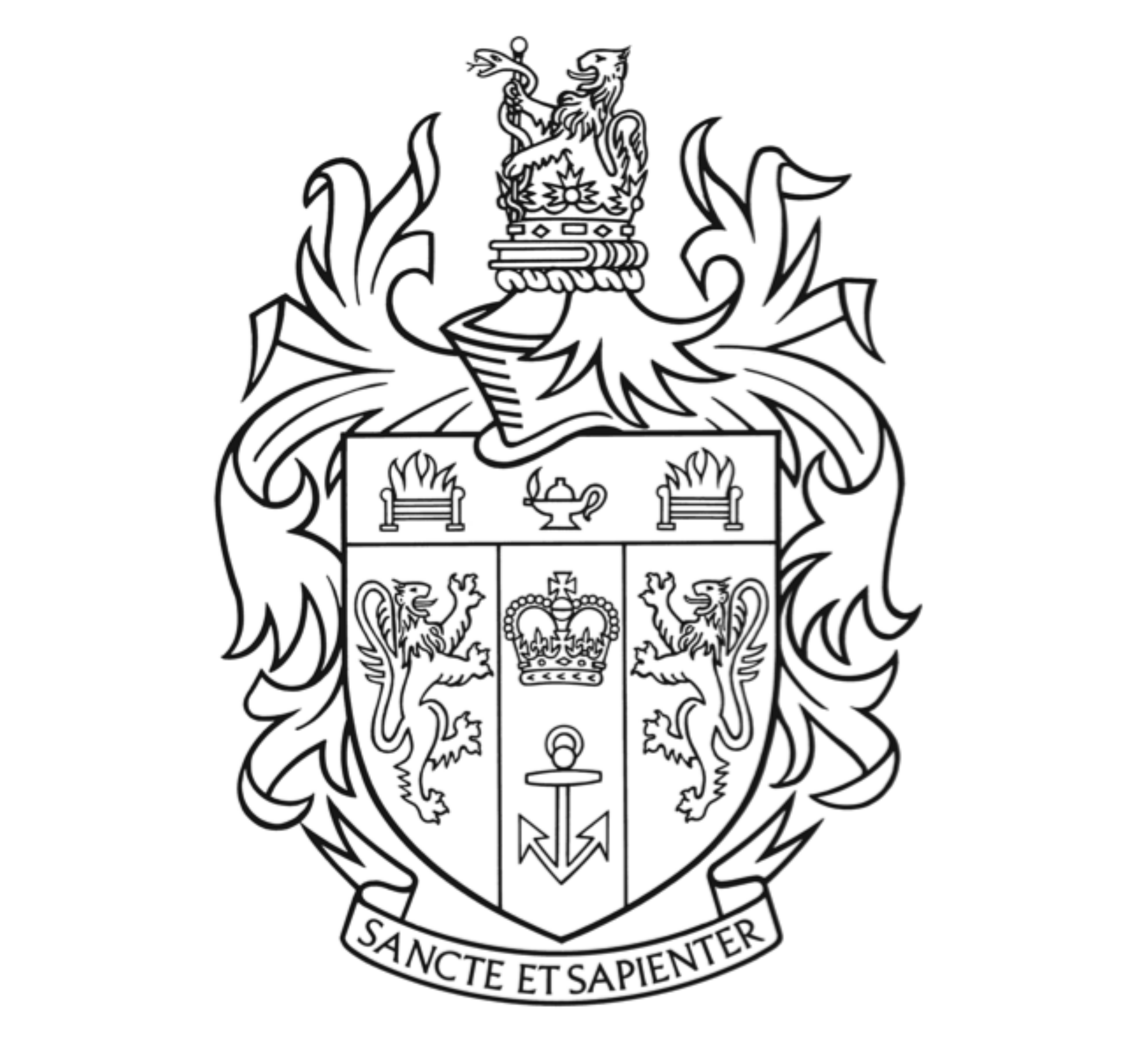}} \par } \vspace*{0mm}}
	\textsc{\LARGE \rm{King's College London }}\\[1.0cm]
	\textsc{ \LARGE \textbf{Gravitino condensation, supersymmetry breaking and inflation}}\\[0.8cm]
	\textsc{ \Large {\textbf {N. Houston}}} \\ \vspace{1.6cm}

	\parbox{320pt}{\begin{center}A thesis submitted to King's College London for the degree of 
	    Doctor of Philosophy in the Department of Physics, 
	    School of Natural and Mathematical Sciences, September 2015\end{center} }
	\vfill
     \end{center}

%% file: Chapters/abstract.tex
Supersymmetry is a well-motivated theoretical paradigm, which, if it exists, must be broken at low energies. 
As such, understanding the origin of this breaking is key in order to make contact with known phenomenology.
Motivated by dualistic considerations of the reality of Cooper pairing in low-temperature superconductivity and quark condensation in quantum chromodynamics, and the connections of supergravity to the exotic physics of string and M-theory, we investigate the dynamical breaking of local supersymmetry via gravitino condensation.
We firstly demonstrate non-perturbative gravitino mass generation via this mechanism in flat spacetime, and from this derive the condensate mode wavefunction renormalisation.
By then calculating the full canonically normalised one-loop effective potential for the condensate mode about a de Sitter background, we demonstrate that, contrary to claims in the literature, this process may both occur and function in a phenomenologically viable manner.
In particular, we find that outside of certain unfortunate gauge choices, the stability of the condensate is intimately tied via gravitational degrees of freedom to the sign of the tree-level cosmological constant.
Furthermore, we find that the energy density liberated may provide the necessary inflation of the early universe via an effective scalar degree of freedom, provided by the condensate, acting as the inflaton. 
However, in so doing we find that simultaneously phenomenologically viable inflation and supersymmetry breaking via this approach are mutually incompatible in the simplest supergravity settings.
As this mechanism takes place in the gravitational sector, relying on the ubiquitous gravitino torsion terms, we argue that it can also enjoy a certain universality in the context of supergravity and string theories, in that it does not rely on specific or arbitrary choices of potential and/or matter content.
This then allows straightforward transplantation of these results into other settings.
We present in detail our findings establishing contact between this scenario and known phenomenology, and discuss future avenues for research.

%% file: Chapters/Declaration.tex
This thesis is the product of my own work.
Where the work of others has been consulted, it is properly attributed as such.
No part of this document has been submitted toward qualification at this or any other institution.
The content presented herein is based upon the research contained in the following papers.
\begin{enumerate}
	\item  J.~Alexandre, N.~Houston and N.~E.~Mavromatos,
	\textit{Dynamical Supergravity Breaking via the Super-Higgs Effect Revisited,}
  Phys.\ Rev.\ D {\bf 88} (2013) 125017,
  [arXiv:1310.4122].
	\item J.~Alexandre, N.~Houston and N.~E.~Mavromatos, 
	\textit{Starobinsky-type Inflation in Dynamical Supergravity Breaking Scenarios,}
  Phys.\ Rev.\ D {\bf 89} (2014) 2,  027703,
  [arXiv:1312.5197].
	\item J.~Alexandre, N.~Houston and N.~E.~Mavromatos,
  \textit{Inflation via Gravitino Condensation in Dynamically Broken Supergravity,}
  Int.\ J.\ Mod.\ Phys.\ D {\bf 24} (2015) 04,  1541004,
  [arXiv:1409.3183].
\end{enumerate}

These articles are references \cite{Alexandre:2013iva,Alexandre:2013nqa,Alexandre:2014lla} in the bibliography.
\vspace{10pt}

N. Houston

King's College London

1st July, 2015

%% file: Chapters/Appendix_A.tex
\section*{Summary of notation and conventions}

\noindent Throughout this thesis we use natural units in which Planck's constant $\hbar$ and the speed of light $c$ are both set to one, with a reduced Planck mass
\begin{align*}
	M_{P}^{-2} \equiv 8\pi G =\kappa^2= \left(2.4 \times 10^{18}\, {\rm GeV} \right)^{-2}\,.
\end{align*}

\noindent We use $t$ and $\tau$  for physical and conformal time respectively, and denote four-dimensional spacetime coordinates by $x^\mu$, three-dimensional spatial coordinates by $x^i$, and three-dimensional vectors by $\textbf{x}$. 

\noindent The metric signature is $(-\,+\,+\,+)$, and our curvature conventions are
\begin{align*}
	R^\rho{}_{\sigma\mu\nu}&=\partial_\mu\Gamma^\rho{}_{\nu\sigma}-\partial_\nu\Gamma^\rho{}_{\mu\sigma}
	+\Gamma^\rho{}_{\mu\tau}\Gamma^\tau{}_{\nu\sigma}-\Gamma^\rho{}_{\nu\tau}\Gamma^\tau{}_{\mu\sigma}\\
	&=e^{\rho a} e_\sigma{}^b\left(\partial_\mu\omega_{\nu ab}-\partial_\nu\omega_{\mu ab}
	+\omega_{\mu ac} \omega_\nu{}^c{}_b-\omega_{\nu ac} \omega_\mu{}^c{}_b\right)\\
	R_{\mu\nu}&=R^\rho{}_{\nu\rho\mu}\,,\quad
	R=g^{\mu\nu}R_{\mu\nu}\,,\quad
	R_{\mu\nu}-\frac{1}{2}g_{\mu\nu}R=\kappa^2T_{\mu\nu}\,.
\end{align*}

\noindent Our Fourier convention is
\begin{align*}
	R_{\textbf{k}} = \int d^3 x \, R\left(\textbf{x}\right) \,e^{i \textbf{k}\cdot \textbf{x}} \,,
\end{align*}
the power spectrum for a statistically homogeneous field is defined by
\begin{align*}
	\langle R_{\textbf{k}\vphantom{'}} R_{\textbf{k}'} \rangle \equiv P_R\left(k\right) \delta\left(\textbf{k} + \textbf{k}'\right) \,,
\end{align*}
and the dimensionless power spectrum is
\begin{align*}
	\Delta_R^2\left(k\right) \equiv \frac{k^3}{2\pi^2} P_R\left(k\right)\,.
\end{align*}

\noindent The Hubble slow-roll parameters are
\begin{align*}
	\epsilon  \equiv -  \frac{\dot H}{H^2}\,, \quad 
	\eta  \equiv \frac{\dot \varepsilon}{H \varepsilon}\,,
\end{align*}
where overdots represent derivatives with respect to physical time $t$, and the slow-roll parameters are
\begin{align*}
	\epsilon_\phi \equiv \frac{M_P^2}{2} \left( \frac{V'}{V}\right)^2\,, \quad 
	\eta_\phi \equiv M_P^2 \frac{V''}{V}\,,
\end{align*}
where primes are derivatives with respect to the inflaton $\phi$, and $V\left(\phi\right)$ is the inflaton potential.

%% file: Chapters/Introduction.tex
Arguably, the preeminent question in modern theoretical physics is how we should connect present and future experimental signatures to the wealth of constructs in high-energy theory, particularly those provided by string and M-theory.
Especially in light of the restart of the Large Hadron Collider, the keystone of this particular problem is supersymmetry; a well-motivated theoretical paradigm, which, if it exists, must be broken at low energies. 
An understanding of the phenomenon leading to this breaking is then crucial in order to connect the known to the conjectured.

We may note that, from Cooper pairing in low-temperature superconductivity and superfluidity, to quark condensation in quantum chromodynamics (QCD), one mechanism leveraged by Nature in numerous settings and to numerous ends is the pairing of fermions into effective bosonic degrees of freedom.
Motivated by the dualistic considerations of this well-understood reality and the connections of local supersymmetry to the exotic physics of string and M-theory, the topic of this thesis is then the investigation of a specific scenario connecting these disparate realms; the dynamical breaking of local supersymmetry via condensation of the gravitino, the fermionic superpartner of the graviton.

This is a process largely analogous to the manner in which chiral symmetry in QCD is broken via quark condensation, whereby fermionic bilinears develop non-trivial vacuum expectation values via the non-perturbative action of gluons, ensuring that the vacuum of the theory is no longer invariant under chiral transformations. 
Gravitino condensation should in principle proceed similarly, inducing a gravitino mass via the super-Higgs effect and in so doing breaking the supersymmetric degeneracy with the massless graviton.

In order to exploit this analogy we note the existence of simplified effective theories, such as the Nambu-Jona-Lasinio (NJL) model, used to help illuminate features of chiral symmetry breaking. 
As we will discuss, the underlying physical linkage between fundamental and effective descriptions in this instance are the non-perturbative gluon configurations which integrate out to give the characteristic four-fermion interactions of the NJL model.

Proceeding analogously, we approach supergravity in the spirit of an NJL-type effective description of some more fundamental theory, seeking to understand gravitino condensation within it.
The propriety of this perspective is of course reinforced by the general non-renormalisability and four-gravitino interactions present in supergravity, not to mention the status of some supergravity theories as low energy effective descriptions of corresponding string theories.

This is a topic which has been explored to a limited extent in the literature, originating with the articles \cite{Jasinschi:1983wr,Jasinschi:1984cx}.
Therein, it was argued that a gravitino condensate could indeed form and break local supersymmetry, albeit via a calculation neglecting gravitational degrees of freedom and the role of the cosmological constant.
As we will demonstrate, beyond linking spin 2 and spin 3/2 degrees of freedom, unbroken local supersymmetry intimately connects the cosmological constant to the mass of the gravitino.
As such, it is unclear if the results of \cite{Jasinschi:1983wr,Jasinschi:1984cx} are necessarily definitive.

Indeed, the impossibility of dynamical local supersymmetry breaking was subsequently asserted in \cite{Buchbinder:1989gi,Odintsov:1988wz}, as an apparent consequence of the presence of gravitational degrees of freedom.
That the mechanism of gravitino condensation has not been further explored is in part a conceivable consequence of these claims, possibly in conjunction with the intrinsic difficulty associated to quantum field theory in curved backgrounds.
Nevertheless there does exist a small body of literature exploring the scenario \cite{Duff:1982yi,Jasinschi:1984mr,Oh:1985bm,HelayelNeto:1986dg,Jasinschi:1986ze,Pollock:1987cu,Kitazawa:2001zi, Katagiri:2002pq,Pollock:2004si,Hatanaka:2006hv,Ellis:2011mz,Ellis:2013zsa,Basilakos:2015yoa}, albeit which does not address these fundamental issues.

By calculating in detail the full one-loop effective potential for the condensate mode, we demonstrate however that, contrary to claims in the literature, this process may both occur and function in a phenomenologically viable manner.
The instability claimed in \cite{Buchbinder:1989gi,Odintsov:1988wz} is found to ultimately trace back to the simple absence in their formalism of the requisite goldstino degrees of freedom for the gravitino to absorb and therefore become massive.

A previously undetected subtlety is also explored, in that even despite the local supersymmetry of the action, the four-gravitino coupling into the scalar condensate channel is ultimately ambiguous at the perturbative level, exemplified in the freedom to perform Fierz transformations between different channels.
As we illustrate, the resultant freedom to vary this coupling is ultimately crucial in order to achieve sufficiently light supersymmetry breaking.

Furthermore, as we will argue this approach can enjoy a number of useful features.
Firstly, the energy density liberated may provide the necessary inflation of the early universe via an effective scalar degree of freedom, provided by the condensate, acting as the inflaton. 
By forcing this mechanism of gravitino condensation to perform Ôdouble duty' one may then simultaneously confront both cosmological and particle physics phenomenology. 

As this approach takes place in the gravitational sector it can also enjoy a certain universality in the context of supergravity and string theories, in that it does not rely on specific or arbitrary choices of potential and/or matter content.
This then allows straightforward transposition of these results into other, more extensive, settings.

It should be however noted that we cannot expect these results in and of themselves to provide a convincingly `natural' rationale for the relative lightness of the electroweak scale.
This is an expected consequence of the limitations of the effective NJL-type description we pursue, where the ruinous effect of quadratic divergences can be absorbed, albeit only to resurge elsewhere in the theory. 

As in the original NJL model, the `naturally' small factors required to safely generate such a hierarchy, thought to be non-perturbative in origin, are invisible at this level.
Regardless, we may appeal to their implicit presence, and speculate as to their potential origin.

\section{Scope and structure} 

Approaching supergravity in the spirit of an NJL-type effective description and making use of the associated formalism, we will then compute the one-loop effective potential and wavefunction renormalisation factor for the condensate mode, which, upon combination, yield the canonically normalised effective potential.
From this, the behaviour of the condensate may be quantitatively understood, and the relevance thereof to supersymmetry breaking and early universe cosmology may be assessed.

The structure of this thesis is as follows.
\begin{itemize}
	\item Chapter 2 comprises a concise introduction to the motivation and methods of early universe inflation, specialising to the aspects of inflationary phenomenology relevant for later discussion.
	\item Chapter 3 analogously introduces the requisite aspects of supersymmetry and supergravity, including the super-Higgs mechanism which is central to the topic at hand.
	\item Chapter 4 motivates and expands upon the central analogy of this thesis in supergravity as an NJL-type theory, based largely on \cite{Alexandre:2014lla}.
	Repurposing some tools from the study of the latter theory, we derive and solve the flat-space gap equation leading to a dynamical gravitino mass, and explore some of related issues pertaining to the role of the coupling constant.
	Using this, we then derive the wavefunction renormalisation for the condensate mode via the Bethe-Salpeter equation and an all-orders resummation of four-gravitino bubble graphs.
	\item Chapter 5 builds upon the preceding chapter in deriving the one-loop effective potential in a de Sitter background, incorporating fully the previously neglected role of gravitational degrees of freedom and the cosmological constant. 
	Given the possible influence of gauge dependence, particular emphasis is placed upon the intricacies of gauge fixing. 
	Leveraging the resultant potential the stability of the condensate is examined, finding that, outside of certain unfortunate gauge choices, the stability of the condensate is directly linked to the sign of the tree-level cosmological constant.
	These results are based on \cite{Alexandre:2013iva}.
	\item Chapter 6 centres on the analysis of the canonically normalised one-loop effective potential derived via the results of chapters 4 and 5,  assessing the resultant suitability of the gravitino condensation mechanism for supersymmetry breaking and early universe inflation. 
	Therein, we demonstrate the expected resurgence of the tuning associated to the lightness of the electroweak scale in the four-gravitino coupling into the scalar condensate channel, so that, given sufficient proximity of the coupling to a critical value, viable supersymmetry breaking may always be engineered.
	We also illustrate the possibility of a suitable inflationary phase, characterised by a negligible tensor to scalar ratio. 
	Notably, we also demonstrate that these circumstances cannot coexist in the basic supergravity setting.
	These results are based on \cite{Alexandre:2013iva,Alexandre:2013nqa,Alexandre:2014lla}, with some modifications arising due to increased understanding and sophistication of approach.
	\item Chapter 7 provides some concluding remarks, and directions for future investigations.
	Principal amongst these is the need to understand via wider contexts the issues raised in chapter 6 regarding the near-criticality of the four-gravitino coupling.
	\item Further technical details including a summary of gamma matrix technology, Fierz transformations and zeta function regularisation are presented in the appendices.
\end{itemize}

%% file: Chapters/Inflation.tex
This chapter constitutes a concise introduction to the various elements of inflationary cosmology necessary for later discussion.
After detailing in section \ref{sec: The early universe} the shortcomings of the standard Hot Big Bang scenario, which motivate the inflationary hypothesis, we demonstrate the resolution of these issues in section \ref{sec: A decreasing Hubble sphere} via an early period where the comoving Hubble radius decreases.
Section \ref{sec: Inflationary theory} then details one manner in which this can be achieved in practice through the time evolution of a scalar field, known as the inflaton.
Finally, section \ref{sec: Primordial perturbations} derives the pertinent consequences of this inflationary mechanism, outlining how we may connect primordial perturbations during inflation to present day observations.

\section{The early universe} \label{sec: The early universe}

As demonstrated by the results of Planck and other experiments, the standard Lambda Cold Dark Matter ($\Lambda$CDM) cosmology is the simplest model in agreement with most aspects of late time cosmology \cite{Adam:2015rua}. 
This is a six parameter model set in a spatially flat Friedmann-Robertson-Walker (FRW) spacetime, with line element
\begin{align}
	ds^2=-dt^2+a^2\left(t\right)\left(\frac{dr^2}{1-kr^2}+r^2d\Omega_2^2\right)\bigg|_{k=0} 
	=a^2\left(\tau\right) \left(-d\tau^2+dr^2+r^2d\Omega_2^2\right)\,,
\end{align}
where $t$ and $\tau$ are cosmological and conformal time, respectively. 
The parameter $k=\left\{-1,0,1\right\}$ denotes the curvature of spacelike 3-hypersurfaces, which for simplicity we will largely set to zero in what follows.
As is well known, the $\Lambda$CDM scenario is incomplete in a number of regards.

To elucidate several of these aspects, it is firstly illuminating to consider the causal structure of the theory.
The total comoving distance covered by a light ray up to some time $t$ is
\begin{align}\label{particle horizon}
	\chi = \int_0^{t} \frac{dt'}{a\left(t'\right)}
	=\int_{0}^{a} \frac{da'}{a'\dot a'}
	=\int_{1}^{\ln a} \frac{d\ln a'}{a'H}\,, \quad
	H\equiv\frac{1}{a}\frac{da}{dt}\,,
\end{align}
where by definition $t=0$ is when the Big Bang occurs.
Since $\chi$ determines whether particles with a given comoving separation could have been causally connected in the past, it is known as the comoving particle horizon.

The comoving Hubble radius $\left(aH\right)^{-1}$ from the final integrand of \eqref{particle horizon} also appears in the continuity equations for a fluid-dominated FRW spacetime, giving
\begin{align}\label{friedmann}
	\left(aH\right)^{-1}=a^{\left(1+3\omega\right)/2}/H_0\,, \quad 
	\omega\equiv p/\rho\,,
\end{align}
where $p$ and $\rho$ are respectively the pressure and energy density of the fluid, and $H_0$ is the value of the present day Hubble parameter.
Conventional matter obeys the Strong Energy Condition (SEC), so that $1+3\omega > 0$ and therefore $\left(aH\right)^{-1}$ is an increasing function of $a$.

\subsection{Horizon problem}
Since $\left(aH\right)^{-1}$ is also necessarily positive, the dominant contribution to the integral in \eqref{particle horizon} therefore comes from late times, when $\left(aH\right)^{-1}$ is largest.
This is acutely problematic as it implies that the vast majority of conformal time elapsed since the Big Bang has occurred after the formation of the Cosmic Microwave Background (CMB). 
This took place at recombination; when the temperature of the universe decreased sufficiently to allow the formation of neutral hydrogen, which we can estimate from the rate of cooling of an expanding universe to have occurred approximately 3.8$\times\,10^5$ years after the initial singularity \cite{Planck:2015xua}.

\begin{figure}[h!]
  	\centering
    	\includegraphics[width=0.8\textwidth]{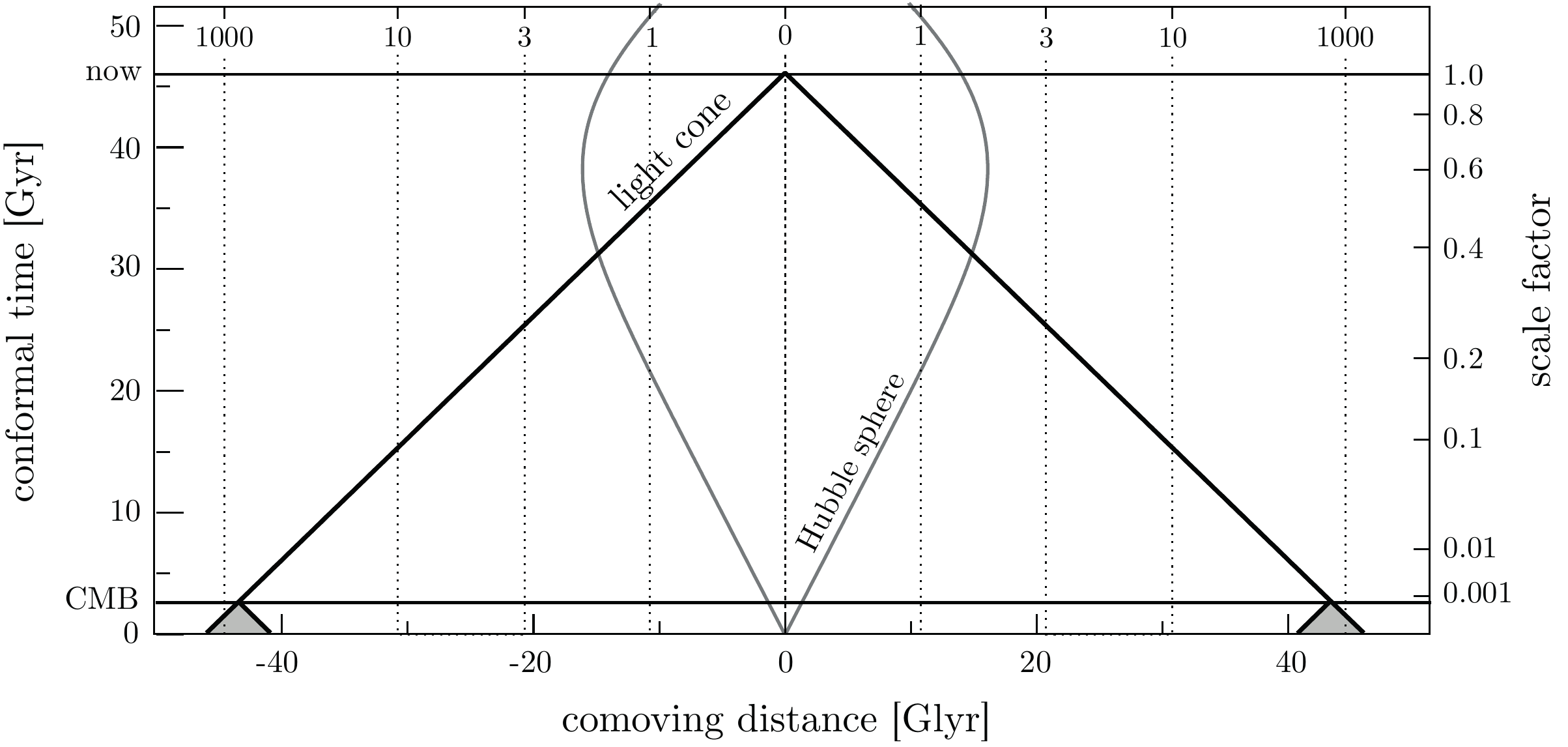}
 	\caption{
	Causal evolution of the FRW spacetime in comoving coordinates centred on our worldline, where the top portion indicating a decreasing comoving Hubble sphere corresponds to the dark energy dominated era we presently inhabit. 
	As can be seen, our current lightcone encloses regions which were causally disconnected at recombination.
	Given no reason for conditions in these regions to be similar, we then expect the CMB to comprise inhomogeneous patches. 
	Figure reproduced from \cite{Baumann:2014nda}, having been adapted originally from \cite{Lineweaver:2003ie}.}
	\label{fig:frw}
\end{figure}

If this is indeed the case, we would then expect most regions of the CMB we observe today to be causally disconnected from each other, as can be seen in Figure \ref{fig:frw}, where any pre-recombination lightcones connecting nearby regions have only minimal conformal time until the Big Bang in which to intersect.
In the standard $\Lambda$CDM cosmology specifically, the CMB comprises $\mathcal{O}\left(10^4\right)$ disconnected patches \cite{Lineweaver:2003ie}.

Precision observations of the CMB however, indicate that these a priori causally disconnected patches of the CMB sky are in fact homogeneous and isotropic to within 1 part in $10^{5}$ \cite{Adam:2015rua}.
This is commonly referred to as the horizon problem.

It is notable however that this argument implicitly rests upon the behaviour of the integral \eqref{particle horizon} arbitrarily close to the initial singularity.
One may expect quantum mechanical effects of gravity to enter in that regime, and conceivably alter physics there such that there is no horizon problem after all \cite{Magueijo:2002xx}.
In the absence of a precise notion of quantum gravity however, this line of reasoning may culminate in something of an impasse.
As such, we will instead investigate approaches which can be mechanistically explored.

\subsection{Flatness problem} 
A further concern may be identified in the natural-units Friedmann equation 
\begin{align}
	H^2=\frac{\rho}{3M_P^2}-\frac{k}{a^2}+\frac{\Lambda}{3}, 
	\label{eq:friedmann}
\end{align}
which, upon dividing through by $H^2$, can be compactly expressed in terms of an effective density parameter $\Omega$
\begin{align}\label{friedmann 2}
	(1-\Omega)=-\frac{k}{\left(Ha\right)^2}\,, \quad
	\Omega\equiv \frac{\rho}{\rho_c}+\frac{\Lambda}{3H^2}\,, \quad
	\rho_{c}\equiv3M_P^2 H^2\,.
\end{align}
where $\Lambda$ is the cosmological constant, and $M_P$ is the reduced Planck mass.

Since the measured value of $\Omega$ is very close to 1, our universe is extremely close to spatial flatness \cite{Planck:2015xua}. 
This is problematic as \eqref{friedmann 2} implies that $(1-\Omega)\left(Ha\right)^2$ is constant throughout the evolution of the universe, whilst as we have seen, $\left(Ha\right)^{-1}$ must necessarily increase over time.

Making use of a subscript to indicate present day quantities, we can write 
\begin{align}
	\left(1-\Omega\right)\left(Ha\right)^2
	=\left(1-\Omega_0\right)\left(H_0a_0\right)^2\, ,
\end{align}
where we have the usual redshift relation $a=a_0/\left(1+z\right)$.
Specialising for simplicity to the case of radiation domination, where $\omega=1/3$, \eqref{friedmann} indicates that $H\propto a^{-2}$.
This yields
\begin{align}
	\left(1-\Omega\right)\left(1+z\right)^{2}
	=\left(1-\Omega_0\right)\, .
\end{align}

Especially given the current experimental limits of $0.995<\Omega_0<1.005$ \cite{Planck:2015xua}, we can then only conclude that as we look backwards in time to higher and higher redshifts, $\Omega$ must have been ever increasingly close to unity in order to counterbalance an increasing $z$.
Equivalently, the condition $\Omega=1$ is very unstable; any departures from flatness present in the early universe are amplified by the evolution of the universe.
This is commonly referred to as the flatness problem.

\section{A decreasing Hubble sphere}
\label{sec: A decreasing Hubble sphere}

Given the suggestive phrasing of these issues in terms of the problems associated with an increasing comoving Hubble radius, it is hopefully unsurprising to see that an elegant common solution may be sought via an early period of decreasing $\left(aH\right)^{-1}$, where
\begin{align}
	\frac{d}{dt} \frac{1}{aH} <0\,.
\end{align}

\begin{figure}[h!]
  \centering
    \includegraphics[width=0.8\textwidth]{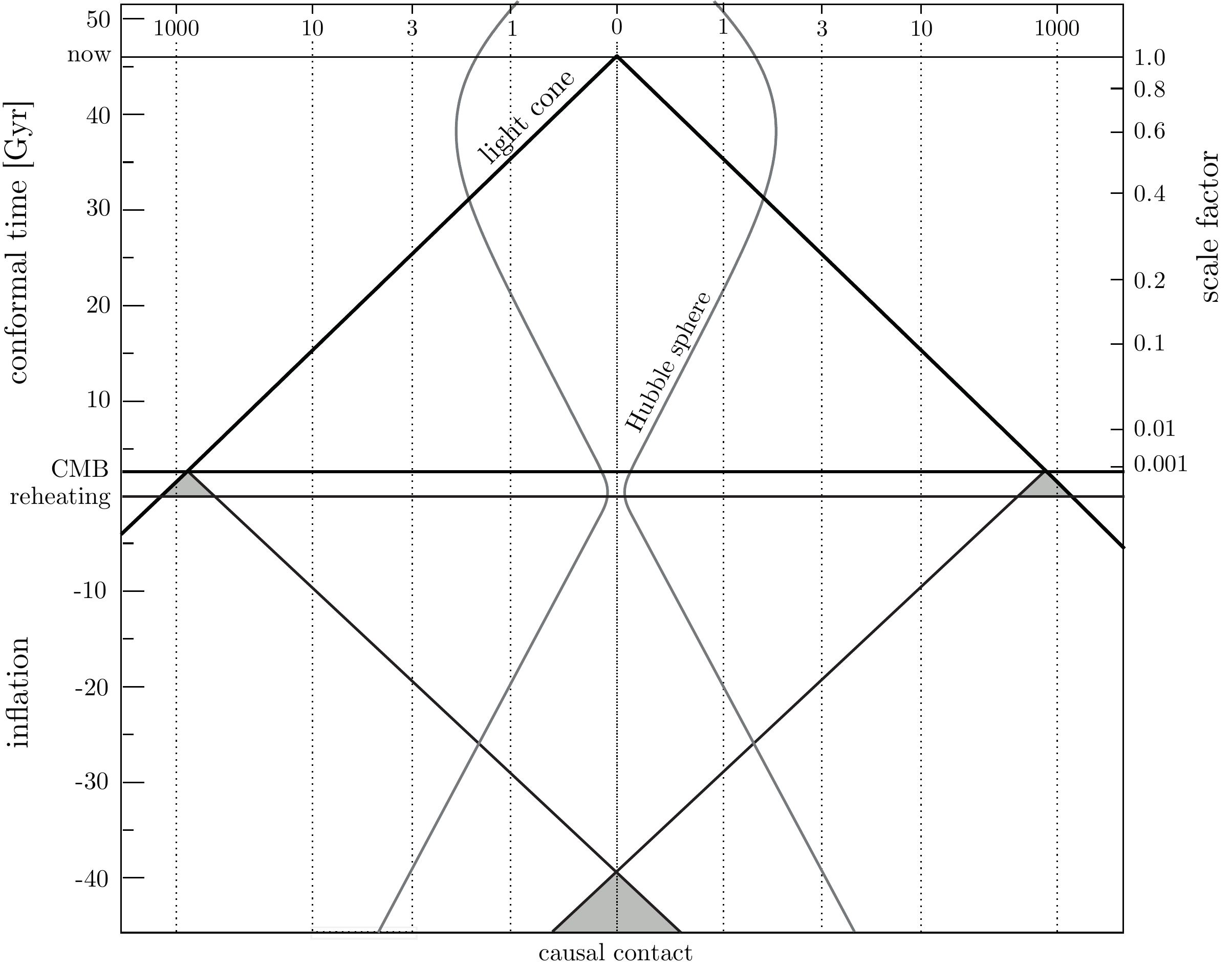}
 	\caption{Extension of Figure \ref{fig:frw} incorporating an earlier period of decreasing comoving Hubble sphere.
	With sufficient negative conformal time spent in this phase, the previously disconnected regions at recombination now are able to share a causal connection in the past.
	This can explain the homogeneity and isotropy of the CMB we observe today. 
	Figure reproduced from \cite{Baumann:2014nda}.}
	\label{fig:inflation}
\end{figure}
To demonstrate the effect of such an epoch, we may substitute \eqref{friedmann} into \eqref{particle horizon} and integrate, yielding 
\begin{align}\label{chi}
	\chi\equiv\tau_0-\tau=\frac{2}{\left(1+3\omega\right)H_0}a'^{\left(1+3\omega\right)/2}\bigg|_{0}^{a_0}\,.
\end{align}
The lower limit of this expression corresponds to the conformal time at which the Big Bang occurs, which, if $\omega>-1/3$, is at $\tau = 0$. 
If however $\omega<-1/3$, as we would expect from \eqref{friedmann} during a period in which $\left(aH\right)^{-1}$ is decreasing, then the Big Bang now takes place at $\tau = -\infty$.

As can be seen via Figure \ref{fig:inflation} this addresses the horizon problem by modifying \eqref{particle horizon} via the addition of an extra period of negative conformal time in the lower limit of the integral.
The previously non-intersecting lightcones influencing CMB observables are then allowed to extend further backwards, so that with sufficient conformal time spent whilst $\left(aH\right)^{-1}$ decreases, the entire CMB we observe today originates from a single causally connected region.

Simultaneously, we may revisit the Friedmann equation \eqref{friedmann 2}, noting that $\left(1-\Omega\right) \propto \left(aH\right)^{-2}$.
Given sufficient conformal time where $\left(aH\right)^{-1}$ decreases, we then expect $\Omega$ to be dynamically driven to one, irrespective of initial conditions.
As we will demonstrate in the following section, we can identify decreasing $\left(aH\right)^{-1}$ with a phase of exponential expansion, in which case this conclusion is entirely sensible; rapid expansion will naturally dilute whatever SEC-satisfying energy density is present away to nothing.
Given sufficient expansion the subsequent phase where $\left(aH\right)^{-1}$ increases now has far less effect on $\Omega_0$, and what we previously regarded as the extremely special degree of flatness we observe today therefore becomes a generic feature. 

\subsection{Inflation}
Given the definition of $H$ in \eqref{particle horizon}, we can see that
\begin{align}\label{decrease}
	\frac{d}{dt} \frac{1}{aH}
	=-\frac{\ddot a}{\left(\dot a\right)^2}\,,
\end{align}
and a decreasing Hubble radius then requires $\ddot a>0$, implying an early period of accelerated expansion, which we may henceforth refer to as inflation.
On the basis of \eqref{friedmann}, we can see that achieving this in practice requires violating the SEC.
Intuitively this is somewhat sensible; conventional matter would be strongly diluted by such rapid expansion and would be unable to drive continued inflation.

Solving the Friedmann equation \eqref{eq:friedmann} for a constant right-hand side (RHS) yields the prototypical de Sitter inflationary spacetime, in which $a \sim e^{Ht}$.
In practice we require the inflationary phase to end, in which case perfect de Sitter is not appropriate. 
However we may instead make use of a quasi de Sitter spacetime for which $e^{Ht}$ is a good approximation to $a$, but with appropriate time dependence such that inflation is finite.
For these reasons we may identify inflation with a period of quasi de Sitter.

In addition to the resolution of several shortcomings of the standard cosmology, the elegant inflationary hypothesis also carries a number of useful consequences.

\subsection{Origin of inhomogeneities}
We would like to be able to explain the primordial perturbations from which the inhomogeneities we observe in the universe arise.
For a perturbation of wavelength $\lambda$, we may make use of comoving Fourier space to write
\begin{align}
	\lambda=\frac{2\pi}{k_\lambda}a\,,
	\label{eq: wavenumber}
\end{align}
for a corresponding wavenumber $k_\lambda$.
Whilst, as expected, the physical wavelength increases during expansion of the universe, in comoving units it is constant.
Assuming, as in the standard cosmology, an increasing comoving Hubble radius, perturbations that are observable now must then have originally existed outside our Hubble sphere, having only crossed over once $k_\lambda\sim2\pi aH$.

Considering the upper limit of \eqref{chi}, we may derive an upper bound on the comoving particle horizon in this context
\begin{align}
	\frac{2}{\left(1+3\omega\right)H_0}a^{\left(1+3\omega\right)/2}
	=\frac{2}{1+3\omega}\left(aH\right)^{-1}\, ,
\end{align}
where we have made use of \eqref{friedmann}.
Since the magnitude of the comoving particle horizon is $\sim\left(aH\right)^{-1}$ at most, we can then conclude that perturbations outside the comoving Hubble sphere will be causally disconnected.
In that case we should not observe coherent fluctuations on super-horizon scales, so that the power spectrum of these perturbations will look like Gaussian noise for $k_\lambda<<2\pi aH$.

Needless to say, this is not what we observe in the CMB; the anisotropies are coherent on angular scales greater than 1 degree, which corresponds to the scale of the horizon at the time of CMB formation \cite{Lineweaver:2003ie}.
Furthermore, the observation of acoustic peaks in the CMB power spectrum betray the presence of cosmological perturbations generated long before horizon crossing \cite{Hu:1996vq}.

\begin{figure}[h!]
  \centering
    \includegraphics[width=0.7\textwidth]{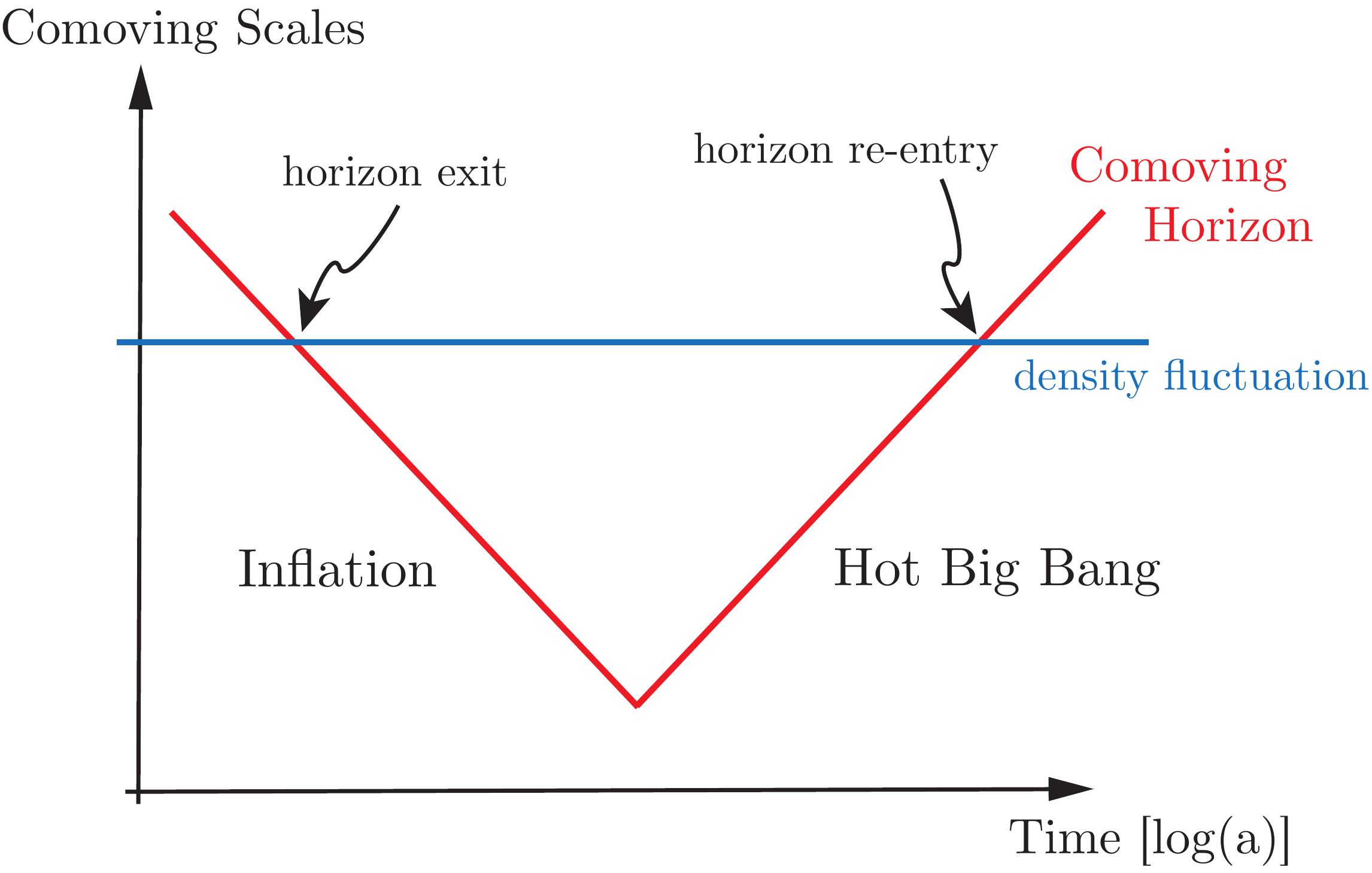}
    	\caption{
	Diagram illustrating the origin of cosmological perturbations. 
	Fluctuations produced during inflation cross the horizon and freeze out ${}^1$, before later reentering to influence cosmological observables.
	As in the previous figure, we can again identify on purely geometrical grounds that we approximately require at least equal amounts of inflation in both the inflationary and Hot Big Bang epochs.
	Figure reproduced from \cite{Baumann:2009ds}.}
	\label{fig:summary}
\end{figure}

It is however clear from Fig \ref{fig:summary} that, as with the Horizon problem, this may be addressed via a period where $\left(aH\right)^{-1}$ decreases.
In this case, causally correlated fluctuations produced in the early universe first exit the horizon, whereupon they no longer evolve \footnote{The non-evolution of perturbations on superhorizon scales is a non-trivial assertion, which we will not prove here. A full derivation can be found in \cite{Weinberg:2008zzc}}.
At some later stage they reenter our Hubble sphere, whereupon they may source the inhomogeneities we ultimately observe today.
Given however the nature of inflation, there is an important nontrivial aspect to this.
At this stage we may make some rather qualitative remarks in this regard, which will be made quantitative later.

To recapitulate, perhaps the most striking feature of the inflationary hypothesis is that it provides an elegant mechanism for generating the initial inhomogeneities in the early universe, from which all the structures we observe in the universe originate from.

However, any sufficient amount of inflation dilutes conventional matter and energy away to nothingness, erasing whatever pre-inflationary conditions existed.
All that can then remain are the intrinsic quantum fluctuations of the vacuum. 
Considering in particular the familiar example of spontaneous vacuum pair production, during rapid spacetime expansion we expect that any pairs produced could become spacelike separated, whereupon, unable to recombine, they must then be thought of as real particles.
The energy debt associated with this process is paid by whatever mechanism is driving inflation.
Quantum fluctuations can thus become real fluctuations via the process of inflation, ultimately to be written across the entire universe.

At no extra cost, this then addresses the implicit issue of initial conditions in the standard cosmology; in inflationary cosmology they are provided by the quantum fluctuations of the vacuum, which needless to say have no preceding cause.
There may remain however an initial condition problem of the inflationary mechanism itself, as we will discuss in the following section.

\subsection{Monopoles}

A further concern, especially from the perspective of ultraviolet (UV) physics, is that of topological relics. 
Monopoles and other exotica are fairly generic predictions of Grand Unified Theories (GUTs), where they arise during symmetry-breaking phase transitions \cite{Vilenkin:1984ib}.
If produced, these relics could carry a significant contribution to the total energy density, and thus ultimately overclose the universe \cite{Preskill:1979zi}.
It is then necessary, if we are to trust GUT physics, to find a rationale for the apparent absence of such a generic feature.

Inflation provides exactly such an argument, under the assumption that any cosmologically dangerous phase transitions occur prior to the end of inflation.
In that case, the energy density of any relics would be rapidly diluted by exponential cosmic expansion.
This also gives a plausible explanation for the non-observation to date of such species; after sufficient inflation we could expect only $\mathcal{O}\left(1\right)$ or less per observable universe.

\section{Inflationary theory}\label{sec: Inflationary theory}

\subsection{Duration}

Having established the utility of an early period of cosmic expansion, it is now natural to consider the requisite characteristics of such an epoch.
A primary consideration is the duration of inflation.

Given the homogeneity and isotropy of the CMB it is first of all necessary that the pre-inflationary Hubble sphere exceeds that of the present day, so that $\left(a_{\rm pre}H_{\rm pre}\right)^{-1}>\left(a_0H_0\right)^{-1}$.
Assuming again radiation domination since the end of inflation, in which case $H\propto a^{-2}$, we can estimate the total amount of post-inflationary expansion as
\begin{align}
	\frac{a_0H_0}{a_{\rm post}H_{\rm post}}
	\sim\frac{a_0}{a_{\rm post}}\left(\frac{a_{\rm post}}{a_0}\right)^2
	=\frac{a_{\rm post}}{a_0}\, .
\end{align}
Since the energy of radiation is proportional to wavelength, we can conclude that $E\propto a^{-1}$ and identify $a_{\rm post}/a_0$ with $T_0/T_{\rm post}$, where $T_{{\rm post}}$ and $T_{0}$ are respectively the temperatures of the universe at the end of inflation, and of the CMB today, $10^{-3}$ eV \cite{Planck:2015xua}.

Although the former is unknown, with an estimate of $T_{{\rm post}}\sim 10^{15}$ GeV we arrive at 
\begin{align}
	\left(a_{\rm pre}H_{\rm pre}\right)^{-1}>\left(a_0H_0\right)^{-1} \sim 10^{27} \left(a_{\rm post}H_{\rm post}\right)^{-1}\, ,
\end{align}
corresponding to at least a factor $10^{27}$ decrease in the Hubble radius during inflation.
Given an approximately constant $H$ during inflation, which we will justify in the next section, $H_{\rm pre}\sim H_{\rm post}$ and we then have
\begin{align}\label{e-folds}
	\ln\left(a_{\rm post}/a_{\rm pre}\right)> 62\, .
\end{align}
We therefore require a minimum amount of inflation on the order of 60 e-folds.
Equivalently, as one may observe from the simple geometry of Figures \ref{fig:inflation} and \ref{fig:summary}, we require at least an equal amount of expansion pre and post-recombination.

Any further inflation beyond this is unobservable to us, unless the size of a causally connected patch happens to be less than the ultimate size of our comoving particle horizon.
In that instance, we would expect the presently homogeneous CMB to develop inhomogeneities at some point in the future, once the size of our comoving particle horizon exceeds that of the pre-inflationary patch the CMB originated from.

\subsection{Hubble slow roll}

Having demonstrated the necessary amount of inflation we require, we may now consider conditions on $H$ such that this can be achieved in practice.
Rephrasing the condition \eqref{decrease}, we have
\begin{align}\label{epsilon definition}
	\frac{d}{dt}\left(aH\right)^{-1}
	=-\frac{a\dot H+\dot a H}{\left(a H\right)^2}
	=\frac{\epsilon-1}{a}, \quad 
	\epsilon\equiv-\frac{\dot H}{H^2}\,,
\end{align}
so that for a decreasing Hubble radius we require $0<\epsilon<1$.

As $H$ evolves to allow $\left(aH\right)^{-1}$ to decrease, $\epsilon$ also varies.
To ensure that the inflationary phase has sufficient duration, we then require that the condition $0<\epsilon<1$ remains satisfied. 
Writing
\begin{align}\label{e-fold definition}
	dN\equiv d\ln a=Hdt\, ,
\end{align}
for $N$ the number of e-folds; each e-fold being an increase in length scale by a factor $e$, we may define by analogy the fractional change in $\epsilon$ per e-fold via
\begin{align}\label{eta definition}
	\eta\equiv\frac{d\ln\epsilon}{dN}
	= \frac{\dot\epsilon}{H\epsilon}\, .
\end{align}
For $|\eta|<<1$, the fractional change in $\epsilon$ per e-fold is small.

To summarise these conditions for later use, we require
\begin{align}	\label{slow roll}
	0<\epsilon<1\,,\quad
	|\eta|<<1\,.
\end{align}
These are known as the Hubble slow roll conditions. As an aside, they justify the assumption made in \eqref{e-folds} that $H$ should be approximately constant during inflation, as this is equivalent to $\epsilon<<1$.

Given the requirements we have established on the evolution of $H$ in order to realise suitable inflation, we now require a mechanism to enact this in practice.

\subsection{The inflaton}

As we have seen earlier, driving inflation ultimately requires violating the SEC in some sense, so that $1+3\omega<0$.
One simple possibility is a cosmological constant, in which case $p=-\rho$ and therefore $\omega=-1$.
Ultimately this cannot work since a cosmological constant will never decay in order for inflation to end, but it does point toward a possible way forward.

A time-dependent cosmological constant may in fact be engineered via the time evolution of a scalar field, with suitable potential.
This can yield a large cosmological constant which ultimately decreases as the field evolves to a minimum of the potential.
Needless to say, a single scalar field evolving in this fashion is only the simplest realisation of this concept.
There exist a number of generalisations which, for reasons of brevity, we will not detail.

The general action for a scalar $\phi$ in a gravitational background is
\begin{align}\label{inflaton action}
	S=\int d^4x\sqrt{-g}\left(\frac{1}{2\kappa^2}R-\frac{1}{2}g^{\mu\nu}\partial_\mu\phi\partial_\nu\phi-V\left(\phi\right)\right)\,,
\end{align}
where $V$ is an unspecified potential and $\kappa^2=8\pi G=M_P^{-2}$.
Non-minimal couplings between $\phi$ and $R$ are possible, but these may be eliminated via field redefinitions.

The field $\phi$, commonly referred to as the inflaton, has stress-energy tensor
\begin{align}
	T_{\mu\nu}=\partial_\mu\phi\partial_\nu\phi-g_{\mu\nu}\left(\frac{1}{2}g_{\alpha\beta}\partial^\alpha\phi\partial^\beta\phi-V\left(\phi\right)\right)\,.
\end{align}
Given that consistency with an FRW background mandates that $\phi$ only be a function of $t$, we can extract a density $T^0{}_0=\rho$ and a pressure $T^i{}_j=-p\delta^i{}_j$ as
\begin{align}\label{rho and P}
	\rho=\frac{1}{2}\dot\phi^2+V\left(\phi\right), \quad
	p=\frac{1}{2}\dot\phi^2-V\left(\phi\right)\,,
\end{align}
where an overdot denotes differentiation with respect to $t$.
Revisiting the condition $1+3\omega<0$, where $\omega\equiv\rho/p$, we can see that $V\left(\phi\right)>>\dot\phi^2/2$ is a sufficient condition for inflation to occur.

With the relations \eqref{rho and P} and $k=0$, the Friedmann equation \eqref{friedmann 2} can be written
\begin{align}\label{friedmann 3}
	H^2=\frac{\rho}{3M_P^2}
	=\frac{1}{3M_P^2}\left(\frac{1}{2}\dot\phi^2+V\right)\, .
\end{align}
Taking a time derivative and making use of the FRW continuity relation $\dot\rho=-3H\left(\rho+p\right)$ gives the relation
\begin{align}
	\dot H=\frac{\dot\rho}{6HM_P^2}
	=-\frac{\rho+p}{2M_P^2}
	=-\frac{1}{2}\frac{\dot\phi^2}{M_P^2}\, ,
\end{align}
which can be substituted back into the time derivative of \eqref{friedmann 3} to arrive at the Klein-Gordon equation
\begin{align}\label{klein gordon}
	\ddot\phi+3H\dot\phi=-V'\, ,
\end{align}
where a prime indicates a derivative with respect to $\phi$.

Revisiting the conditions \eqref{slow roll} we may now reinterpret them as conditions on $\phi$, where
\begin{align}\label{slow roll phi}
	\epsilon&\equiv-\frac{\dot H}{H^2}
	=\frac{1}{2}\frac{\dot\phi^2}{M_P^2H^2}\,,\nonumber\\
	\eta&\equiv\frac{\dot\epsilon}{H\epsilon}
	=2\frac{\ddot\phi}{H\dot\phi}+2\epsilon=2\left(\epsilon-\delta\right)\,,
\end{align}
and we have defined the dimensionless acceleration per Hubble time 
\begin{align}
	\delta\equiv-\frac{\ddot\phi}{H\dot\phi}\,.
\end{align}
\subsection{Slow roll}
In the instance where the conditions \eqref{slow roll} are satisfied, commonly known as the slow roll approximation, we may derive simplified expressions for the various quantities outlined in the previous section. 

Noting firstly that $\epsilon <<1$ implies via \eqref{friedmann 3} and \eqref{slow roll phi} that $\dot\phi^2<<V$, the Friedmann equation \eqref{friedmann 3} becomes
\begin{align}
	H^2\approx\frac{V}{3M_P^2}\,.
\end{align}
Similarly, additionally requiring that $|\eta|<<1$ mandates that $|\ddot\phi /H\dot\phi |<<1$, so that the Klein-Gordon equation \eqref{klein gordon} becomes
\begin{align}
	3H\dot\phi\approx-V'\,. 
\end{align}
Inserting these relations into \eqref{slow roll phi} yields
\begin{align}\label{slow roll parameters}
	\epsilon&=\frac{1}{2}\frac{\dot\phi^2}{M_P^2H^2}
	\approx\frac{M_P^2}{2}\left(\frac{V'}{V}\right)^2\equiv\epsilon_\phi\,, \nonumber\\
	\delta+\epsilon&=-\frac{\ddot\phi}{H\dot\phi}-\frac{\dot H}{H^2}
	\approx M_P^2\frac{V''}{V}\equiv\eta_\phi\,.
\end{align}

\begin{figure}[h!]
  \centering
    \includegraphics[width=0.8\textwidth]{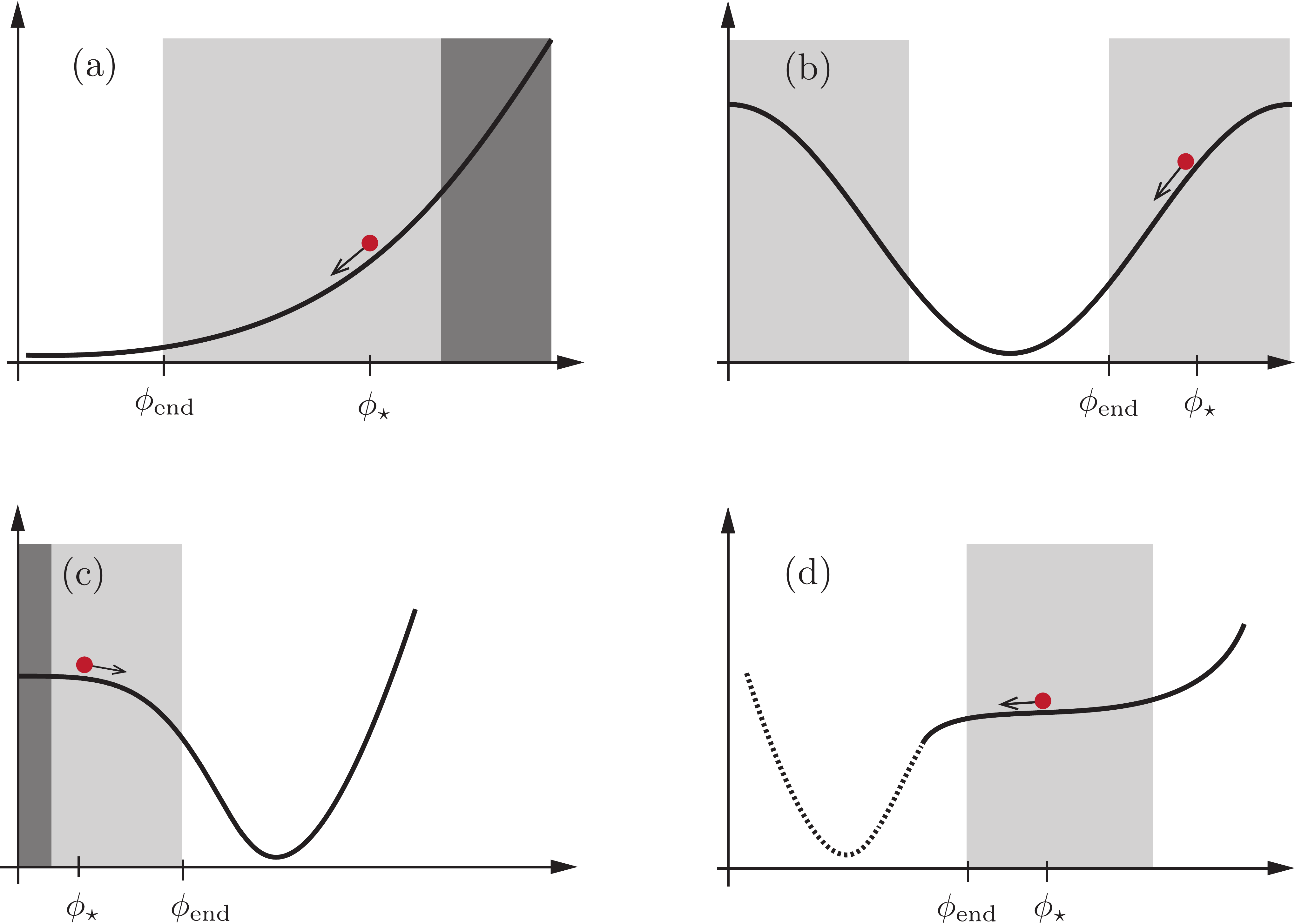}
 	\caption{
	Examples (not to scale) of potentials satisfying the conditions \eqref{slow roll 2}. 
	Grey delimits regions where slow roll can occur, with darker areas indicating where we expect some regions of the universe to inflate eternally.
	Figure reproduced from \cite{Baumann:2014nda}.}
	\label{fig:SRexamples}
\end{figure}

To distinguish $\{\epsilon_\phi,\eta_\phi\}$ from $\{\epsilon,\eta\}$, the former are generally referred to as the slow roll parameters, whilst the latter are the Hubble slow roll parameters.
The Hubble slow roll conditions \eqref{slow roll} are implied by the slow roll conditions
\begin{align}	\label{slow roll 2}
	0<\epsilon_\phi<1\,,\quad
	|\eta_\phi|<<1\,,
\end{align}
which can broadly be interpreted as requiring that the speed and acceleration of the inflaton should be small relative to Hubble scales.
Needless to say, as demonstrated in Figure \ref{fig:SRexamples}, there is in general no unique choice of inflaton potential satisfying these conditions.

The number of e-folds of inflation may be similarly approximated in this regime. 
Given $dN\equiv d\ln a$ we may compute
\begin{align}
	N\equiv\int_{a_*}^{a_{\rm end}}d\ln a
	=\int_{t_*}^{t_{\rm end}} H dt
	=\int_{\phi_*}^{\phi_{\rm end}} \frac{H}{\dot\phi} d\phi
	\approx \int_{\phi_*}^{\phi_{\rm end}} \frac{d\phi}{\sqrt{2\epsilon_\phi M_P^2}}\, ,
\end{align}
where $\phi_*$ and $\phi_{\rm end}$ are the boundaries in field space of an interval satisfying \eqref{slow roll 2}, and we have made use of 
\begin{align}
	 \frac{H}{\dot\phi}\approx -\frac{3H^2}{V'}
	 \approx -\frac{V}{V'M_P^2}
	 =\frac{1}{\sqrt{2\epsilon_\phi M_P^2}}\,.
\end{align}

Despite the successes of the inflationary paradigm in addressing the issue of initial conditions in the standard cosmology, we can now see that, at least in the simplest instance of scalar field inflation one may argue that the problem has simply been transplanted to that of the initial conditions for the inflaton and flatness of the inflationary potential.
If we consider the flatness of space to be less fundamental than the flatness of an inflaton potential, then this is arguably progress.
Indeed, it is long established that supersymmetry, a conjectural fundamental symmetry, can lead naturally to flat scalar potentials which would then ensure this spatial flatness \cite{Ellis:1982ed}.
However, in the absence of a concrete model for the inflaton, there is little more to say at this point.

\section{Primordial perturbations}
\label{sec: Primordial perturbations}

As qualitatively outlined already, inflation can provide the primordial seeds required to explain the inhomogeneities present in the universe.
In this section we will make these notions explicit. 

Our starting point is a decomposition of the inflaton and metric of the previous section into background and fluctuation terms
\begin{align}
	\phi\left(t,\boldsymbol{x}\right)\to\phi\left(t\right)+\delta\phi\left(t,\boldsymbol{x}\right)\, ,\quad
	g_{\mu\nu}\left(t,\boldsymbol{x}\right)\to g_{\mu\nu}\left(t\right)+\delta g_{\mu\nu}\left(t,\boldsymbol{x}\right)\, .
\end{align}
The task in hand is now to track the evolution of these perturbations during the inflationary process.

We can decompose further into scalar, tensor, and vector components, which to first order are uncoupled.
The vector components can additionally be neglected in the present context as they contribute only decaying solutions \cite{Baumann:2009ds}.

As previously outlined, the nature of inflation is such that we expect the effects of quantum mechanics to be significant.
After deriving the classical behaviour of the scalar and tensor perturbations, we will then quantise appropriately to see this first hand.

Since there is in general no unique way to define these perturbations, it is necessary to gauge fix to eliminate non-physical degrees of freedom.
A priori we have one scalar field perturbation $\delta\phi$ and the four metric perturbations $\delta g_{00}$, $\delta g_{0i}$, $\delta g_{ii}$ and $\delta g_{ij}$. 
Two of these modes are removed by the Einstein constraint equations, and two are associated with gauge invariances $t\to t+\delta t$ and $x_i\to x_i+\delta x_i$, leaving only one physical mode.

A convenient gauge choice is comoving gauge, for which the inflaton perturbation $\delta\phi$ vanishes.
In this gauge the gravitational sector perturbation $\delta g_{ij}$ also has the relatively straightforward form 
\begin{align} 
	\delta g_{ij}=a^2\left(1-2\zeta\right)\delta_{ij}+a^2\overline{h}_{ij}^\perp\, ,
	\label{eq: delta g}
\end{align}
where $\overline{h}_{ij}^\perp$ is transverse traceless.
To understand the dynamics of $\overline{h}_{ij}^\perp$ and $\zeta$, we firstly need to compute their respective field equations. 

An important result which we will quote without proof is that on superhorizon scales, $\zeta$ is constant \cite{Weinberg:2008zzc}. 
This is crucial if we are to relate the inflationary formalism to observations, as it ensures that the character of primordial scalar perturbations is preserved until horizon recrossing irrespective of whatever physics occurs in the intervening period. 

\subsection{Tensor perturbations} 

Perhaps counterintuitively, it can be more straightforward to compute the behaviour of the tensor rather than scalar perturbations. 
Substitution of \eqref{eq: delta g} into the Einstein-Hilbert action gives, to second order in $\overline{h}_{ij}^\perp$
\begin{align}\label{tensor action}
	S=\frac{1}{8\kappa^2}\int d^4x\, a^3\left(\left(\dot {\overline{h}}_{ij}^{\perp}\right)^2-\frac{1}{a^2}\left(\nabla \overline{h}_{ij}^\perp\right)^2\right)\, .
\end{align}
As we will ultimately only be concerned with Gaussian fluctuations, which are characterised by their two point function, we needn't go beyond quadratic order.

Inserting the Fourier representation of a transverse traceless tensor
\begin{align}
	\overline{h}_{ij}^\perp\left(t,\boldsymbol{x}\right)=\int\frac{d^3\boldsymbol{k}}{\left(2\pi\right)^{3/2}}\sum_{\gamma=+,\times}\epsilon_{ij}^\gamma h_{\boldsymbol{k}}^\gamma\left(t\right)e^{-i\boldsymbol{k}\cdot\boldsymbol{x}}\,,\quad
	\epsilon_{ij}^{\gamma'}\epsilon_{ij}^\gamma=2\delta^{\gamma'\gamma}\,, \quad
	\epsilon_{ii}^\gamma=k^i \epsilon_{ij}^\gamma=0\,,
\end{align}
where $h_{\boldsymbol{k}}^\gamma$ represents the two possible helicity eigenstates $\left(+,\times\right)$ of the graviton, \eqref{tensor action} becomes
\begin{align}
	S=\frac{1}{4\kappa^2}\sum_{\gamma=+,\times}\int dt\,d^3\boldsymbol{k} \,a^3\left(\left(\dot h_{\boldsymbol{k}}^\gamma\right)^2-\frac{\boldsymbol{k}^2}{a^2}\left(h_{\boldsymbol{k}}^\gamma\right)^2\right)\,.
\end{align}
Shifting to conformal time and the canonically normalised field via 
\begin{align}\label{canonical tensor}
	dt=a\left(t\right) d\tau\,, \quad
	v_{\boldsymbol{k}}^\gamma\equiv \frac{aM_P}{2} h_{\boldsymbol{k}}^\gamma\, ,
\end{align} 
we then have
\begin{align}
	S=\frac{1}{2}\sum_{\gamma=+,\times}\int d\tau\,d^3\boldsymbol{k}\left(\left(v'{}_{\boldsymbol{k}}^\gamma\right)^2-\left(\boldsymbol{k}^2-\frac{a''}{a}\right)\left(v_{\boldsymbol{k}}^\gamma\right)^2\right)\,,
	\label{eq: tensor canonical action}
\end{align}
where primes indicate derivatives with respect to conformal time.

\subsection{Scalar perturbations}

To derive the action for $\zeta$ we insert $\delta g_{00}$ and $\delta g_{0i}$ into the inflaton action \eqref{inflaton action} and expand to quadratic order in $\zeta$, yielding
\begin{align}
	S=\frac{1}{2}\int d^4x\,a^3\frac{\dot\phi^2}{H^2}\left(\dot\zeta^2-\delta_{ij}\frac{1}{a^2}\partial^i\zeta\partial^j\zeta\right)\,,
\end{align}
where we have made use of the background FRW metric. 
Transforming again to conformal time and the canonically normalised field
\begin{align}\label{canonical normalisation}
	v\equiv z\zeta\,, \quad
	z^2\equiv \frac{a^2\dot\phi^2}{H^2}
	=2a^2M_P^2\epsilon\,,
\end{align}
we arrive at the action
\begin{align}
	S=\int d\tau\,d^3\boldsymbol{x}\left(\left(v'\right)^2+\frac{z''}{z}v^2-\delta_{ij}\partial^i v\partial^j v\right)\,.
\end{align}
Fourier transforming to establish similarity with \eqref{eq: tensor canonical action} yields
\begin{align}\label{scalar action}
	S=\int d\tau\,d^3\boldsymbol{k}\left(\left(v'_{\boldsymbol{k}}\right)^2-\left(k^2-\frac{z''}{z}\right)v_{\boldsymbol{k}}^2\right)\,.
\end{align}

The associated field equation for both the scalar and tensor actions is the \\Mukhanov-Sasaki equation \cite{Mukhanov:2005sc, Sasaki:1986hm}
\begin{align}\label{mukhanov-sasaki}
 	v''_{\boldsymbol{k}}+w^{s,t}_{\boldsymbol{k}}\left(t\right)^2 v_{\boldsymbol{k}}=0\,,
\end{align}
from which we may recognise that for both the scalar fluctuations and for each polarisation of the tensor fluctuations we have a harmonic oscillator with a time dependent mass
\begin{align}
	\omega^s_{\boldsymbol{k}}\left(t\right)^2\equiv\left(\boldsymbol{k}^2-\frac{z''}{z}\right)\,, \quad
	\omega^t_{\boldsymbol{k}}\left(t\right)^2\equiv\left(\boldsymbol{k}^2-\frac{a''}{a}\right)\,.
\end{align}

A general solution to \eqref{mukhanov-sasaki} can be expressed via the mode expansion
\begin{align}\label{mode expansion}
	v_{\boldsymbol{k}}\equiv a_{\boldsymbol{k}}v_k\left(t\right)+a_{\boldsymbol{k}}^\dagger v_k^*\left(t\right)\,,
\end{align}
where $v_k$ and $v_k^*$ are linearly independent solutions to \eqref{mukhanov-sasaki}, dependent only on $|\boldsymbol{k}|$.

\subsection{Quantisation}

To begin the process of quantisation we can promote $v\left(\tau, \boldsymbol{x}\right)$ to an operator and impose the usual equal time commutation relations on $\hat v$ and the conjugate momentum $\hat\pi$
\begin{align}
	\left[\hat v\left(\tau,\boldsymbol{x}\right),\hat\pi\left(\tau,\boldsymbol{y}\right)\right]
	=i\delta\left(\boldsymbol{x}-\boldsymbol{y}\right)\,, \quad
	\left[\hat v\left(\tau,\boldsymbol{x}\right),\hat v\left(\tau,\boldsymbol{y}\right)\right]
	=\left[\hat\pi\left(\tau,\boldsymbol{x}\right),\hat\pi\left(\tau,\boldsymbol{y}\right)\right]
	=0\,.
\end{align}
Fourier transforming gives the appropriate momentum space condition
\begin{align}
	\left[\hat v_{\boldsymbol{k}}\left(\tau\right),\hat\pi_{\boldsymbol{k'}}\left(\tau\right)\right]
	=i\delta\left(\boldsymbol{k}+\boldsymbol{k'}\right)\,,
\end{align}
so that we may substitute \eqref{mode expansion} to find
\begin{align}	
	W\left[v_k,v_k^*\right]\times\left[\hat a_{\boldsymbol{k}},\hat a_{\boldsymbol{k}'}^\dagger\right]=\delta\left(\boldsymbol{k}+\boldsymbol{k'}\right)\,,\quad
	W\left[v_k,v_k^*\right]\equiv v_k'v_k^*-v_k v_k^*{}'\,,
\end{align}
where $W$ is the Wronskian of the mode functions.
Normalising the mode functions $v_k$ so that $W\left[v_k,v_k^*\right]=-i$ then implies that 
\begin{align}
	\left[\hat a_{\boldsymbol{k}},\hat a_{\boldsymbol{k}'}^\dagger\right]=i\delta\left(\boldsymbol{k}+\boldsymbol{k'}\right)\,,
\end{align}
and we can then use these operators to construct states in the usual fashion.

There remains however the thorny issue of the choice of vacuum in a time dependent spacetime.
Specifically, we may rescale $a_{\boldsymbol{k}}$ and $v_k$ simultaneously such that the general solution $v_{\boldsymbol{k}}$ from \eqref{mode expansion} remains unchanged, but the condition defining the vacuum, $\hat a_{\boldsymbol{k}}|0\rangle\equiv0$, is no longer satisfied.
Indeed, in a general background there may not be a unique vacuum. 
However, in the present context we can make use of boundary conditions to fix an appropriate state.

At early times we expect all modes of interest to lie deep inside the horizon, in which case they are not sensitive to the curvature of spacetime, and we expect $\omega_{\boldsymbol{k}}\left(\tau\right)^2\to k^2$.
This is the result familiar from flat space, suggesting that we necessarily solve the Mukhanov-Sasaki equation with the Minkowski initial condition
\begin{align}
	 v_k\left(\tau\right)\bigg|_{\tau\to-\infty}=\frac{1}{\sqrt{2k}}e^{-ik\tau}\,.
\end{align}
This resolves the ambiguity via a preferred set of mode functions
\begin{align}
	v_k=\frac{1}{\sqrt{2k}}e^{-ik\tau}\left(1-\frac{i}{k\tau}\right)\,,
\end{align} 
and thus a unique vacuum; the Bunch-Davies vacuum \cite{Bunch:1978yq}.

\subsection{Power spectra}
With this information, we are now in a position to calculate the power spectrum of zero point fluctuations, $\langle v_{\boldsymbol{k}}v_{\boldsymbol{k'}}\rangle$. 
Inserting \eqref{mode expansion} gives
\begin{align}
	\langle0|\left(\hat a_{\boldsymbol{k}}v_k+a_{\boldsymbol{k}}^\dagger v_k^*\right)\left(\hat a_{\boldsymbol{k'}}v_{k'}+\hat a_{\boldsymbol{k'}}^\dagger v_{k'}^*\right)|0\rangle
	=v_{{k}}v_{{k'}}^*\langle0|[\hat a_{\boldsymbol{k}},\hat a_{\boldsymbol{k'}}^\dagger] |0\rangle
	=|v_k|^2\delta\left(\boldsymbol{k}+\boldsymbol{k'}\right)\,.
\end{align}
This yields the power spectrum $P_v\left(k\right)\equiv |v_k|^2$.
The power spectra for the fields $\zeta$ and $\overline{h}_{ij}^\perp$ can then be obtained via rescaling by their respective canonical normalisation factors.

It should be noted that in perfect de Sitter space $P_\zeta\left(k\right)$ will then be ill defined, as the normalisation factor of $z=2a^2 M_p^2 \epsilon$ from \eqref{canonical normalisation} vanishes.
This is however inconsequential, as perfect de Sitter space has no time dependence and is therefore inappropriate for our purposes. 
The small deviation we require into quasi-de Sitter space to ensure time dependence is parametrised by $\epsilon$.

As previously mentioned, perturbations do not evolve on superhorizon scales. 
This suggests that our object of interest is the power spectrum at horizon crossing, when $k=aH$ and subsequent evolution is halted.
Taking for convenience the superhorizon limit $k\tau\to0$, we find
\begin{align}
	P_v\left(k\right)\equiv |v_k|^2
	\xrightarrow{k\tau\to0}\frac{1}{2k^3}\left(aH\right)^2\,.
\end{align}
In the scalar case the canonical normalisation factor $z=2a^2 M_p^2\epsilon$ gives the dimensionless scalar power spectrum 
\begin{align}
	\Delta_s^2\left(k\right)\equiv \frac{k^3}{2\pi^2}P_s=\frac{1}{8\pi^2}\frac{H^2}{M_P^2}\frac{1}{\epsilon}\bigg|_{k=aH}\,,
\end{align}
whilst \eqref{canonical tensor} analogously yields the tensor equivalent
\begin{align}
	\Delta_t^2\left(k\right)=\frac{2}{\pi^2}\frac{H^2}{M_P^2}\bigg|_{k=aH}\,,
\end{align}
with a factor of two coming from the sum over polarisations.
Notably the tensor mode power spectrum is sensitive only to the ratio $H/M_P$, thus encoding the scale at which inflation occurs.

Since these quantities are in principle time dependent, a useful observable derived therefrom are the spectral indices, which measure the deviation from scale invariance during inflation.
Near a fiducial reference scale $k^*$ we expect the power spectra to have a power law dependence 
\begin{align}
	\Delta_s^2\left(k\right)=A_s\left(\frac{k}{k^*}\right)^{n_s-1}\,,\quad
	\Delta_t^2\left(k\right)=A_t\left(\frac{k}{k^*}\right)^{n_t}\,,
\end{align}
with numerical coefficients $A_s$ and $A_t$, and following the slightly awkward common convention for defining $n_s-1$ and $n_t$. 
We may then isolate the scalar and tensor spectral indices
\begin{align}\label{spectral indices}
	n_s-1\equiv\frac{d\ln \Delta_s^2}{d\ln k}\,,\quad
	n_t\equiv\frac{d\ln \Delta_t^2}{d\ln k}\,,
\end{align}
where the RHS are evaluated at $k^*$.
Inflation predicts that $n_s \sim 1$ and $n_t\sim0$, with the deviation from the exact scale invariance arising because $H$ is only approximately constant during realistic inflation.

To connect \eqref{spectral indices} to our slow roll parameters, we may write
\begin{align}
	\frac{d\ln \Delta_s^2}{d\ln k}
	=\frac{d\ln \Delta_s^2}{dN}\frac{dN}{d\ln k}
	=\left(2\frac{d\ln H}{dN}-\frac{d\ln\epsilon}{dN}\right)\frac{dN}{d\ln k}\,.
\end{align}
By virtue of the definitions $\epsilon\equiv-\dot H/H^2$, $dN\equiv d\ln a=Hdt$ and $\eta\equiv d\ln\epsilon/dN$ the first term in brackets is just $-2\epsilon$ and the latter is $-\eta$. 
Furthermore
\begin{align}
	\frac{d\ln k}{dN}
	\xrightarrow{k=aH}\left(1+\frac{d\ln H}{dN}\right)
	\approx 1-\epsilon\,,
\end{align}
so to first order in the Hubble slow roll parameters we then find
\begin{align}
	n_s-1=-2\epsilon-\eta\,.
\end{align}

Similarly, for the tensor spectral index
\begin{align}
	n_t=\frac{d\ln \Delta_t^2}{d\ln k}
	=\frac{d\ln \Delta_t^2}{dN}\frac{dN}{d\ln k}
	=2\frac{d\ln H}{dN}\frac{dN}{d\ln k}\approx -2\epsilon\,.
\end{align}

A convenient normalisation for the power spectra is expressed via 
\begin{align}
	r\equiv\frac{\Delta_t^2}{\Delta_s^2}
	=16\epsilon\,,
\end{align}
known as the tensor to scalar ratio. 
$r$ gives insight into whether inflationary gravitational fluctuations had sufficient amplitude to be inferred from future CMB observations.

Furthermore, given that $\Delta_s^2$ has been successfully measured, and $\Delta_t^2$ is sensitive to $H^2\sim V$, we may write
\begin{align}
	V^{1/4}\sim\left(\frac{r}{0.01}\right)^{1/4}10^{16}\, \rm{GeV}\,.
\end{align}
A measurement of the value of $r$ would therefore also allow us to infer the characteristic energy scale of inflation.

Whilst the observation of a non-zero scalar spectral index is good evidence in favour of inflation, a spectrum of tensor fluctuations is unavoidably predicted, so their inferred observation would constitute even greater evidence in favour of the inflationary hypothesis.

\subsection{Concluding remarks}

\begin{figure}[h!]
  \centering
    \includegraphics[width=0.8\textwidth]{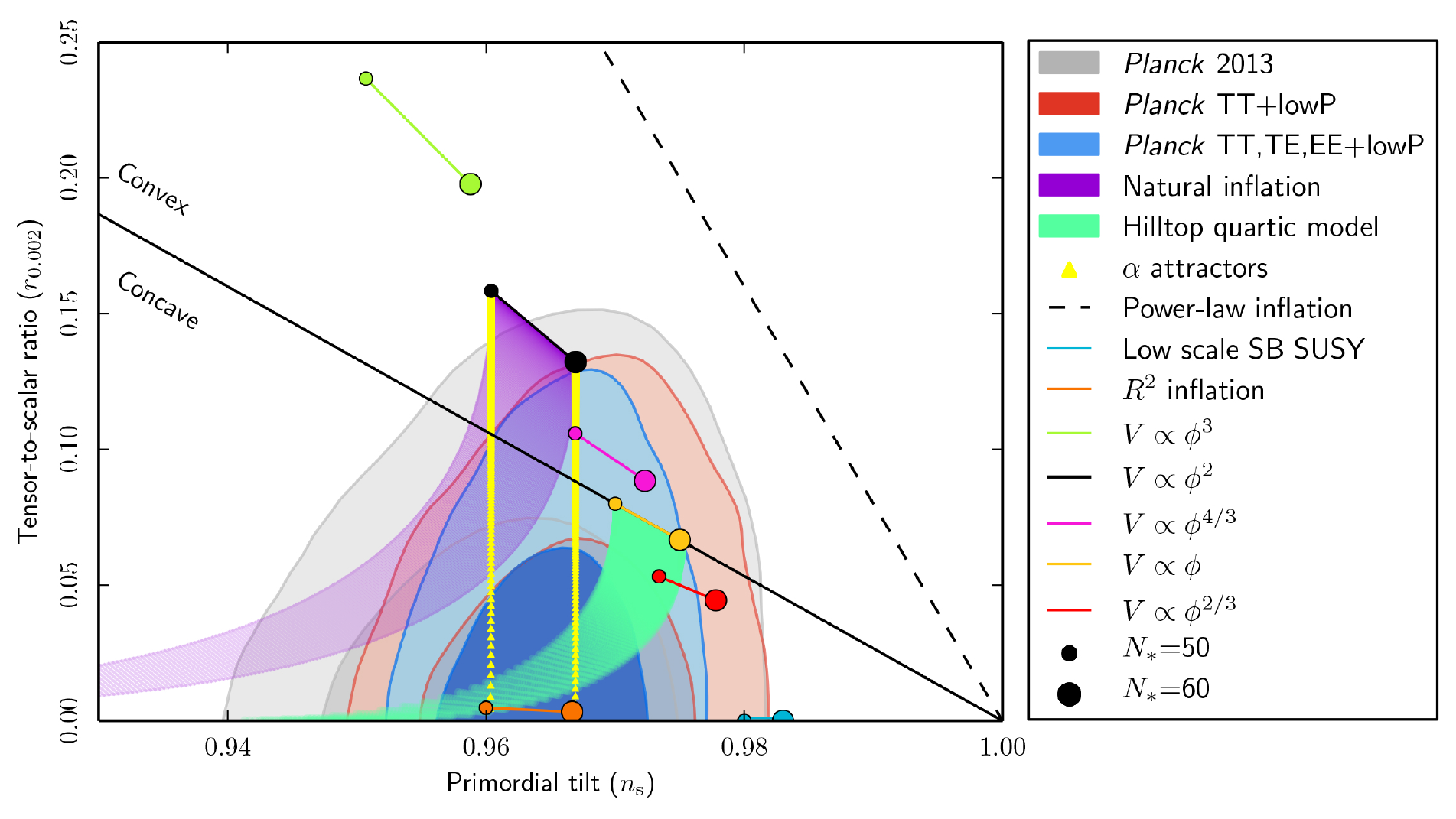}
 	\caption{Planck 2015 68\% and 95\% marginalised confidence levels for $n_s$ and $r$, reproduced from \cite{Ade:2015lrj}.}
	\label{fig:Planck_inflation}
\end{figure}

Having motivated and developed the necessary elements for inflation, we now conclude with the experimental constraints on inflationary physics relevant for the remainder of this thesis.
Primary amongst these is Figure \ref{fig:Planck_inflation}, which gives the acceptable regions $n_s$, $r$ plane, overlaid with the predictions from a number of common inflationary models.
By identifying the slow roll parameters \eqref{slow roll parameters} with the Hubble slow roll parameters, this allows a straightforward evaluation of the inflationary suitability of a given potential $V\left(\phi\right)$.

For completeness, some remarks are also in order about the conclusion of the inflationary phase.
Ultimately the inflationary epoch must not only terminate, but segue smoothly into the standard cosmology. 
Since inflation vastly dilutes conventional energy and matter, whatever process concludes the inflationary phase must then reheat the universe back to pre-inflationary temperatures.

Given this dilution, during inflation most of the total energy of the post Big Bang universe is stored in the inflaton potential.
Once the slow roll conditions \eqref{slow roll 2} are no longer satisfied, typically because the inflaton has evolved out of a sufficiently flat region of $V$, it will begin to move quickly on Hubble scales as this potential is converted into kinetic energy.
In particular, upon reaching a minimum of the potential the inflaton will begin to oscillate. 
From there, this kinetic energy is converted into the standard model particles which populate the universe, via the decays of the inflaton.
This era, known as reheating, is accordingly characterised in part by the couplings of the inflaton to these particles.
Once these decay products thermalise, the standard Hot Big Bang epoch can begin.

%% file: Chapters/Supersymmetry.tex
It goes without saying that symmetries are central to modern physics.
Principally, these take the form of either internal or spacetime symmetries.
It is however natural to ask if these categories are mutually exclusive; are symmetries possible which encapsulate both spacetime and internal degrees of freedom?

In the early days of gauge theory it was precisely this line of thought that led to the Coleman-Mandula theorem, which demonstrates that, under some reasonable assumptions, any theory possessing such an exotic symmetry must be internally inconsistent, or trivial \cite{Coleman:1967ad}.
This seemingly conclusive obstruction rests however on the assumption of only `bosonic' symmetries; those whose generators satisfy commutation relations.

Supersymmetry is then the primary circumvention of this result, having generators instead satisfying anti-commutation relations.
To preserve the usual \\(anti)commutation relations for bosons and fermions, the action of this symmetry must then exchange integer and half-integer spin states.
It should be noted however that the net result is still a purely spacetime symmetry, and furthermore, the supersymmetric generalisation of the Poincar\'e algebra is the only known `reasonable' circumvention of the result of Coleman-Mandula; a statement codified in the Haag-Lopuszanski-Sohnius theorem \cite{Haag:1974qh}.

It can then be asserted that supersymmetry occupies a privileged position in the context of `fundamental' extensions to known physics.
As a purely theoretical endeavour, this is arguably interesting in its own right.
However, in the context of particle physics there a number of known problems which may be addressed to a greater or lesser extent by supersymmetric approaches.
\begin{itemize}
	\item The measured value of the Higgs mass requires an extremely unnatural degree of fine tuning in the Standard Model \cite{Aad:2015zhl}.
	By imposing a symmetry between bosons and fermions, supersymmetry can cancel the quadratic divergences driving this effect, allowing the value of the Higgs mass to be technically natural.
	\item Whilst the notion of unification in physics has been fruitful thus far, the renormalisation group extrapolation of the Standard Model gauge couplings to high energies seems to preclude the ultimate unification of the electromagnetic, weak and strong forces.
	In the presence of supersymmetry however, these couplings can in fact unify into a single coupling at high energies \cite{Schwarz:1998ny}
	\footnote{Going beyond this simple extrapolation by incorporating the contribution of the massive states associated to the unification threshold may of course modify this conclusion.}.
	\item The anomalous rotation curves of galaxies suggest that the majority of their mass comes from particles which are `dark'
	\footnote{Although `transparent' would in reality be a more accurate characterisation.}, in that they do not interact electromagnetically and are not part of the Standard Model \cite{Bertone:2004pz}. 	
	Supersymmetry naturally provides a plethora of new particle species, as each observed particle must possess a superpartner, some of which could fulfil such a role.
	\item The Standard Model has thus far excelled in describing particle physics, albeit with the glaring omission of gravity.
	As we will see in what follows, local supersymmetry not only allows gravity to be incorporated into particle physics in an elegant fashion, but necessarily requires that it be present. 
	\item In the direction of more theoretical endeavours, there are further reasons to study supersymmetry, and particularly local supersymmetry.
Primarily, it is known that supergravity theories can constitute low-energy limits of corresponding string and M theories, which are thought to provide consistent paths to the quantisation of gravity in concert with known facets of particle physics \cite{Freedman:2012zz}.
\end{itemize}

With these motivations established, we may now proceed to outline the quantitative aspects of the theory relevant for later discussion.
These comprise the basics of global and local supersymmetry, in sections \ref{sec: global supersymmetry} and \ref{sec: local supersymmetry} respectively, the mechanics of supersymmetry breaking and the super-Higgs effect in \ref{sec: supersymmetry breaking}, and some of the consequences of supersymmetry for inflationary physics in \ref{sec: inflation in supergravity}.

\section{Global supersymmetry} 
\label{sec: global supersymmetry}

Broadly speaking we can understand supersymmetry as a symmetry generated by some operator $Q$ which exchanges bosons and fermions via
\begin{align}
	Q|B\rangle\sim|F\rangle, \quad 
	Q|F\rangle\sim|B\rangle\;,
	\label{Q action}
\end{align}	
grouping bosons and fermions into supermultiplets under the action of $Q$.

Given the usual Poincar\'e algebra, the action of Q can be incorporated via addition of the relations
\begin{align} 
	\left\{Q_\alpha,\overline Q_\beta\right\}=-\frac{1}{2}\left(\gamma^\mu\right){}_{\alpha\beta}P_\mu\,, \quad
	\left[Q_\alpha, P_\mu\right]=0\,, \quad
	\left[Q_\alpha,M^{\mu\nu}\right]=-\frac{1}{2}\left(\gamma^{\mu\nu}\right){}_{\alpha}{}^\beta Q_\beta\,,
	\label{super-poincare}
\end{align}
where $P_\mu$ and $M_{\mu\nu}$ are respectively the generators of translations, boosts and rotations, $\alpha,\beta$ are spinor indices, and $\mu,\nu$ are spacetime indices. 
This defines the Super-Poincar\'e algebra.

A single $Q_\alpha$ generates $\mathcal{N}=1$ supersymmetry, which may be extended via additional generators up to a presumed upper limit of $\mathcal{N}=8$, arising from the assumption that there can be no massless particles with helicity $|\lambda|>2$.
However, considerations of chirality render extended supersymmetries phenomenologically uninteresting in the present context \footnote{This is a consequence of all renormalisable $\mathcal{N}>2$ theories having $\lambda=\pm1$ particles transforming in the adjoint representation, which is non-chiral.
The $\lambda=\pm1/2$ particles in the same multiplets would transform similarly, whereas the Standard Model contains only chiral matter.
$\mathcal{N}=2$ multiplets lacking $\lambda=\pm1$ particles can evade this argument, however these multiplets must then consist of $\lambda=1/2$ and $\lambda=-1/2$.
These multiplets are not chiral either, as they cannot be left/right asymmetric.}.
As such, the most viable candidate for a realisation of supersymmetry in nature is of the $\mathcal{N}=1$ variety, broken at some sufficiently low scale to ensure naturalness of the Higgs mass \footnote{Extended supersymmetries may be partially broken, in the sense that an $\mathcal{N}>1$ theory could be broken down to $\mathcal{N}=1$ at some unknown scale, which then subsequently breaks at the requisite $\mathcal{O}$(100 GeV) \cite{Kiritsis:1997ca}.
Alternatively one could engineer the lightness of the Higgs via other methods, such as extra-dimensional scenarios \cite{ArkaniHamed:1998rs}, so that realisations of supersymmetry need not be constrained to break in an `acceptable' fashion.}.

\subsection{The goldstino} \label{sec: The goldstino}
By virtue of relating the supercharges to $P_\mu$, an important consequence of \eqref{super-poincare} is that we can express
\begin{align}
	H\equiv P_0=\frac{1}{4}\sum_\alpha Q^2_\alpha\,.
\end{align}
For a supersymmetric vacuum state $|\Omega\rangle$, $Q_\alpha|\Omega\rangle=0$, and therefore $H|\Omega\rangle=0$.
Global supersymmetry is then spontaneously broken if and only if the vacuum energy is non-vanishing.

This condition may be interpreted as a condition on the expectation value of the energy-momentum tensor $\langle0|T_{\mu\nu}|0\rangle=Eg_{\mu\nu}$, which in a supersymmetric theory can be expressed in terms of the supercurrent $S_{\nu \beta}$ via 
\begin{align}	
	T_{\mu\nu}=\left(\gamma_\mu\right)^{\alpha\beta}\left\{Q_\alpha, S_{\nu \beta}\right\}\,, \quad
	Q_\alpha=\int d^3x\,S_{0\alpha}\,.
\end{align}

The corresponding supercurrent Ward identity is then
\begin{align}	
	\partial^\sigma\langle0|T\left(S_{\sigma \alpha}\left(x\right)\overline{S}_{\nu \beta}\left(0\right)\right)|0\rangle\sim\delta^4\left(x\right)\left(\gamma^\mu\right)_{\alpha\beta}\langle T_{\mu\nu}\rangle\,,
	\label{supercurrent expectation}
\end{align}
with $T$ denoting time-ordering.
Integrating both sides to remove the delta function, positivity of the vacuum energy implies that if supersymmetry is spontaneously broken the integral of the left-hand side (LHS) must be non-vanishing.

Given that as an integrand it is a total divergence this can only be the case in the presence of boundary contributions, in which case $\langle0|T\left(S_{\sigma \alpha}\left(x\right)\overline{S}_{\nu \beta}\left(0\right)\right)|0\rangle$ must then vanish exactly as $1/|x|^3$ for large $|x|$.
On purely dimensional grounds the Fourier transform of this correlator must then fall off as $1/|p|$ for small $|p|$; exactly the asymptotic behaviour associated with the propagation of a single massless fermion.
By analogy with the Goldstone boson associated with the spontaneous breaking of continuous `bosonic' symmetries, this fermion is known as the Goldstino.

\subsection{Auxiliary fields}\label{sec: Auxiliary fields}
A straightforward consequence of \eqref{Q action} is that the fields making up a supermultiplet must have equal numbers of degrees of freedom. 
However, as numbers of bosonic and fermionic degrees of freedom shift differently when going off shell we must necessarily introduce auxiliary fields to ensure that the supersymmetry algebra always closes.

Given the simple example of a multiplet containing a complex scalar field and a spin 1/2 Majorana fermion, we can realise supersymmetry explicitly via
\begin{align} \label{wess-zumino}
	\mathcal{L}=\partial^\mu\phi^*\partial_\mu\phi-m^2\phi^*\phi+\frac{1}{2}\overline{\psi}\left(i\gamma^\mu\partial_\mu-m\right)\psi\,,\quad
	\phi\equiv\frac{1}{\sqrt{2}}\left(A+iB\right)\,,
\end{align}
which is invariant (up to a total divergence) under the transformations
\begin{align}\label{susy transforms}
	\delta A=\overline{\epsilon}\psi\,,\quad
	\delta B=i\overline{\epsilon}\gamma_5\psi\,,\quad
	\delta \psi=-\left(\left(i\gamma^\mu\partial_\mu+m\right)\left(A+iB\gamma_5\right)\right)\epsilon\,,
\end{align}
where $\epsilon$ is an infinitesimal spinor.
In general a complex scalar has two real degrees of freedom, whilst a spin 1/2 Majorana fermion has four. 
The Dirac equation fixes two of these, so that on-shell the number of degrees of freedom match.

Off shell however, they do not.
To remedy this we may introduce two extra degrees of freedom via an auxiliary complex scalar field $F$, replacing $m^2\phi^*\phi$ in \eqref{wess-zumino} with the auxiliary Lagrangian 
\begin{align}
	\mathcal{L}_{\rm{aux}}=F^*F+m\left(F\phi+F^*\phi^*\right)\,,
\end{align}
and modifying the transformation rules \eqref{susy transforms} accordingly.
Applying the equation of motion for $F$ then returns \eqref{wess-zumino}.

We can identify spontaneous supersymmetry breaking via a non-zero vacuum expectation value of the variation under supersymmetry of some field. 
Considerations of Lorentz invariance restrict the variation in question to be that of a fermion, as only scalar vacuum expectation values are admissible.
The fermionic transformation rule \eqref{susy transforms}, modified to incorporate $F$, reads
\begin{align} 
	\delta\psi\sim \left(F+\gamma^\mu\partial_\mu\phi\right)\epsilon\,.
\end{align}
Spontaneous supersymmetry breaking can then be associated with the development of non-zero vacuum expectation value for the auxiliary field $F$. 
The superpartner of this non-vanishing $F$ component is the Goldstino of the previous section.

Generic scalar potentials may also contain another class of auxiliary fields, traditionally labelled $D$, in the presence of gauge symmetries.
Similar logic applies to these fields, so in the interests of economy of presentation we will restrict attention to $F$-type breaking.

With respect to the discussion that follows in later chapters of this thesis, a particularly salient point is that despite being a symmetry between bosons and fermions, vacuum expectation values of elementary (i.e. non-auxiliary) scalar fields cannot break supersymmetry.
This is a consequence of the general criterion that supersymmetry is spontaneously broken if and only if the anticommutator of some operator $X$ with a supercharge is non-vanishing
\begin{align}
	\langle 0| \left\{X, Q\right\}|0 \rangle\neq0\,.
\end{align}

Elementary scalars can never be expressed as $\left\{X, Q\right\}$, so they may develop vacuum expectation values without issue.
This is of course reassuring in that it allows global symmetries to be broken independently of supersymmetry breaking, but also suggestive in that composite scalars need not obey any such constraint.
We will return to this observation later.

\section{Local supersymmetry} 
\label{sec: local supersymmetry}

In analogy with gauge theory, we can of course promote supersymmetry from a global to a local symmetry.
More precisely, we may specialise to theories which are locally invariant under supersymmetry transformations parametrised by spinors which are local functions of spacetime coordinates.

One key consequence is then encoded in 
	\begin{align}
		\left[\overline{\epsilon}_1Q,\overline{\epsilon}_2Q\right]=2 \overline{\epsilon}_1\left(x\right)\gamma^\mu\epsilon_2\left(x\right)P_\mu\,,
	\end{align}
the commutator of supersymmetry transformations.
Successive infinitesimal supersymmetry transformations then can be seen to yield infinitesimal coordinate transformations.
Succinctly, local supersymmetry necessarily mandates gravity.

Furthermore, the converse is also true; any supersymmetric theory incorporating gravity requires that the parameters $\epsilon$ must be local functions of spacetime coordinates.
This is a consequence of the constraint that in a generally covariant scenario there can be no constant spinors, as, like a constant vector field, this would be incompatible with the underlying diffeomorphism symmetry.

\subsection{The gravitino}
Upon making supersymmetry local, we expect new transformations of the form $\delta\left(\dots\right)\sim\partial_\mu\epsilon\left(x\right)$, which suggests we require a field carrying both a spinor (spin 1/2) and a Lorentz (spin 1) index.
This is the spin 3/2 gravitino, the superpartner of the graviton. 

To understand the gravitino in detail, we may firstly classify the representations of the Lorentz group via the double cover SU$\left(2\right)\times\,$SU$\left(2\right)$.
They are
\begin{itemize}
	\item $\left(0,0\right)$ representation, corresponding to scalars.
	\item $\left(\frac{1}{2},0\right)\oplus\left(0,\frac{1}{2}\right)$ representation, or either of the two irreducible parts taken individually.
	This corresponds to spin 1/2 Dirac or Majorana fermions, or in the latter case, Weyl fermions.
	\item $\left(\frac{1}{2},\frac{1}{2}\right)$ representation, corresponding to vectors.
	These would be, for example, the familiar gauge bosons of the Standard Model.
	\item $\left(1,0\right)\oplus\left(0,1\right)$ representation, or either of the two irreducible parts taken individually.
	This corresponds formerly to two-form fields such as the electromagnetic field strength tensor, and latterly to two-forms satisfying an (anti) self-duality condition.
	\item $\left(1,1\right)\oplus\left(0,0\right)$ representation, corresponding to tensors. 
	This gives a rank two traceless symmetric tensor and a scalar, corresponding to the graviton and its trace.
\end{itemize}

To form a spin 3/2 representation we can combine the spin 1/2 and spin 1 parts
\begin{align}
	\left(\left(\frac{1}{2},0\right)\oplus\left(0,\frac{1}{2}\right)\right)\otimes\left(\frac{1}{2},\frac{1}{2}\right)
	=\left(\left(\frac{1}{2},0\right)\otimes\left(\frac{1}{2},\frac{1}{2}\right)\right)\oplus \left(\left(0,\frac{1}{2}\right)\otimes\left(\frac{1}{2},\frac{1}{2}\right)\right)\,,
\end{align}
and make use of the SU$\left(2\right)$ Clebsch-Gordan decomposition
\begin{align}
	V^n\otimes V^m=V^{n+m}\oplus V^{n+m-1}\oplus\dots\oplus V^{|n-m|}\,,
\end{align}
where $V^n$ and $V^m$ are two irreducible representations, to arrive at
\begin{align}
	\left(\left(1,\frac{1}{2}\right)\oplus\left(0,\frac{1}{2}\right)\right)\oplus\left(\left(\frac{1}{2},1\right)\oplus\left(\frac{1}{2},0\right)\right)\,.
\end{align}

In addition to the spin 3/2 piece we expect, we inevitably then also have a spin 1/2 part.
This need not be problematic however; in the absence of interactions we expect massless spin 3/2 fields to obey the Rarita-Schwinger equation
\begin{align}
	\gamma^{\mu\nu\rho}\partial_\nu\psi_\rho=0\,, \quad
	\gamma^{\mu\nu\rho}\equiv\gamma^{[\mu}\gamma^\nu\gamma^{\rho]}\,,
\end{align}
which is invariant under $\psi_\mu\to\psi_\mu+\partial_\mu\epsilon$. 
In fixing this gauge symmetry, we can remove exactly this spin 1/2 component.

To perform the degree of freedom counting necessary for supersymmetry, we may firstly note that general solutions of the Dirac equation have eight real components in $D=4$. 
Taking left or right chiral projections yield Weyl spinors, which accordingly satisfy the condition $P_{L/R}\psi=\psi$ and have four real components in $D=4$.
Imposing instead the reality condition $\psi^{\rm{C}}=\psi$, where C denotes complex conjugation, yields Majorana spinors, which also have four real components in $D=4$. 
Since Majorana and Weyl spinors can be used to construct Dirac spinors, we can arguably consider them as `more' fundamental. 

Considering then for simplicity a Majorana vector-spinor, in $D=4$ we expect sixteen real degrees of freedom.
Four of these correspond to local supersymmetry transformations, which are removed by gauge fixing, leaving twelve degrees of freedom off shell.

On the gravitational side, the presence of gravity and fermions necessitates the use of the Cartan formalism of general relativity, whereby, instead of the familiar metric tensor we employ a frame field satisfying the relation
\begin{align}
	g_{\mu\nu}\left(x\right)=e_\mu{}^a\left(x\right)\eta_{ab} e^b{}_\nu\left(x\right)\,,
\end{align}
where roman characters indicate tangent-space indices.
By virtue of transforming locally under the Lorentz group, this field furnishes the spinor representations we require.
Since the frame field is not symmetric under index exchange, it has more degrees of freedom than the $g_{\mu\nu}$.
These are however cancelled by the local Lorentz symmetry it additionally provides, so that the number of physical degrees of freedom remains unchanged. 

Specifically, the frame field also carries sixteen degrees of freedom, four of which correspond to general coordinate and six to local Lorentz transformations.
Gauge fixing then leaves six degrees of freedom off shell.

Given the mismatch in off shell degrees of freedom, we again require auxiliary fields to ensure the supersymmetry algebra always closes.
The choice of these fields isn't unique however, in the simplest $D=4$ supergravity common choices are the `old' and `new' minimal sets.
These differ in their use, respectively, of a complex scalar or an antisymmetric gauge invariant tensor.
Classically these choices are equivalent, however it is worth noting that the theories may differ at the quantum level 
\footnote{For example, in their anomaly coefficients \cite{Mangano:1988kf}.}.

\subsection{Cartan formalism}

At the most basic level the $\mathcal{N}=1$ on-shell supergravity multiplet consists of the frame field $e_\mu{}^a$ carrying the gravitational degrees of freedom, along with a Majorana  vector-spinor $\psi_\mu$, the gravitino. 
In the `old minimal' formulation, these are supplemented respectively by vector, scalar and pseudo-scalar auxiliary fields $A_\mu$, $S$, and $P$.

The frame field firstly allows us to interconvert between spacetime and local Lorentz indices via
\begin{align}
	x^a=e^a{}_\mu x^\mu\,.
\end{align}
More generally however, it allows all the familiar machinery of Riemannian geometry to be re-expressed without reference to $g_{\mu\nu}$.

This is achieved by treating the local Lorentz symmetry they introduce analogously to a non-Abelian gauge symmetry, with a spin connection $\omega_\mu{}^{ab}$ substituting the usual Yang-Mills connection $A_\mu$. 
This allows an interpretation of the spin connection as the gauge field of local Lorentz transformations, and furthermore provides a spinor-compatible formulation of physics in curved spacetimes.

To demonstrate the necessity of this approach, we may consider the two form \footnote{With wedge product $dx^\mu\wedge dx^\nu\equiv dx^\mu\otimes dx^\nu-dx^\nu\otimes dx^\mu$.} 
\begin{align}
	de^a=\frac{1}{2}\left(\partial_\mu e^a{}_\nu-\partial_\nu e^a{}_\mu\right)dx^\mu\wedge dx^\nu\,.
\end{align} 
Given the usual symmetry properties of Christoffel symbols this is equivalent to an antisymmetrised covariant derivative, so it must transform as a $(0,2)$ tensor under general coordinate transformations. 
It does not however transform as a local Lorentz vector, given that
\begin{align}
	de'^a=d\left(\left(\Lambda^{-1}\right)^a{}_b e^b\right)
	=d\left(\Lambda^{-1}\right)^a{}_b \wedge e^b+\left(\Lambda^{-1}\right)^a{}_b de^b\,.
\end{align}
To remedy this, we need to account for the connection $\omega_\mu{}^{ab}$ in the usual fashion.
This is achieved by modifying the derivative into
\begin{align}
	de^a+\omega_a{}^b\wedge e_b\equiv T^a\,,
\end{align}
where $T^a$ is known as the torsion two-form. 
Local Lorentz covariance is then preserved if the spin connection transforms as
\begin{align}
	\omega'^{a}{}_{b}=\left(\Lambda^{-1}\right)^a{}_c\, d\Lambda^c{}_b+\left(\Lambda^{-1}\right)^a{}_c \,\omega^c{}_d\Lambda^d{}_b\,.
\end{align}
Needless to say, these are the transformation properties of an $O\left(D-1,1\right)$ Yang-Mills connection.

In general it is convenient to decompose $\omega_{\mu ab}$ into a torsion free connection and a contortion tensor $K_{\mu ab}$ via 
\begin{align}\label{contortion tensor}
	\omega_{\mu ab}=\omega_{\mu ab}\left(e\right)+K_{\mu ab}\,,\quad
	K_{\mu[\nu\rho]}\equiv-\frac{1}{2}\left(T_{[\mu\nu]\rho}-T_{[\nu\rho]\mu}+T_{[\rho\mu]\nu}\right)\,,
\end{align}
where we have made use of a coordinate basis to write $T^\alpha=T_{\mu\nu}{}^\alpha dx^\mu\wedge dx^\nu$.
The torsion-free connection $\omega_{\mu ab}\left(e\right)$ is then the familiar Levi-Civita connection.
In most circumstances the contortion tensor vanishes, however coupling certain types of matter to gravity can render it non-zero.
As we will see, this is precisely what occurs in supergravity.

Furthering the analogy with Yang-Mills theory, the Riemann tensor can then be defined as the spin connection field strength
\begin{align}
	R_{\mu\nu ab}\equiv\partial_\mu \omega_{\nu ab}-\partial_\nu \omega_{\mu ab}+\left[\omega_\mu,\omega_\nu\right]_{ab}\,,
\end{align}
from which we can construct the usual curvature quantities as required.

Noting that under an infinitesimal Lorentz transformation parametrised by $\theta_{ab}$, a fermion $\psi$ transforms as
\begin{align}
	\delta\psi=-\frac{1}{4}\theta_{ab}\gamma^{ab}\psi\,,\quad
	\gamma^{ab}\equiv\gamma^{[a}\gamma^{b]}\,,
\end{align}
we can also define a locally Lorentz covariant fermion derivative
\begin{align}
	D_\mu\psi_\nu\equiv\partial_\mu\psi_\nu+\frac{1}{4}\omega_{\mu ab}\gamma^{ab}\psi_\nu\,,
\end{align}
to be contrasted with the usual `coordinate' covariant derivative
\begin{align}
	\nabla_\mu \psi_\nu\equiv\partial_\mu\psi_\nu-\Gamma^\rho{}_{\mu\nu}\psi_\rho\,.
\end{align}
Whilst the former transforms as a local Lorentz spinor but not a tensor, and the latter a spinor and $(0,2)$ tensor, both yield a $(0,2)$ tensor under antisymmetrisation, albeit which differ by a torsion term
\begin{align}
	\nabla_\mu\psi_\nu-\nabla_\nu\psi_\mu
	=D_\mu\psi_\nu-D_\nu\psi_\mu-T_{\mu\nu}{}^\rho\psi_\rho\,.
\end{align}

\subsection{Action}

With these elements we can assemble the universal part of the supergravity action
\begin{align}
	S=\int d^4x\,e\left(\frac{1}{2\kappa^2}e^{a\mu}e^{b\nu}R_{\mu\nu ab}\left(\omega\right)-\frac{1}{2}\overline{\psi}_\mu\gamma^{\mu\nu\rho}D_\nu\psi_\rho\right)\,,\quad
	e\equiv\det\left(e_\mu{}^a\right)\,,
	\label{supergravity action}
\end{align}
written on-shell following the conventions of \cite{Freedman:2012zz}, and
invariant under the transformations
\begin{align}\label{supergravity transformations}
	\delta\psi_\mu=\frac{1}{\kappa}D_\mu \epsilon\,,\quad
	\delta e^m{}_\mu=\frac{1}{2}\kappa\overline{\epsilon}\gamma^m\psi_\mu\,.
\end{align}

Solving for the spin connection via the requirement that the transformations \eqref{supergravity transformations} preserve the action \eqref{supergravity action} yields the unique solution for the torsion tensor
\begin{align}
	T^\mu{}_{ab}=\frac{\kappa^2}{2}\overline{\psi}_a\gamma^\mu\psi_b\,,
\end{align}
from which the contortion tensor \eqref{contortion tensor} can be written 
\begin{align}
	K_{\mu\nu\rho}=-\frac{\kappa^2}{4}\left(\overline{\psi}_\mu\gamma_\rho\psi_\nu-\overline{\psi}_\nu\gamma_\mu\psi_\rho+\overline{\psi}_\rho\gamma_\nu\psi_\mu\right)\,.
\end{align}

This allows an alternative form for the supergravity action \eqref{supergravity action}, in which the four-gravitino terms are explicit and the curvature terms are functions of the torsion free connection
\begin{align}
	&S=\int d^4x\,e\left(\frac{1}{2\kappa^2}e^{a\mu}e^{b\nu}R_{\mu\nu ab}\left(e\right)-\frac{1}{2}\overline{\psi}_\mu\gamma^{\mu\nu\rho}D_\nu\psi_\rho+\mathcal{L}_{\rm torsion}\right)\,,\\
	&\mathcal{L}_{\rm torsion}
	=-\frac{\kappa^2}{32}\left(\left(\overline{\psi}^\rho\gamma^\mu\psi^\nu\right)\left(\overline{\psi}_\rho\gamma_\mu\psi_\nu+2\overline{\psi}_\rho\gamma_\nu\psi^\mu\right)
	-4\left(\overline{\psi}_\rho\gamma_\mu\psi^\mu\right)\left(\overline{\psi}^\rho\gamma_\mu\psi^\mu\right)\right)\,.
\end{align}

The torsion terms are firstly simplified via the gauge choice $\gamma_\mu\psi^\mu=0$.
As we will see in the following, they can be usefully manipulated further via use of Fierz identities.

\subsection{Cosmological constant}
\label{sec: cosmological constant}

A primary extension of \eqref{supergravity action} is via the addition of a cosmological constant 
\begin{align}
	S=\int d^4x\,e\left(\frac{1}{2\kappa^2}\left(e^{a\mu}e^{b\nu}R_{\mu\nu ab}\left(\omega\right)+2\Lambda\right)-\frac{1}{2}\overline{\psi}_\mu\gamma^{\mu\nu\rho}D_\nu\psi_\rho-\sqrt{\frac{\Lambda}{3}}\overline{\psi}_\mu\gamma^{\mu\nu}\psi_\nu\right)\,,
	\label{cosmological constant supergravity}
\end{align}
which now requires the modified transformations \cite{Townsend:1977qa}
\begin{align}
	\delta\psi_\mu=\frac{1}{\kappa}D_\mu \epsilon-\frac{1}{\kappa}\sqrt{\frac{\Lambda}{12}}\gamma_\mu\epsilon\,,\quad
	\delta e_\mu{}^m=\frac{1}{2}\kappa\overline{\epsilon}\gamma^m\psi_\mu\,.
\end{align}

As can be seen, in the presence of a cosmological constant $\Lambda$, local supersymmetry mandates the presence of an apparent mass term for the gravitino.
Interpreting this requires a degree of caution however, as the notion of mass in a curved background is not straightforward.

The gravitino in this instance still has the same numbers of degrees of freedom as in the massless case, and the correct interpretation we should thus ascribe is not of a massive gravitino, but a massless gravitino propagating in a curved background.
Comparison with the canonical Einstein-Hilbert Lagrangian $\left(R-2\tilde\Lambda\right)$ indicates that for $\Lambda>0$, required for reality of the apparent mass term, the appropriate vacuum is then anti de Sitter space.

As we will see in the following section, the breaking of local supersymmetry is necessarily accompanied by the development of a gravitino mass.
Since the action is still invariant under the modified transformations, supersymmetry is unbroken and we can be assured that the mass term is not to be interpreted literally.

One notable advantage to a locally supersymmetric formulation can then be seen in that supersymmetry breaking is assured whenever the degeneracy between the cosmological constant and gravitino mass terms is not respected.
A desirable theory with broken supersymmetry via a massive gravitino and zero cosmological constant is then possible.

This is to be contrasted with the analogous situation in globally supersymmetric theories, where we necessarily must have a positive vacuum energy $\langle F\rangle$ related to the scale at which supersymmetry breaks. 
Given the measured value of the vacuum energy density of $\left(10^{-3}\rm{eV}\right)^4$ \cite{Riess:1998cb, Perlmutter:1998np}, even GeV scale breaking could not then be accommodated.

\subsection{Effective theory status}

As might be expected, the coupling constant of supergravity is inherited from General Relativity.
Since this has negative mass dimension the theory must then be perturbatively non-renormalisable, suggesting that we interpret it as an effective description of some more fundamental theory incorporating gravity
\footnote{This logic could conceivably be circumvented in Weinberg's asymptotic safety scenario, where the renormalisation group evolution of the couplings drives them to a non-trivial fixed point in the ultraviolet \cite{Weinberg:1980gg}.
In this case only a finite number of counterterms would be required to renormalise the theory, and it could be interpreted as a viable microscopic theory.}
\footnote{A further loophole exists in that $\mathcal{N}=8$ supergravity may be UV finite, and so too could function as a fundamental theory \cite{Bern:2006kd}.}.

Indeed, this is precisely what string and M-theory suggest; it has been long established that there exists an eleven-dimensional supergravity theory describing the low energy limit of M-theory, which can be dimensionally reduced to give the ten-dimensional IIA and IIB supergravities capturing the low-energy dynamics of their counterpart string theories \cite{Freedman:2012zz}. 
The remaining extra dimensions are then presumably compactified further to give a four dimensional supergravity theory, which describes the universe we commonly observe.

Whilst there are many possible variants of resultant supergravity theories, in the phenomenologically relevant $\mathcal{N}=1$ case they crucially all share a common gravitational sector.
As such, the analysis of the later chapters will centre entirely on this aspect of the theory, with the aim of providing an effective description enjoying a wide regime of applicability.

\section{Supersymmetry breaking} 
\label{sec: supersymmetry breaking}

Given that supersymmetry in any form is not observed in nature, we are forced to conclude that if it exists it is thus broken by some unknown mechanism. 
Primarily we will be interested in supersymmetry breaking taking place in some `hidden' sector, involving fields that are not part of the Standard Model.

This is largely motivated by the difficulties associated with reconciling known phenomenology with the consequences imposed by breaking supersymmetry in some `visible' supermultiplet \cite{Binetruy:2006ad}.
In the example of the $F$-term breaking discussed in section \ref{sec: Auxiliary fields} we would postulate a non-zero vacuum expectation value for some hidden sector auxiliary field $F$, which will then be communicated to the visible sector.

In supergravity, the breaking of local supersymmetry must be accompanied by a non-zero gravitino mass.
This may be seen by noting firstly that were the gravitino to become massive, it would break the supersymmetric degeneracy with the massless graviton.
On the other hand, unbroken supersymmetry strictly constrains the form of S-matrix elements, implying that if the gravitino is massless, supersymmetry is unbroken \cite{Grisaru:1976vm}.

As established previously, we also expect a massless fermion in the presence of broken supersymmetry.
Since this is the analogue of the Goldstone mode associated with the Higgs effect, it is then natural to consider the analogous effect in the instance of supersymmetry breaking.

\subsection{Super-Higgs effect} 

As demonstrated in section \ref{sec: The goldstino}, if the supersymmetry within some supermultiplet coupled to supergravity is spontaneously broken we expect a non-zero vacuum energy density, accompanied by a massless fermion; the Goldstino. 
To encompass the possibility of the super-Higgs effect, we may firstly incorporate this field via the non-linear Volkov-Akulov Lagrangian 
\begin{align}
	\mathcal{L}_\lambda = -f^2\det\left(\delta^\mu{}_\nu + i\overline{\lambda} \gamma^\nu \partial_\mu \lambda/2f^2 \right)
	=- f^2 - \frac{1}{2} i \overline{\lambda} \gamma^\mu \partial_\mu \lambda + \dots\,,
	\label{eq: goldstino action}
\end{align}
where the ellipsis denotes non-linear $\lambda$-dependent terms which ultimately will be gauged away \cite{Volkov:1973ix}.

The constant $f$ characterises the scale of global supersymmetry breaking associated to the Goldstino, with a resulting non-linear realisation of global supersymmetry 
\begin{equation}\label{trnl}
	\delta \lambda =\sqrt{2} \epsilon f + i \overline{\epsilon} \gamma^\mu \lambda \partial_\mu \lambda/f\,,
\end{equation}
where $\epsilon$ is an infinitesimal spinor.

As discussed in \cite{Deser:1977uq}, this may be incorporated into a locally supersymmetric context by allowing the parameter $\epsilon$ to depend on space-time coordinates, and coupling \eqref{eq: goldstino action} to $\mathcal{N}=1$ supergravity in such a way that the combined action is invariant under the transformations
\begin{align}
	\delta \lambda &= \sqrt{2}f \epsilon\left(x\right) + \dots \,, \nonumber \\
	\delta e^a{} _\mu  & = -i \kappa \overline{\epsilon}\left(x\right) \gamma^a \psi_\mu\,, \nonumber \\
	\delta \psi_\mu & =  - 2  \partial_\mu \epsilon\left(x\right)/\kappa + \dots\,,
	\label{eq: super-Higgs transformations}
\end{align}
where the ellipsis in the $\lambda$ transformation again denotes non-linear $\lambda$-dependent terms. 
The action that changes by a total divergence under these transformations is the standard $\mathcal{N}=1$ supergravity action plus
	\begin{equation}
		\mathcal{L}_\lambda = - f^2 - \frac{i}{2}\overline{\lambda} \gamma^\mu \partial_\mu \lambda - \frac{i\,f}{\sqrt{2}} \overline{\lambda} \gamma^\nu \psi_\nu + \dots \,,
		\label{va2b}
	\end{equation}
which contains the coupling of the Goldstino to the gravitino. 

Noting from \eqref{eq: super-Higgs transformations} that the Goldstino is shifted by a constant under supersymmetry transformations we may freely make the gauge choice $\lambda=0$, or equivalently implement a suitable redefinition of the gravitino field and the frame field such that the Goldstino is absorbed.
This then leaves behind a negative cosmological constant term, $-f^2$, so the total Lagrangian after these redefinitions is
	\begin{equation}\label{va3}
		\mathcal{L}_{\rm total} = -f^2 +\mathcal{L}_{\rm SG}\,, 
	\end{equation}
where $\mathcal{L}_{\rm SG}$ is the supergravity Lagrangian given in \eqref{supergravity action}. 
Given this absorption of the Goldstino, imposing the gauge condition 
	\begin{equation}
		\gamma^\mu\psi_\mu  = 0\,,
		\label{gravinogauge}
	\end{equation}
for the gravitino then results in it possessing the correct number of degrees of freedom for a massive field.
This is the super-Higgs effect \cite{Deser:1977uq}.

A possible concern may be found in the argument given in section \ref{sec: The goldstino} for the existence of the Goldstino, where the Ward identity associated to the two-point function of supersymmetry currents imply the existence of a massless fermion.
If $\lambda$ is gauged away, then this logic seemingly no longer applies.
However, one should note that when supersymmetry is made local there is necessarily another fermion which can mediate this two-point function; the gravitino.
The zero eigenvalue previously associated to the masslessness of the Goldstino can then be interpreted instead as resulting from the reduction of rank of the fermion mass matrix.

\section{Inflation in supergravity}
\label{sec: inflation in supergravity}

Given that supersymmetry is incompatible with the positive cosmological constant associated to de Sitter backgrounds, it may seem that it has little role to play regarding inflation. 
Despite this, there are however a number of arguments for pursuing a supersymmetric setting for early universe physics.
Primary amongst these is the inherent ultraviolet sensitivity of inflationary physics.

Qualitatively, we may understand this as arising from the tight constraints placed on inflationary potentials by the slow roll conditions of the previous chapter.
It is first of all necessary that radiative corrections do not spoil the requisite flatness of the potential, and furthermore required that even for radiatively stable models, neither do the higher-dimensional operators induced by renormalisation.

In both instances we can address the problem via either symmetry considerations, or fine tuning.
Given that the latter is somewhat unsatisfying, the former may be preferable, then requiring that whatever symmetries are necessary survive in an ultraviolet completion.

This sensitivity constitutes a key challenge for theories of inflation.
Needless to say, it also provides a window into physics at scales which are otherwise inaccessible.
Realising inflation in a supersymmetric setting can then allow the precision early universe phenomenology afforded by recent CMB observations to shed light on how supersymmetry is realised in the universe, in a complementary fashion to collider-based searches.
Accordingly, there exists a large body of literature pertaining to inflation in supergravity, exemplified in the review \cite{Yamaguchi:2011kg}.

\subsection{$\eta$ problem}
\label{sec: eta problem}

As is well known from the hierarchy problem of electroweak physics, scalar masses typically receive large corrections from loop effects \cite{Susskind:1978ms}.
In the absence of protective symmetries, we then expect scalar mass terms to be driven to the cutoff scale $M_\Lambda$.
This is acutely problematic in the present context as the flatness requirement
\begin{align}
	|\eta|\equiv M_P^2\frac{|V''|}{V}<<1\,, 
	\label{eq: eta definition}
\end{align}
implies that we need the inflaton to be light relative to the scale of inflation.
The correction we expect to this relation is
\begin{align}
	\Delta\eta\sim\frac{M_\Lambda^2}{H^2}\gtrsim1\,,
	\label{Delta eta}
\end{align}
accounting for $H$ as the characteristic scale of the inflationary potential.
This is known as the $\eta$ problem.
It is often presented as an issue afflicting inflation in supergravity specifically, however as the above presentation suggests it is in fact more general, affecting all effective descriptions of inflation.

Supersymmetry can of course allow for light scalars to be technically natural, however as already observed, the positive vacuum energy associated with de Sitter space is an obstruction to this.
However, even broken supersymmetry can ameliorate this problem to some extent.
For modes that are deep inside the horizon, or equivalently have sufficiently large energies to be insensitive to the curvature of spacetime, supersymmetric cancellations can still take place.

Modes with frequencies that are below the Hubble scale will however not enjoy such a benefit.
There will be incomplete cancellations between bosons and fermions, arising from $\mathcal{O}\left(H\right)$ mass splittings within supermultiplets.
Corrections to the inflaton mass will then be of order the Hubble scale, with the effect of reducing the naive estimate of \eqref{Delta eta} to $\Delta\eta\sim 1$.
Needless to say however, this is still too large.

Two approaches exist to address this problem; further symmetries and fine-tuning.
In the latter instance, we can gain some intuition from the limit $\epsilon<<|\eta|$, where the analysis of the previous chapter implies the approximate relation
\begin{align}
	|\eta|\approx n_s-1\,.
\end{align}
For the Planck 2015 best fit value of $n_s\sim0.96$ \cite{Planck:2015xua} we then have $\eta\sim0.04$, implying a percent-level tuning.

As far as the former is concerned, radiative stability can be engineered via global symmetries which prevent the harmful corrections driving the inflaton mass to the cutoff scale.
One such example is the shift symmetry $\phi\to\phi+c$, slightly broken by a small mass term.
Loop corrections are then scaled by the symmetry breaking parameter, which conveniently is the mass itself, so that $\Delta m^2\sim m^2$.
Shift symmetries are notable in that they typically feature in the axionic sector of string compactifications, allowing the construction of many string-theoretic inflationary models leveraging this property \cite{Baumann:2014nda}. 

Whether this useful radiative stability is preserved in the ultraviolet is however a non-trivial question, and one which cannot be definitively answered in the context of an effective approach.
Indeed, it is often assumed on general grounds that quantum theories of gravity do not respect global symmetries \cite{Banks:2010zn}.
One facet of this may be seen in the thermodynamic behaviour of black holes, where conservation of baryon number may be violated by absorption of baryonic matter and subsequent evaporation into non-baryonic species.

More precisely, integrating high momentum degrees of freedom may in principle always introduce irrelevant operators such as
\begin{align}
	\mathcal{O}_6=V\frac{\phi^2}{M_\Lambda^2}\,,
	\label{eq: dimension 6 O}
\end{align}
which yield only a small correction to the potential $V$, but have a significant effect on the inflaton mass.
Substitution into \eqref{eq: eta definition} leads to a contribution
\begin{align}
	\Delta\eta\sim2\frac{M_P^2}{M_\Lambda^2}\,,
\end{align}
allowing the $\eta$ problem to recur.

Whilst these problematic issues may affect any model of inflation, ultraviolet sensitivity is even further enhanced in models of inflation capable of generating an observable tensor to scalar ratio, $r$. 

\subsection{Lyth bound} 

A particularly useful result in this context, of which the aforementioned ultraviolet sensitivity is a consequence, is the Lyth bound
\begin{align} 
	\frac{\Delta\phi}{M_P}\gtrsim\left(\frac{r}{0.01}\right)^{\frac{1}{2}}\,.
	\label{eq: Lyth bound}
\end{align}
This implies the requirement that in the simple models of inflation the inflaton must undergo trans-Planckian excursions in field space in order to produce $r\gtrsim0.01$ \cite{Lyth:1996im}, which conveniently is the expected experimental sensitivity to $r$ of experiments in the near future \cite{Verde:2005ff}.
Models capable of exceeding the bound are typically known as `large field' scenarios, whilst by extension those which cannot are known as `small field'.

Needless to say, in going beyond simple inflationary scenarios there exist a number of generalisations and workarounds to \eqref{eq: Lyth bound} \cite{Baumann:2014nda}.
These may offer a route to an observable $r$ without trans-Planckian inflaton excursions, but are beyond the scope of the present discussion.

Although the prospect of trans-Planckian physics may seem like dangerous territory, from the perspective of an effective description this need not be problematic.
It is first of all important to note that even if field values exceed the Planck scale in appropriate units, the requirement of sufficient hierarchy between $M_P$ and the scale of the inflationary potential from the previous chapter implies that the energy densities in question can still lie safely below the Planck scale, even if the field values are not. 
Furthermore, in the presence again of an approximate shift symmetry, the arguments of the previous section again indicate that radiative mass corrections can remain under control.

In contrast however, from an ultraviolet perspective exceeding the Lyth bound can be particularly dangerous.
Suppressed contributions such as \eqref{eq: dimension 6 O} in particular now have the potential to compound the difficulties they previously created.
Specifically, there is no strong reason for these operators not to introduce sub-Planckian structure to the inflaton potential, in stark contrast to the inherent large field requirement of flatness on super-Planckian scales.

Preventing these contributions may again require some additional symmetry structure, the ultraviolet complete realisation of which is a subtle and non-trivial task.
This points to the necessity of a complete theory of quantum gravity, such as string theory, in order to fully understand inflation and the symmetries it requires, and thus the utility of the effective description that supergravity provides thereof.

%% file: Chapters/Condensation.tex
Having established the requisite aspects of early universe inflation and supersymmetry breaking, we may now proceed to concretely explore a particular scenario in which these phenomena can be realised.
As we will establish in this chapter, the development of a non-trivial vacuum expectation value for the gravitino bilinear $\langle\overline\psi_\mu\psi^\mu\rangle$ may break local supersymmetry, whilst simultaneously providing an inflaton candidate in the effective scalar degree of freedom it describes. 

We will argue that this approach to supersymmetry breaking can exhibit certain useful properties.
By virtue of taking place in the gravitational sector, it can firstly enjoy a certain universality amongst supergravity theories, in that it does not rely on specific or arbitrary choices of potential, matter content, gauge group or representation.

Furthermore, by making use of a single composite field to both break local supersymmetry and inflate the early universe we can simultaneously confront both cosmological and particle physics phenomenology.

The first sections of this chapter are intended to serve as a comparatively qualitative introduction to this topic, focussing on the analogy between supersymmetry breaking in supergravity and chiral symmetry breaking in QCD, and simplified effective descriptions thereof.
After detailing these considerations in section \ref{sec: chiral symmetry breaking}, we leverage in section \ref{sec: gravitino condensation} some of the tools from the study of the NJL model to demonstrate dynamical gravitino mass generation via the corresponding gap equation.

Since these arguments rely on a suitably strong gravitino coupling into the scalar condensate channel, in section \ref{sec: coupling constant} we furthermore demonstrate that, despite the requirements of local supersymmetry, this coupling is in fact unfixed.
It is also shown that a coupling which is sufficiently strong to induce scalar condensation does not necessarily also lead to unwanted pseudoscalar and pseudovector condensates.

We also make use of the gap-equation analysis to derive the wavefunction renormalisation for the condensate mode in section \ref{sec: wavefunction renormalisation}, via the Bethe-Salpeter equation and an all-orders resummation of gravitino bubble graphs.
This then allows the full effective potential for the condensate mode, computed in the next chapter, to be canonically normalised.

The results presented herein are largely based on the article \cite{Alexandre:2014lla}.

\section{Chiral symmetry breaking in QCD}
\label{sec: chiral symmetry breaking}

In order to develop some intuition for gravitino condensation, it may be helpful to firstly explore a more down to earth example in the $\langle \overline{q}_L q_R\rangle$ quark condensates known to break chiral symmetry in QCD \cite{Gasser:1982ap}.

In addition to the $SU\left(3\right)$ colour group, the fermionic sector of QCD in the two-flavour limit enjoys the approximate global symmetry
\begin{align}
	U\left(2\right)_L\times U\left(2\right)_R
	\sim SU\left(2\right)_L\times SU\left(2\right)_R\times U\left(1\right)_V\times U\left(1\right)_A\,,
\end{align}
where subscripts denote left, right, vector, and axial, respectively.
As is well known however, this approximate symmetry is not preserved by the vacuum of the theory.

\begin{figure}[h!]
  \centering
        \includegraphics[width=0.4\textwidth]{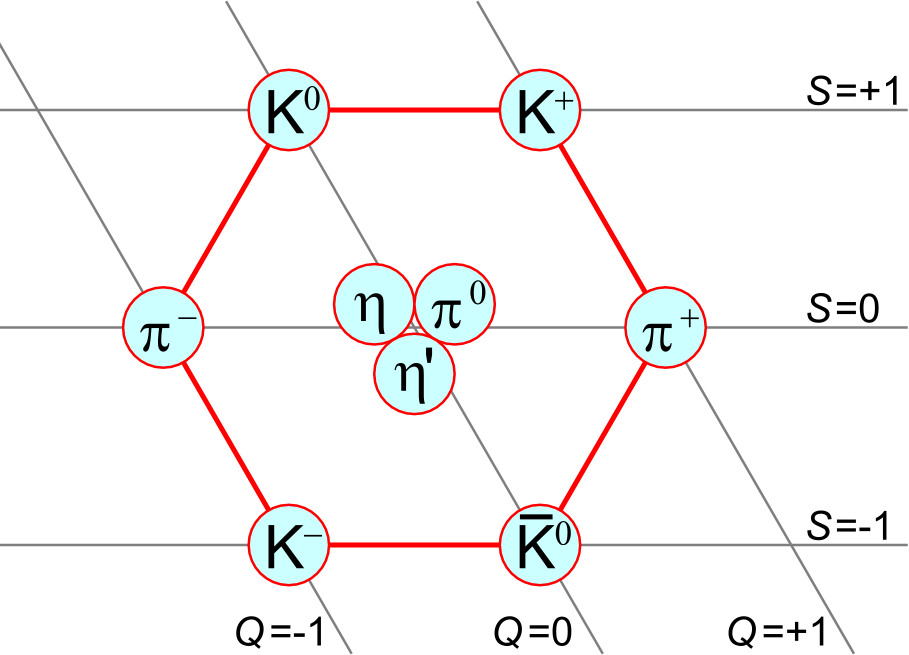}
    \caption{The pseudoscalar meson nonet represented in electric charge and strangeness, comprising the three pions, four kaons and the $\eta$ and $\eta'$.}
    \label{fig: Meson nonet}
\end{figure}

More precisely, it is spontaneously broken to a remnant $SU\left(2\right)_{ISO}$ isospin symmetry, with the three associated pseudo-Goldstone bosons identified with the pions $\pi^{\pm}$ and $\pi^0$ occupying the zero-strangeness axis of the pseudoscalar meson nonet shown in figure \ref{fig: Meson nonet}
\footnote{The pions are only pseudo-Goldstone bosons since chiral symmetry is only exact in the massless quark limit. 
Since the up and down quarks are much lighter than the characteristic scale of two-flavour QCD, $M_{QCD}\sim 300$ MeV, the pions should be, and are, accordingly light ($m_\pi\sim135$ MeV) \cite{Fritzsch:2012wq}.}.
Since the vast majority of baryonic mass arises from the effect of this chiral symmetry breaking, reflected in the effective masslessness of the up and down quarks relative to the proton and neutron, this generates most of the visible mass of the universe, independently of the Higgs mechanism.

Whilst the $U\left(1\right)_V$ symmetry ensuring baryon number conservation is unaffected by chiral symmetry breaking, chiral condensates such as $\langle \overline{q}_L q_R\rangle$ should however break the $U\left(1\right)_A$ symmetry. 
One of the puzzles of early QCD phenomenology was then the absence of the associated pseudo-Goldstone boson, the so-called $U\left(1\right)_A$ problem. 

As demonstrated by `t Hooft however, rather than being spontaneously broken by the vacuum of the theory, the global $U\left(1\right)_A$ is actually anomalous, in that whilst the action of QCD is invariant under the action of the symmetry, the functional measure is not \cite{'tHooft:1986nc}.
As such, the theory was never invariant under $U\left(1\right)_A$ in the first place, and there can be no associated pseudo-Goldstone boson.

Suggestively, in the massless quark limit the divergence of the associated Noether current would be
\begin{align}
	\partial_\mu j_A^\mu=\frac{g_s^2}{16\pi^2}\epsilon_{\alpha\beta\delta\gamma}\Tr\left({F}^{\delta\gamma}F^{\alpha\beta}\right)\,,
	\label{eq: chiral divergence}
\end{align}
where $g_s$ is the characteristic coupling of QCD.
Integrating both sides yields a relation for the obstruction to charge conservation
\begin{align}
	\Delta Q = \frac{g_s^2}{16\pi^2}\int d^4x\epsilon_{\alpha\beta\delta\gamma}\Tr\left({F}^{\delta\gamma}F^{\alpha\beta}\right)\,,
	\label{eq: charge non-conservation}
\end{align}
which is clearly integer-valued in an appropriate normalisation for the axial charge. 
The RHS is also suitably discrete in that it is a topological invariant, the second Chern class 
\footnote{Equivalently known as the Pontryagin index, or the winding number.}.
We must then conclude that perturbative effects cannot source the anomaly evident in \eqref{eq: chiral divergence}, as the relation given cannot depend smoothly on the coupling constant $g_s$.

Furthermore, in the correct normalisation where $\int\partial_\mu j_A^\mu\in\mathbb{Z}$, tree-level terms go as $g_s^{-2}$.
The discrete nature of the relation \eqref{eq: charge non-conservation} then implies that there always exists a renormalisation scheme where \eqref{eq: chiral divergence} is one-loop exact, as it can only carry a factor of $g_s^0$.

Thus, although the effect driving the chiral anomaly cannot be seen in perturbation theory, its signature may be detected at one-loop.
Even in the limit of strong coupling, this one-loop exactness further guarantees that the effect cannot be cancelled by higher orders in perturbation theory.

As an aside, the importance of this operator suggests that it should be included in the QCD Lagrangian via the so-called theta term
\begin{align}
	\mathcal{L}_\theta=-\frac{g_s^2\theta_{QCD}}{32\pi^2}\epsilon_{\alpha\beta\delta\gamma}\Tr\left({F}^{\delta\gamma}F^{\alpha\beta}\right)\,.
\end{align}
Since this violates CP symmetry in the absence of any QCD-sector experimental signatures thereof, such as a non-zero neutron electric dipole moment \cite{Baker:2006ts}, we must find an explanation for the smallness of the $\theta_{QCD}$ parameter.
This is the as-of-yet unresolved strong CP problem \cite{Peccei:1977hh}.

Nonetheless, the nature of the anomalous chiral current \eqref{eq: chiral divergence} suggests a concrete mechanism by which $\langle \overline{q}_L q_R\rangle$ may develop a non-trivial vacuum expectation value.
Since the bilinear vacuum expectation value $\langle \overline{q}_L q_R\rangle$ is zero to all orders in perturbation theory, we expect a solution of accordingly non-perturbative character.

\subsection{The role of instantons}

The Euclidean Yang-Mills equations for an $SU\left(N\right)$ gauge theory are automatically satisfied if the field-strength is (anti)self-dual
\begin{align}
	F_{\mu\nu}=\pm\tilde{F}_{\mu\nu}
	\equiv\pm\frac{1}{2}\epsilon_{\mu\nu\alpha\beta}F^{\alpha\beta}\,.
\end{align}
Finite action configurations satisfying this condition are necessarily localised in Euclidean time, leading to the instanton terminology \cite{Dorey:2002ik}.

\begin{figure}[h!]
 \begin{minipage}[t]{0.5\textwidth}
  \mbox{}\\[-\baselineskip]
	\caption{Simplified schematic detailing the origin of instantons. 
	Given a field $\phi$ with multiple topologically inequivalent vacua $\tau_i$, we naturally expect quantum mechanical tunnelling to occur within some characteristic timescale.
	Solving the Euclidean field equations for the vacuum behaviour of the field then yields solutions which occupy different vacua at different times, with smooth transitions in-between.
	Since Euclideanisation places space and time on the same footing, the resulting profile should also be a valid description in the spatial dimensions.
	The non-trivial field configurations associated to these tunnelling events then provide a background which cannot be seen in perturbation theory.
	In the instance of QCD, $\phi$ would be identified with the gluon field, and chiral symmetry breaking is a consequence of the interaction of quarks with this instanton background.}
	\label{fig: Instantons}
  \end{minipage}
  \begin{minipage}[t]{0.5\textwidth}
  \mbox{}\\[-\baselineskip]
    \includegraphics[width=\textwidth]{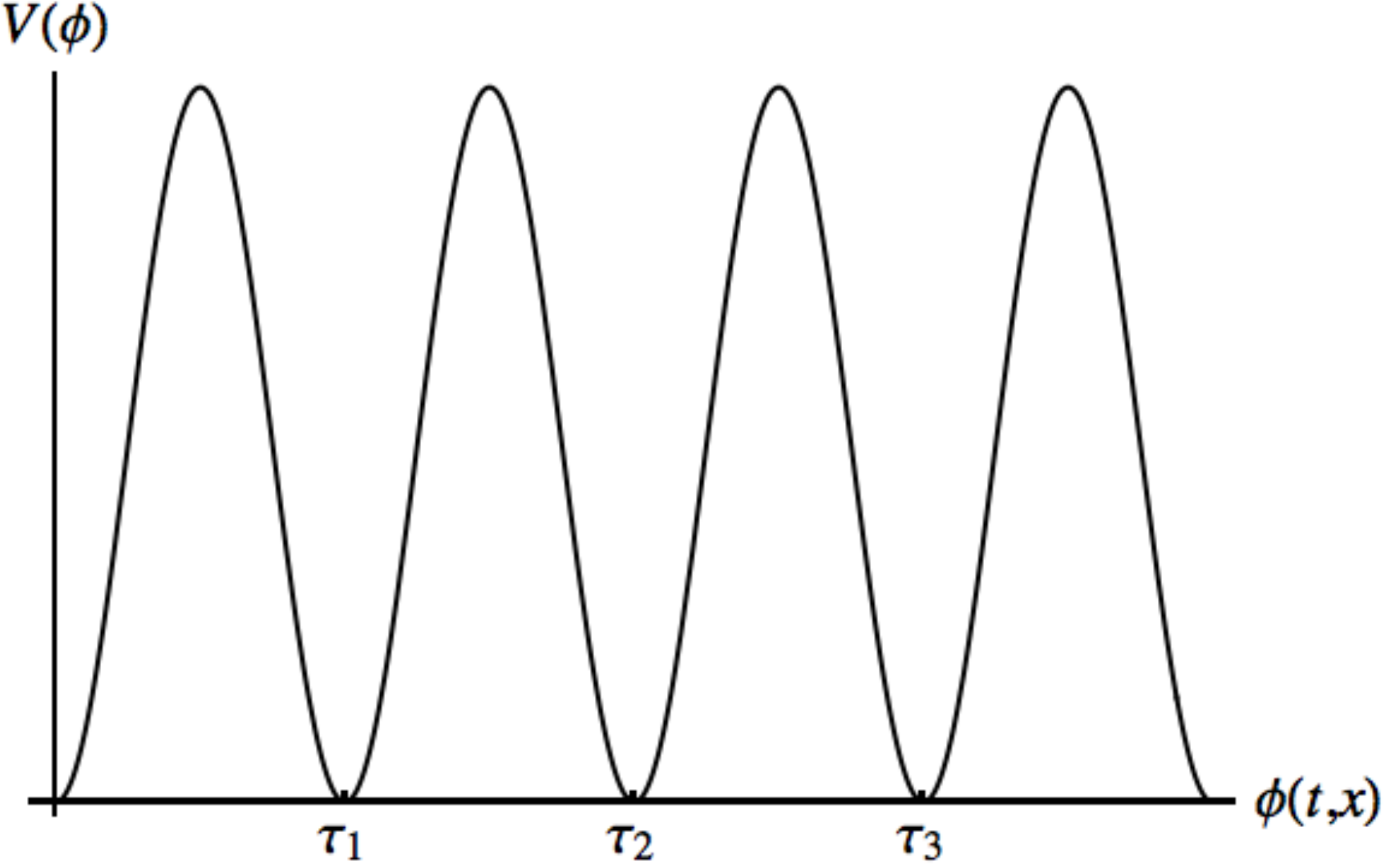}
    \includegraphics[width=\textwidth]{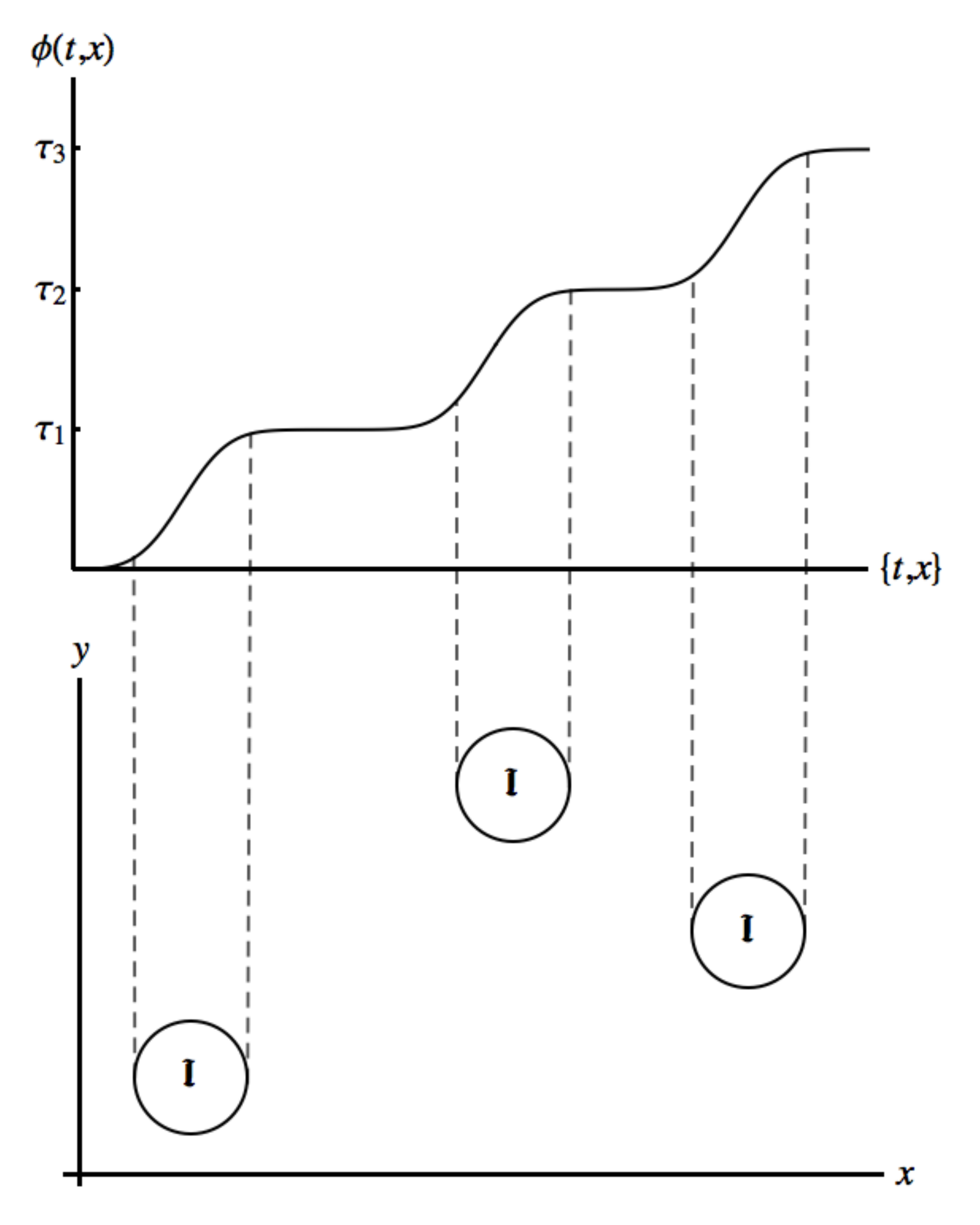}
  \end{minipage}\hfill
\end{figure}

These solutions are associated with tunnelling between topologically distinct vacua, indexed by the invariant \eqref{eq: charge non-conservation}.
To gain a little intuition into the role they play we may make use of some simplified pictorial logic, outlined in Figure \ref{fig: Instantons}.

The effect of these non-trivial field configurations may be seen by integrating them out, resulting in an effective four-fermion interaction $\left(\overline q_Lq_R\right)\left(\overline q_Rq_L\right)$ \cite{'tHooft:1976fv}.
The dimensionful coupling constant required on dimensional grounds is then inherited from the characteristic scale associated to the instanton configurations.
Chiral symmetry breaking can then be understood at the level of the effective theory as a consequence of the non-invariance of such a term under chiral transformations.

Given the difficulty associated with the strongly coupled, non-perturbative nature of QCD, it is natural to explore simplified effective descriptions which hopefully encapsulate some of the essence of the problem, absent some of the intractable features.

\subsection{The Nambu-Jona-Lasinio model}
\label{sec: the nambu-jona-lasinio model}

Exactly such a scenario of relevance to the present context is the Nambu-Jona-Lasinio (NJL) model of chiral fermions coupled via the Lagrangian
\begin{align}
	\mathcal{L}_{\rm NJL}=i\overline\psi_a\gamma_\mu\partial^\mu\psi^a+\frac{g_{\rm NJL}}{N}\left(\left(\overline\psi_a\psi^b\right)\left(\overline\psi_b\psi^a\right)-\left(\overline\psi_a\gamma^5\psi^b\right)\left(\overline\psi_b\gamma^5\psi^a\right)\right)\,,
	\label{eq: NJL Lagrangian}
\end{align}
where $a, b=1\dots N$ are flavour indices \cite{Nambu:1961tp, Nambu:1961fr}.
As in QCD with $N$ flavours, this is invariant under the $SU\left(N\right)_L\times SU\left(N\right)_R\times U\left(1\right)_V\times U\left(1\right)_A$ symmetry group.

Given that we can parametrise the effect of QCD instantons in terms of four-fermion interaction vertices, the form of \eqref{eq: NJL Lagrangian} as an effective model for chiral symmetry breaking follows from the logic of the previous section.
It should be noted however that the NJL model actually preceded this insight by more than a decade; it was originally intended to model pion behaviour by treating them as bound states of nucleons and their antiparticles.
In hindsight however, the actual underlying physical linkage is clear.

Of particular relevance to later discussion is the resulting identification
\begin{align}
	g_{\rm NJL} \sim \frac{\rho^2}{g_s^2}\exp\left(\mathcal{O}\left(10\right)\right)\,,
	\label{eq: g_{NJL}}
\end{align}
where $\rho$ is a scale associated to the size of the instantons \cite{'tHooft:1986nc,'tHooft:1976fv}.
Given the context, that \eqref{eq: g_{NJL}} arises in turn from the characteristic non-perturbative factors of $\exp\left(-8\pi^2/g_s^2\right)$ should not be surprising.

One should note that, in contrast to QCD, the NJL model does not exhibit confinement.
Furthermore, the $U\left(1\right)_A$ axial symmetry is broken spontaneously rather than being anomalous, resulting in an additional pseudo-Goldstone mode.
Despite these obstacles however, there exists a large body of literature making use of the simplified perspective the NJL model provides on QCD   \cite{Hatsuda:1994pi}.

As we would hope from a more tractable treatment, the breaking of chiral symmetry in the NJL model can be seen relatively straightforwardly.
One means to achieve this is via a mean field theory approach, where linearising the scalar channel interaction via
\begin{align}
	\frac{g_{\rm NJL}}{N}\left(\overline\psi_a\psi^a\right)^2\sim-\sigma^2+2\sigma\sqrt{\frac{g_{\rm NJL}}{N}}\left(\overline\psi_a\psi^a\right)\,,
\end{align}
and integrating out the fermions yields an effective potential for the auxiliary scalar $\sigma$.
The vacuum expectation value $\langle\sigma\rangle$ can then be taken as an avatar for the vacuum expectation value of the chiral symmetry violating bilinear $\left\langle\overline\psi_a\psi^a\right\rangle$.
Given a sufficiently strong coupling into this channel, an entirely sensible assumption in light of \eqref{eq: g_{NJL}}, we would then expect bound states such as these to form.

As we will see in the following section, these considerations of chiral symmetry breaking may also be usefully imported into the supersymmetric setting.

\section{Gravitino condensation in supergravity}
\label{sec: gravitino condensation}

Commonly, supersymmetry breaking is achieved simply via the assumption of some hidden-sector superpotential which breaks supersymmetry at tree level.
The utility of this approach notwithstanding, it is nonetheless desirable to have an explicit mechanism to hand: whilst flexible, an arbitrary choice of superpotential and matter content can necessarily lead to a degree of arbitrariness in the corresponding low energy phenomenology.
To this end we will instead explore a concrete scenario, which also espouses a certain degree of flexibility.

As emphasised in the previous chapter, invariance under local supersymmetry necessarily requires four-fermion terms.
Given the status of supergravity as an effective description of a string theoretic picture, exemplified in the perturbatively non-renormalisable character of these four-gravitino interactions, a straightforward approach is to proceed in the spirit of the NJL approach outlined in the previous section.

Specifically, we are free to pursue the supersymmetry breaking bilinear $\langle\overline\psi_\mu\psi^\mu\rangle$ by linearising the interaction term via
\begin{align}
	\kappa^2\left(\overline\psi_\mu\psi^\mu\right)^2\sim-\sigma^2+2\sigma\kappa\left(\overline\psi_\mu\psi^\mu\right)\,,
\end{align}
where $\kappa^2=8\pi G$, and the identification $\sim$ follows at the level of the action as a consequence of the equation of motion for $\sigma$, 
\begin{align}
	\sigma=\kappa\left(\overline\psi_\mu\psi^\mu\right)\,.
	\label{eq: sigma eom}
\end{align}

If the auxiliary field $\sigma$ then develops a non-trivial vacuum expectation value, we may interpret $\langle\sigma\rangle\left(\overline\psi_\mu\psi^\mu\right)$ as an effective mass term for the gravitino.
Accompanying this is a $\langle\sigma\rangle^2$ contribution to the cosmological constant, which can be associated to the energy density associated to the symmetry breaking phase transition. 
The analogy to the chiral symmetry breaking of the previous section is clear.

A further benefit of the gravitational context provided by supergravity is that the effective scalar degree of freedom, which we identify via \eqref{eq: sigma eom} with the composite field $\overline{\psi}_\mu\psi^\mu$, can function as the inflaton.
This then provides the possibility of simultaneously confronting inflationary and particle physics phenomenology.
			 
\subsection{Gravitational instantons}
\label{sec: Gravitational instantons}

Given this analogy with chiral symmetry breaking in QCD, it is interesting to speculate about which gravitational field configurations may fulfil the role of the gluonic instanton configurations in this picture.
Since the Euclidean Einstein field equations in vacuo are automatically satisfied in the presence of (anti)self-dual curvature
\begin{align}
	R_{abcd}=\pm\tilde{R}_{abcd}
	\equiv\pm\frac{1}{2}\epsilon_{cdef}R_{ab}{}^{ef}\,,
	\label{eq: self-duality condition}
\end{align}
there exists a very natural analogy between the Yang-Mills instantons of the previous section, and so-called gravitational instantons satisfying \eqref{eq: self-duality condition} \cite{Eguchi:1980jx}.
Given the possibility explored in the previous chapter of formulating gravity as a Yang-Mills theory associated to local Lorentz transformations, this may not be entirely surprising.

In addition to the (anti)self-duality conditions, for the topic at hand we may further stipulate that we require solutions which  
\begin{itemize}
	\item Are non-compact, to enable definition of the S-matrix.
	\item Admit spin structure, to allow for the presence of fermions.
	\item Are nonsingular and asymptotically locally Euclidean, with zero cosmological constant.
	This is to allow compatibility with appropriate boundary conditions. 
\end{itemize}
Furthermore, to break local supersymmetry we must give the gravitino a mass, corresponding to supplying two extra zero-modes to the two-point function $S_{\mu\nu}\left(x,y\right)$.
This accounts for the normalisable zero modes that exist only for nonzero $m$, when
\begin{align}
	S_{\mu\nu}\left(x,y\right)=
	\frac{\left |\psi^0_\mu\left(x\right)\right\rangle\left\langle\psi^0_\nu\left(y\right)\right|}{im}
	+\sum_{\lambda\neq0}\frac{\left |\psi^\lambda_\mu\left(x\right)\right\rangle\left\langle\psi^\lambda_\nu\left(y\right)\right|}{\lambda+im}\,.
\end{align}
Since we must treat $\psi_\mu$ and $\overline\psi_\mu$ independently in Euclidean signature, this requires two zero-modes overall.

As might be inferred from the association of instantons with tunnelling between topologically distinct vacua, only solutions with a non-vanishing Hirzebruch signature 
\begin{align}
	\tau\equiv\frac{1}{48\pi^2}\int d^4x\sqrt{g}\tilde R_{abcd}R^{abcd}\,,
\end{align}
can supply exactly the requisite number of helicity 3/2 zero modes \cite{Hawking:1979zs}.

A candidate instanton satisfying these constraints is the Eguchi-Hanson metric 
\begin{align}
	ds^2=\left(1-\left(\frac{a}{r}\right)^4\right)^{-1}dr^2+\frac{r^2}{4}\left(\sigma_1^2+\sigma_2^2+\left(1-\left(\frac{a}{r}\right)^4\right)\sigma_3^2\right)\,,
	\label{eq: Eguchi-Hanson metric}
\end{align}
where $\sigma_i$ are the left-invariant one-forms on $S^3$
\begin{align}
	\sigma_1 &= \sin \Xi \, d \theta - \cos \Xi \sin \theta \, d \phi\,,\nonumber\\
	\sigma_2 &= \cos \Xi \, d \theta + \sin \Xi \sin \theta \, d \phi\,,\nonumber\\
	\sigma_3 &= d \Xi + \cos \theta \, d \phi\,,
\end{align}
with coordinate ranges $0\leq\Xi<2\pi$, $0\leq\theta<\pi$, and $0\leq\phi<2\pi$ \cite{Eguchi:1978gw}.

Whilst asymptotically locally Euclidean, the line element \eqref{eq: Eguchi-Hanson metric} is not asymptotically globally Euclidean, in that it is actually asymptotic to the quotient $\mathbb{R}^4/\mathbb{Z}_2$.
Indeed, it is a well known corollary of the positive energy theorem that there are no four-dimensional asymptotically globally Euclidean gravitational instantons in Einstein gravity \cite{Witten:1981mf}.

This is a reassuring result as it guarantees the stability of flat space, however in the present context, one may be concerned that this implies incompatibility with appropriate boundary conditions.
That said, we need not limit consideration to purely gravitational instantons.
Notably, there exist asymptotically globally Euclidean string-theoretic instantons which are presumed to exist in the ultraviolet complete theory of which supergravity is an effective description \cite{Rey:1989xj}.

Ultimately, working at the level of an effective description cannot point definitively to the field configurations responsible.
It remains interesting to speculate however.

A subtle point exists however in that whilst the NJL model may be `derived' via integrating out QCD instantons, leading to the characteristic four-fermion interaction \cite{'tHooft:1976fv}, the identification of supergravity as a low energy effective description of string theory occurs at the level of the spectrum \cite{Freedman:2012zz}, without necessarily proceeding via non-perturbative effects.
Indeed, given the supersymmetric cancellation of non-zero modes, the absence of spin 1/2 zero modes for the Eguchi-Hanson instanton then implies that it cannot generate a superpotential  \cite{Hawking:1979zs, Hanson:1978uv}.

That said, as we will explore in section \ref{sec: coupling constant}, the microphysics controlling the coupling strength into the scalar condensate channel is unclear in the mean field theory context, and may conceivably be sourced in an analogously non-perturbative manner.

\subsection{Field strength condensation}
\label{sec: Field strength condensation}

Notably, another approach to non-perturbative breaking of local supersymmetry in the gravitino sector has been explored in the literature \cite{Konishi:1988mb}.
In the interests of completeness we will briefly review this alternative scenario.

In the simplest example of $\mathcal{N}=1$ supergravity coupled to a single chiral multiplet $\left(z, \Omega\right)$, the chiral $U\left(1\right)$ current enjoys the divergence relation \cite{Delbourgo:1972xb, Eguchi:1976db}
\begin{align}
	D_\mu j^\mu= -\frac{1}{384\pi^2}\tilde{R}_{abcd}R^{abcd}+\dots\,, 
\end{align}
where ellipsis indicates higher order terms unimportant for the present discussion, and the analogy with \eqref{eq: chiral divergence} is clear. 
Varying this relation under supersymmetry yields the anomalous transformation law
\begin{align}
	\delta\left(\overline\Omega z\right) =-\frac{\kappa^2}{384\pi^2}\overline{\psi}_{\mu\nu}\psi^{\mu\nu}+\dots\,,\quad
	\psi_{\mu\nu}\equiv D_{[\mu}\psi_{\nu]}\,,
	\label{eq: anomalous variation}
\end{align}
suggestive of $\left\langle\overline{\psi}_{\mu\nu}\psi^{\mu\nu}\right\rangle$ as a possible order parameter for supersymmetry breaking.

This is the analogue of a super Yang-Mills relation for the gaugino field $\chi$
\begin{align}
	\delta\left(\overline\Omega z\right)=\frac{g^2}{32\pi^2}\overline\chi\chi+\dots\,,
\end{align}
which leads to the gaugino condensation scenario, explored at length in the literature \cite{Ferrara:1982qs}, where these fields condense to give $\langle\overline\chi\chi\rangle\neq0$.

As in the case of both chiral symmetry breaking in QCD and gaugino condensation, we expect the bilinear vacuum expectation value $\left\langle\overline{\psi}_{\mu\nu}\psi^{\mu\nu}\right\rangle$ to be zero to all orders in perturbation theory.
We must then turn to non-perturbative effects, such as gravitational instantons, to catalyse supersymmetry breaking in this manner.

As explored in \ref{sec: Gravitational instantons}, the simplest configuration of relevance is the Eguchi-Hanson metric.
To investigate, the one-loop vacuum expectation value of $\overline{\psi}_{\mu\nu}\psi^{\mu\nu}$ has been computed about this background, integrating out the various instanton parameters to yield the vacuum expectation value
\begin{align}
	\left\langle\overline{\psi}_{\mu\nu}\psi^{\mu\nu}\right\rangle\sim \mu^5\,,
\end{align}
where $\mu$ is the regularisation scale.
The appearance of $\mu$ in the final answer may seem unusual, however it is a simple consequence of the non-renormalisable character of supergravity theories.
This vacuum expectation value then breaks supersymmetry by virtue of \eqref{eq: anomalous variation}, with some possible consequences thereof explored in \cite{Konishi:1989em, Mangano:1988kf}.

Whilst this approach enjoys similar model-independence to the `conventional' gravitino condensation scenario outlined earlier in this chapter, the absence of any generated superpotential, as outlined in the previous section, hinders further exploration of this particular scenario.
As such, we will instead return to gravitino, rather than gravitino field strength, condensation, with the aim of deriving the corresponding effective potential and concretely elucidating various aspects of the theory therefrom.

In line with the overall reasoning of this chapter we may firstly repurpose some of the flat-space tools used in the NJL model, postponing the full curved-space calculations to the following chapter.

In order to elicit suitably non-perturbative behaviour, we firstly require an appropriately non-perturbative starting point. 
It has long been established that the correlation functions of any quantum field theory inherit a family of interrelations from the Euler-Lagrange equations, known in turn as the Schwinger-Dyson equations.
Generally these take the form of an infinite tower of functional differential equations, which are truncated suitably to ensure a manageable result.

\subsection{Gap equations}
\label{sec: gap equations}

To make use of these relations we may firstly take the standard step \cite{Langfeld:2003ye} of identifying the scalar condensate with the trace of the full propagator 
\begin{align}
	\langle\overline\psi_\mu\psi^\mu\rangle=-i\eta^{\mu\nu}\lim_{x\to 0}\Tr P_{\mu\nu}\left(x\right)\,,
\end{align}
where a suitable choice for the momentum space propagator for the massive gravitino in a flat space-time \cite{Freedman:2012zz} is
\begin{align}
	P_{\mu\nu}=-\frac{i}{2}\gamma_\mu\frac{ \gamma_\rho p^\rho+m_{\rm dyn}}{p^2-m_{\rm dyn}^2}\gamma_\nu\,.
	\label{eq: gravitino propagator}
\end{align}
Since the trace of an odd number of $\gamma$ matrices vanishes, it is straightforward to see that in the supersymmetric limit $m_{\rm dyn}\to 0$ the trace of the propagator must vanish, as expected.
This also guarantees trivially that the trace of the massless bare propagator must vanish.

By then identifying the dynamical gravitino mass with the scalar condensate, $m_{\rm dyn}$ at the condensation scale is then a solution of the self-consistent gap equation 
\begin{align}
	m_{\rm dyn}=-\lambda_{\rm S}\eta^{\mu\nu}\lim_{x\to0}\Tr P_{\mu\nu}\left(x\right)
	=8i\lambda_{\rm S} \int \frac{d^4p}{(2\pi)^4}\frac{m_{\rm dyn}}{p^2-m_{\rm dyn}^2}\,,
	\label{eq: gap}
\end{align}
where the dimensionful coupling $\lambda_{\rm S}$, whose role will be elaborated upon fully in the following section, is unfixed, and the trace is both over indices and momentum modes.
The RHS of \eqref{eq: gap} diverges quadratically, yielding upon Euclideanisation the solution
\begin{align}
	m_{\rm dyn}=\frac{\lambda_{\rm S} m_{\rm dyn}}{2\pi^2}\left(M_\Lambda^2-m_{\rm dyn}^2\ln\left(\frac{M_\Lambda^2}{m_{\rm dyn}^2}\right)\right)~,
\end{align}
regulated by a cut off $M_\Lambda$.
\begin{figure}[h!]
  \centering
    \includegraphics[width=0.5\textwidth]{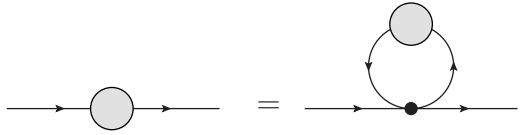}
    \caption{Diagrammatic representation of the gap equation \eqref{eq: gap}.}
    \label{fig: Gap equation}
\end{figure}

One may recognise an avatar of non-perturbativity in Figure \ref{fig: Gap equation} in that we are equating tree-level and loop corrections.
As we will see, non-perturbative character may also be noted in the non-analyticity of the solutions we will find to \eqref{eq: gap}.

Given that $m_{\rm dyn}<M_\Lambda$, when the dimensionful coupling is too small
\begin{align}
	\lambda_{\rm S}<\lambda_{\rm S}\big|_{\rm crit}=\frac{2\pi^2}{M_\Lambda^2}\,,
\end{align}
only $m_{\rm dyn}=0$ is a solution to the gap equation \eqref{eq: gap}. 
Conversely, if $\lambda_{\rm S}>\lambda_{\rm S}\big|_{\rm crit}$, then there exists a non-trivial solution satisfying
\begin{align}
	\omega^2\ln(\omega^2)=\frac{1}{g}-1~<0\,,\quad 
	\omega\equiv \frac{m_{\rm dyn}}{M_\Lambda}\,,\quad
	g\equiv\frac{\lambda_{\rm S} M_\Lambda^2}{2\pi^2}>1\,.
	\label{eq: Gap 2}
\end{align}
For the gap equation \eqref{eq: Gap 2} to be solved with a non-trivial dynamical mass we may then infer the condition $1/g-1<-e^{-1}$. 
The dimensionless coupling constant $g$ must then satisfy
\begin{align}
	1<g\leq\frac{1}{1-e^{-1}}\simeq1.58\,,
	\label{finetuning}
\end{align}
which will be assumed in the following.

Equation \eqref{eq: Gap 2} may be solved exactly via the transcendental Lambert W-function, defined as the set of functions $W$ which satisfy 
\begin{align}	
	z=W\left(z\right)e^{W\left(z\right)}\, \forall z \in \mathbb{C}\, ,
\end{align}
yielding the relation for the dimensionless dynamical mass.
\begin{align}
		\omega^2=e^{W\left(g^{-1}-1\right)}\,,
		\label{musolution}
\end{align}
where we note that since \eqref{eq: Gap 2} admits multiple solutions for a given value of $g$ (e.g. for $g\to1$, we may have $\omega\to1$ or $\omega\to0$), \eqref{musolution} must also necessarily be multivalued.
This mandates consideration of both the principal and lower branches of $W\left(z\right)$, which we denote $W_{0}\left(z\right)$ and $W_{-1}\left(z\right)$ respectively. 

\begin{figure}[h!]
  \centering
    \includegraphics[width=0.5\textwidth]{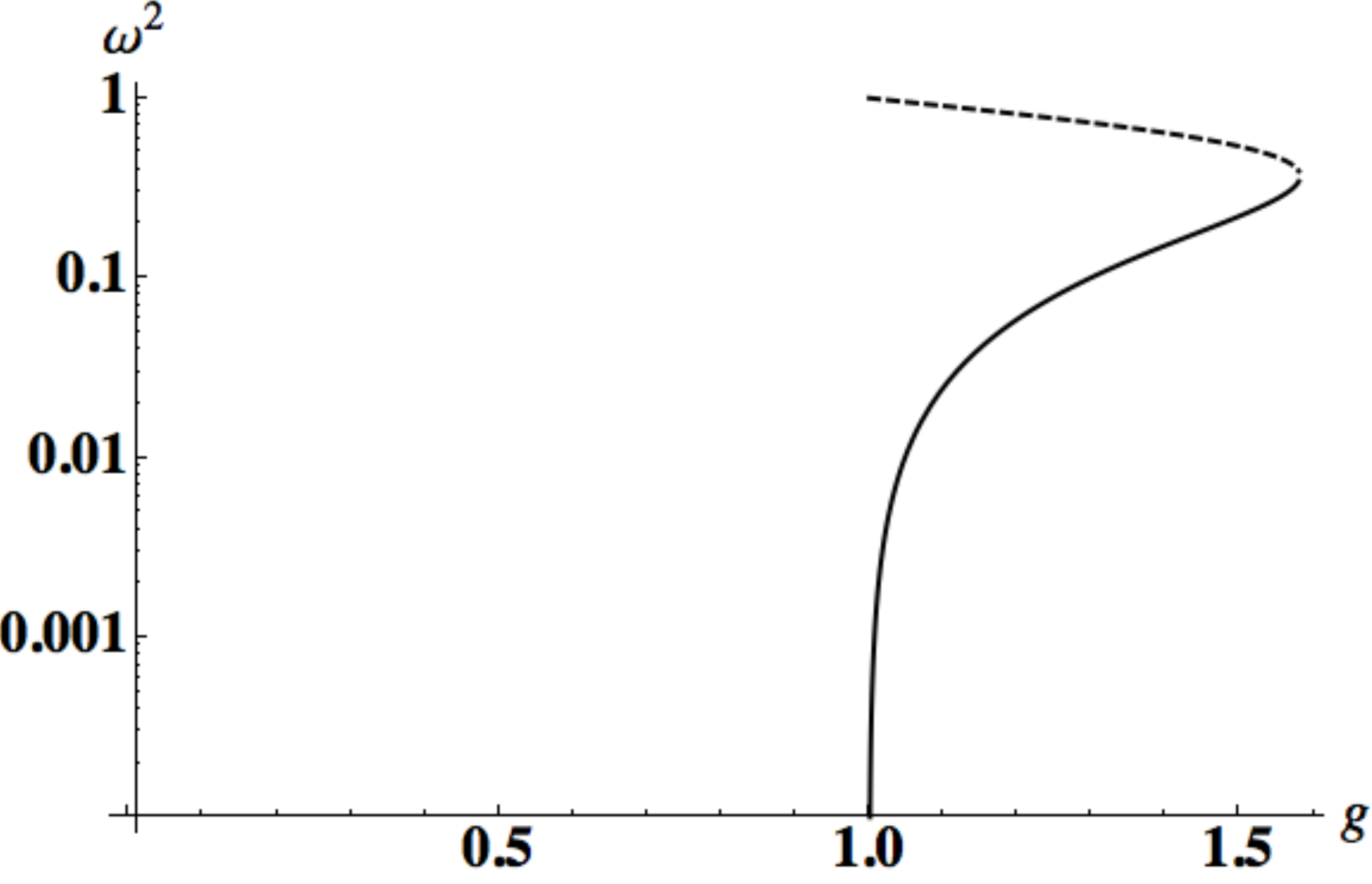}
    \caption{The principal (dashed) and lower branches of \eqref{musolution}. 
    As can clearly be observed, the non-trivial dynamical mass cannot be reached via the weak coupling regime.
    Furthermore, the non-analytic structure of the non-perturbative contributions is reflected in the multivalued nature of the dimensionless dynamical mass.}
    \label{fig: muplot}
\end{figure}

We also note that as shown in Figure \ref{fig: muplot}, for the dynamical mass to be small compared to the cut off, it is necessary that $g\simeq1$, or equivalently that $\lambda_{\rm S}\simeq \lambda_{\rm S}\big|_{\rm crit.}$.
This is the familiar tuning that we expect to reconcile scalar masses below a cutoff scale in the presence of a quadratic divergence.

In what follows we will typically assume just such a sufficiently near-critical coupling, despite the fact that in so doing we will be stretching the validity of the gap-equation based approach.
This is largely done in order to ultimately explore how far the results of this analysis can be taken, and in the knowledge that the investigation we perform here is more at the level of proof of concept rather than serious phenomenology.

\section{Matters of the coupling constant}
\label{sec: coupling constant}

As is the case of QCD, condensation phenomena are generally associated with strongly coupled physics.
One concern is then that the comparative weakness of gravitational interactions may be insufficient to realise behaviour of this type in supergravity.

It is first of all relevant to note that we expect the strength of gravitational interaction to increase as one moves closer to the Planck scale \cite{Dvali:2001gx}.
This may be seen via simple considerations of gravity coupled to a field of mass $M$.
As one probes distances smaller than $M^{-1}$ we expect positive and negative energy quanta of the field to be produced, creating a gravitational dipole distribution.
As is familiar from the analogous case of quantum electrodynamics, this has the effect of anti-screening the gravitational interaction, enhancing the associated field strength.

In the absence of knowledge about the matter species existing between the observable and Planck scales, not to mention the possible influence of extra-dimensional scenarios, the status of the strength of gravitational coupling at sufficiently high energies is thus somewhat unclear \cite{Dvali:2001gx, Calmet:2008tn,Calmet:2008rv}.

This is furthermore compounded by a notable aspect of the present context.
As we will discuss in the following, whilst the value of the overall four-fermion coupling in supergravity is related by local supersymmetry to the gravitational coupling, and thereby to the requirement that it flow to Newton's constant at low energies, there exists a known ambiguity in the context of mean field theory affecting the coupling into the scalar channel we are interested in \cite{Jaeckel:2002rm}.

That this coupling can indeed be sufficiently strong to realise condensation phenomena is the topic of the following section.

\subsection{Fierz ambiguity}

As outlined previously, one perspective on gravitino condensation arises from linearising the four-fermion interactions via suitable auxiliary fields, e.g.	
\begin{align}
	\kappa^2\left(\overline\psi_\mu\psi^\mu\right)^2\sim-\sigma^2+2\sigma\kappa\left(\overline\psi_\mu\psi^\mu\right)\,,
	\label{eq: linearisation}
\end{align}	
where the non-zero scalar vacuum expectation value $\langle\sigma\rangle$ can then source a dynamical mass $m_{\rm dyn}$. 

These four-fermion terms arise in the supergravity action from chapter 3, where
\begin{align}
	&\mathcal{L}_{\rm torsion}
	=-\frac{\kappa^2}{32}\left(\left(\overline{\psi}^\rho\gamma^\mu\psi^\nu\right)\left(\overline{\psi}_\rho\gamma_\mu\psi_\nu+2\overline{\psi}_\rho\gamma_\nu\psi^\mu\right)
	-4\left(\overline{\psi}_\rho\gamma_\mu\psi^\mu\right)\left(\overline{\psi}^\rho\gamma_\mu\psi^\mu\right)\right)\,,
	\label{eq: torsion lagrangian}
\end{align}
with the latter terms vanishing in the gauge $\gamma_\mu\psi^\mu=0$.
In order to make contact with \eqref{eq: linearisation} we may firstly rewrite $\mathcal{L}_{\rm torsion}$ as scalar, pseudoscalar and pseudovector-squared terms via the Fierz identities given in appendix C, enabling the bilinear $\left(\overline{\psi}^\rho\gamma^\mu\psi^\nu\right)\left(\overline{\psi}_\rho\gamma_\mu\psi_\nu\right)$ to be expanded into
\begin{align}
	-\left(\overline\psi_\nu\psi^\nu\right)\left(\overline\psi_\rho\psi^\rho\right)
	-\left(\overline\psi_\nu\gamma^5\psi^\nu\right)\left(\overline\psi_\rho\gamma^5\psi^\rho\right)
	-\frac{1}{2}\left(\overline\psi_\nu\gamma^5\gamma^\alpha\psi^\nu\right)\left(\overline\psi_\rho\gamma^5\gamma_\alpha\psi^\rho\right)\,,
\end{align}
where other possibilities such as vector-squared vanish due to the Majorana nature of the gravitino.

There is however a further Fierz identity \eqref{eq: trilinear} for Majorana fields
\begin{align}\label{trilinear}
	\left(\overline\lambda_{\mu}\lambda^\mu\right)\left(\overline\lambda_{\nu}\lambda^\nu\right)
	=-\left(\overline\lambda_{\mu}\gamma^5\lambda^\mu\right)\left(\overline\lambda_{\nu}\gamma^5\lambda^\nu\right)
	=\frac{1}{4}\left(\overline\lambda_{\mu}\gamma^5\gamma_\alpha\lambda^\mu\right)\left(\overline\lambda_{\nu}\gamma^5\gamma^\alpha\lambda^\nu\right)\,,
\end{align}
which allows these terms to be rotated into one another.
Choosing, for example, to rewrite the four-fermion interactions entirely in the scalar-squared form, we arrive in appendix C at
\begin{align}
	\mathcal{L}_{\rm torsion}=-\frac{3}{16}\kappa^2\left(\overline\psi_\mu\psi^\mu\right)^2\,.
\end{align}
Equivalent pseudoscalar-squared and pseudovector-squared forms are also equally admissible, as are a continuum of possible mixed forms interpolating between these edge cases.

Given the importance of the prefactor of each term and the inherent ambiguity associated to their value, we then must make use of the general parametrisation
\begin{align}
	\mathcal{L}_{\rm torsion}=
	\lambda_{\rm S}\left(\overline\psi_\mu\psi^\mu\right)^2
	+\lambda_{\rm PS}\left(\overline\psi_\mu\gamma^{5}\psi^\mu\right)^2 
	+\lambda_{\rm PV}\left(\overline\psi_\mu\gamma^{5}\gamma^\nu\psi^\mu\right)^2\,,
\end{align}
with the couplings $\lambda_{\rm S}$, $\lambda_{\rm PS}$  and $\lambda_{\rm PV}$ into each channel ultimately unfixed.

This Fierz ambiguity is inherent to the linearisation approach, or equivalently mean-field theory in general, in which we distribute the original four-fermion vertex into assumed-to-be-independent scalar, pseudoscalar and pseudovector channels \cite{Jaeckel:2002rm}.
Concrete knowledge of the actual relative coupling strengths into these channels would require knowledge beyond the pointlike limit, and thus beyond the supergravity approximation.

One would then expect some resolution via an embedding within some string-theoretic framework.
However an exact renormalisation-group analysis, in the spirit of those applied to NJL models may also allow reduction of the associated ambiguity \cite{Jaeckel:2002rm}.

As such, we will approach $\lambda_{\rm S}$ as a free parameter.
One resulting concern is then the possibility that a sufficiently strong $\lambda_{\rm S}$ may necessarily require similar behaviour in $\lambda_{\rm PS}$ and/or $\lambda_{\rm PV}$, leading to the formation of undesirable pseudoscalar and pseudovector condensates.
We may address this issue in the following way. 

\subsection{Unwanted condensates}
\label{sec: unwanted condensates}

Noting firstly that since the couplings $\lambda_{\rm S}$, $\lambda_{\rm PS}$ and $\lambda_{\rm PV}$ must resum to yield the unambiguous expression of the previous chapter, they can only span a two-dimensional parameter space.
As demonstrated via the identities of appendix C, this implies the relation
\begin{align}\label{couplings}
	\left(\lambda_{\rm S}-\lambda_{\rm PS}+4\lambda_{\rm PV}\right)
	=-\frac{3}{16}\kappa^2\,,
\end{align}
from which it is clear that e.g. if all pseudoscalars and pseudovectors are expressed as scalars, providing zero coupling into those channels, then $\lambda_{\rm S}=-3/16 \kappa^2$.

To fully circumvent the Fierz ambiguity we then require two further constraints on these couplings, which we may compute in flat space and shall hope to hold in generality. 

It is first of all sensible to expect that looking solely in the scalar channel we find a non-zero vacuum expectation value $\langle\sigma\rangle$.
For a non-zero and phenomenologically desirable gravitino mass (i.e. $0<m_{\rm dyn}/M_{P}<<1$), the results of the previous section then provide a first constraint in the form 
\begin{align}
	\lambda_{\rm S}\simeq	\lambda_{\rm S}\big|_{\rm crit.}=\frac{2\pi^2}{M_\Lambda^2}\,.
	\label{eq: scalar coupling}
\end{align}

\begin{figure}[h!]
  \centering
      \includegraphics[width=0.7\textwidth]{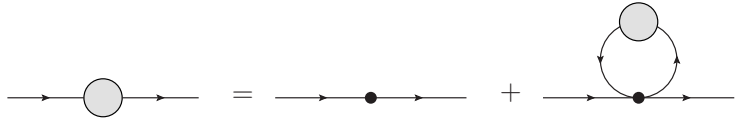}
    \caption{Schwinger-Dyson equation corresponding to \eqref{SD}.}
    \label{fig:SD graphs}
\end{figure}

We may also make use of the lowest order Schwinger-Dyson equation discussed previously and shown diagrammatically in Figure \ref{fig:SD graphs},
\begin{align}
	G_{\rm F}^{\mu\nu}{}^{-1}=G_{\rm F0}^{\mu\nu}{}^{-1}+\Sigma_{\rm F}^{\mu\nu}\,,
	\label{SD}
\end{align}
where, again working around the condensation scale, $G_{\rm F}$ is the full fermion propagator, $G_{\rm F0}$ the free propagator, and $\Sigma_{\rm F}$ the self-energy.

On-shell we have $i\gamma^{\mu\nu\rho}p_\rho+\gamma^{\mu\nu}m_{\rm dyn}=0$, which implies via \eqref{SD} that
\begin{align}
	i\gamma^{\mu\nu\rho}p_\rho=-\gamma^{\mu\nu}m_{\rm dyn}
	=-\Sigma_{\rm F}^{\mu\nu}\,,
\end{align}
providing a relation for $m_{\rm dyn}$ in terms of the self-energy $\Sigma_{\rm F}^{\mu\nu}$, which can itself be expressed again via the trace of the full propagator.
Assuming this holds in generality at the condensation scale, then taking into account the scalar, pseudoscalar and pseudovector channels, as they all contribute to the self energy, this implies a gap equation which is a simple generalisation of \eqref{eq: gap} incorporating all three couplings. 

In the rainbow approximation, when the dressed vertices are replaced by their tree-level values, this takes the form
\begin{align}
	m_{\rm dyn}=\frac{\left(\lambda_{\rm S}+\lambda_{\rm PS}-\lambda_{\rm PV}\right)}{2\pi^2}\int^{M_\Lambda}_0 p^3 dp \frac{m_{\rm dyn}}{p^2+m_{\rm dyn}^2} \,,
	\label{SD1}
\end{align}
where anti-commutation of gamma matrices yields the relative sign difference of $\lambda_{\rm PV}$, and the same caveats discussed at the end of section 4.2.3 surrounding the use of \eqref{eq: gap} again apply.
The requirement again of nontrivial and phenomenologically desirable solutions to this equation then provides an analogous relation for the couplings into all three channels
\begin{align}
	\left(\lambda_{\rm S}+\lambda_{\rm PS}-\lambda_{\rm PV}\right)\simeq
	\left(\lambda_{\rm S}+\lambda_{\rm PS}-\lambda_{\rm PV}\right)\big|_{\rm crit.}
	=\frac{2\pi^2}{M_\Lambda^2}\, ,
	\label{eq: critical coupling}
\end{align}
which we may also assume to be satisfied.
With three relations in three variables, this can then excise the last vestiges of Fierz ambiguity by suitably fixing the relative values of the couplings.

We may conclude from the assumption of a suitably near-critical coupling, in conjunction with the requirement of supersymmetry of the action, the presence of a condensate in the scalar channel, and that a lowest-order Schwinger-Dyson equation is satisfied, the relations \eqref{eq: scalar coupling}, \eqref{eq: critical coupling} and \eqref{couplings} imply 
\begin{align}
	\lambda_{\rm S}\simeq\frac{2\pi^2}{M_\Lambda^2}\,,\quad
	\lambda_{\rm PS}\simeq\lambda_{\rm PV}\,,\quad 
	\lambda_{\rm PS}\simeq-\frac{\kappa^2}{16}-\frac{2\pi^2}{3M_\Lambda^2}\,.
\end{align} 
Given these reasonable criteria we then have the favourable scenario of an attractive coupling into the scalar channel, and repulsive couplings into the pseudoscalar and pseudovector channels, eliminating the possibility of undesirable pseudoscalar or pseudovector condensates.

We also find that, despite the requirements of local supersymmetry, the four-fermion coupling into the scalar channel is actually independent of considerations relating to the measured value of Newton's constant.

\section{Wavefunction renormalisation}
\label{sec: wavefunction renormalisation}

With the notion of dynamical gravitino condensation concretely established, we may now turn to the computation of the wavefunction renormalisation of this scalar mode.
This is a necessary step in deriving the effective Lagrangian describing physics below the condensation scale, but is furthermore of particular importance if we are to make use of $\langle\overline\psi_\mu\psi^\mu\rangle$ in an inflationary context.

This arises because the earlier derivation of inflationary constraints was in the case of a canonically normalised potential, where the kinetic term for a real scalar field has coefficient 1/2.
Given an effective Lagrangian describing the gravitino bound state 
\begin{align}
	\mathcal{L}_{\rm eff}=\frac{Z\kappa^2}{2}\partial_\mu\sigma\partial^\mu\sigma-V_{\text{eff}}(\sigma)\,,
\end{align}
we must rescale via $\tilde\sigma\equiv\kappa\sqrt{Z}\sigma$ to arrive at the canonically normalised Lagrangian
\begin{align}
	\tilde{\mathcal{L}}_{\rm eff}=\frac{1}{2}\partial_\mu\tilde\sigma\partial^\mu\tilde\sigma-\tilde V_{\text{eff}}(\tilde\sigma)\,,
\end{align}
where the coupling constants in the potential $\tilde V_{\text{eff}}$ are defined as
\begin{align}
	\tilde V_{\text{eff}}^{(n)}(0)\equiv\frac{V_{\text{eff}}^{(n)}(0)}{Z^{n/2}\kappa^n}\,.
	\label{Vn}
\end{align}
The latter normalisations ultimately yield the slow roll parameters 
\begin{align}
	\epsilon_\sigma=\frac{1}{Z\kappa^2}\frac{M_{P}^2}{2}\left(\frac{V_{\text{eff}}'}{V_{\text{eff}}}\right)^2\,,\quad
	\eta_\sigma=\frac{1}{Z\kappa^2}M_{P}^2\frac{V_{\text{eff}}''}{V_{\text{eff}}}\,.
	\label{eq: modified slow roll}
\end{align}

As is evident, a desirable suppression of the slow roll parameters may be realised for a large wavefunction renormalisation.
This occurs, as will be demonstrated shortly, for the lower branch of \eqref{eq: gap} only.

Particular salience may be attached in the present context to the computation of the composite wavefunction renormalisation as, unlike fundamental fields, composite fields are not constrained by unitarity to have $0\leq Z\leq1$ \cite{Higashijima:2003et}.
Even outside of the present context this is perhaps a kernel of a useful idea in that a composite inflaton can evade the $\eta$ problem and achieve viable inflation in general via non-trivial wavefunction renormalisation.

One may further recollect that as discussed previously, the inflationary hypothesis supplants the initial value problem of the standard cosmology with another issue.
Namely, the flatness of the inflaton potential, and the question of why the inflaton initially in just such an appropriate region of field space to take advantage of this flatness.
Regarding the latter question we may note that there is no initial condition problem in the present context, in that by associating the inflaton to a symmetry breaking phase transition it is natural for it to originate at the origin of its potential.

\subsection{Bethe-Salpeter equation}

With a view to deriving the wave function renormalisation $Z$ of the gravitino bound state $\langle\overline\psi_\mu\psi^\mu\rangle$, we may leverage methodology from the authors of \cite{Bardeen:1989ds, Miransky:1988xi, Miransky:1989ds} in the context of top quark condensation.
More specifically, we may describe the existence of the bound state via the Bethe-Salpeter equation, for the scalar bound state propagator $\Gamma$. 
As the basis for the covariant treatment of bound states, there exists unsurprisingly a wide body of literature on the use of the Bethe-Salpeter equation in the context of condensation processes, exemplified in the review \cite{Hoyer:2014gna}.

At root it may be schematically derived by declaring a Schwinger-Dyson type equation for the $2\to2$ scattering Greens's function of the form
\begin{align}
	G_{2\to2}=K+G_{2\to2} S K\,,
\end{align}
which is valid if the interaction kernel $K$ satisfies 
\begin{align}
	K=\left(1+G_{2\to2}S\right )^{-1}=G_{2\to2}-G_{2\to2} S G_{2\to2}+\dots\,,
	\label{eq: interaction kernel}
\end{align}
for some propagator function $S$ connecting successive $2\to2$ scatterings, which can then be chosen to enable ease of exact computation of $K$.

Since the existence of bound states cannot be seen at any fixed order in perturbation theory, this choice is important as it will be therefore necessary to resum the infinite series of graphs implicit in \eqref{eq: interaction kernel} in order to detect the presence of the condensate.
This will be most easily achieved in a simplified approximation where we consider only a subclass of graphs.

\begin{figure}[h!]
  \centering
      \includegraphics[width=0.85\textwidth]{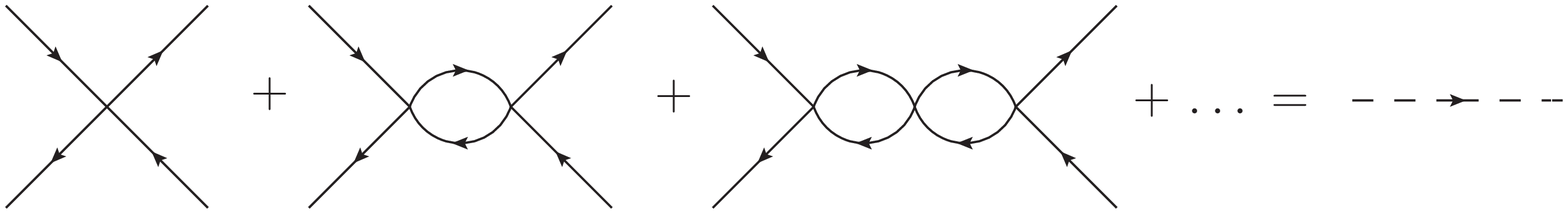}
    \caption{Bubble sum generated by the four-fermion interaction, which resums to yield the scalar bound state propagator.}
    \label{fig: Bubble sum}
\end{figure}

Presently, we will primarily make use of the rainbow-ladder approximation, where we use only tree-level, rather than dressed, vertices and consider only diagrams of ladder topology. 
These simplifications will then reduce the infinite series of graphs to those shown in Figure \ref{fig: Bubble sum}, which may be resummed straightforwardly.

We will furthermore implicitly make use of the fermionic bubble approximation in circumstances with more fields than just a single gravitino present, and thus neglect the loop contributions of all other species.
As an aside, given $N$ species of fermion propagating in the loops this approximation would then become exact in the large $N$ limit.

The relevant Bethe-Salpeter equation is expressed in terms of the gravitino propagator $P_{\mu\nu}$ defined in \eqref{eq: gravitino propagator} in a self-consistent manner via
\begin{align}
	\Gamma=-\frac{\lambda_S}{2}+i\frac{\lambda_S}{2}\Tr\int P_{\mu\nu}\Gamma P^{\nu\mu}\,,
\end{align}
which leads by iteration to the geometric series of bubble graphs in shown in Figure \ref{fig: Bubble sum}, created from $B\equiv i(\lambda_S/2)\Tr\int P_{\mu\nu}P^{\nu\mu}$. 
These may be resummed as a geometric series to yield 
\begin{align}
	\Gamma=-\frac{\lambda_S}{2}\left(1+B+B^2+B^3+\cdots\right)=-\frac{\lambda_S/2}{1-B}\,,
	\label{eq: BS}
\end{align}
which we can assume to be valid around the condensation scale.

Each bubble graph is calculated for both particles with momentum $k_\mu/2$, where $k_\mu$ is the centre of mass momentum, which leads to
\begin{align}
	B\left(k\right)&=\frac{i\lambda_S}{8}~\Tr\int\frac{d^4p}{\left(2\pi\right)^4}\frac{\gamma_\mu\left( \left(p- k/2\right)_\rho\gamma^\rho-m_{\rm dyn}\right)\gamma_\nu\gamma^\nu\left(\left(p+ k/2\right)_\delta\gamma^\delta-m_{\rm dyn}\right)\gamma^\mu}{\left[\left(p-k/2\right)^2-m_{\rm dyn}^2\right]\left[\left(p+k/2\right)^2-m_{\rm dyn}^2\right]}\nonumber\\
	&=2i\lambda_S\int\frac{d^4p}{\left(2\pi\right)^4}\frac{4p^2-k^2+4m_{\rm dyn}^2}{\left(\left(p-k/2\right)^2-m_{\rm dyn}^2\right)\left(\left(p+k/2\right)^2-m_{\rm dyn}^2\right)}\nonumber\\
	&=2i\lambda_S\int\frac{d^4p}{\left(2\pi\right)^4}\frac{4\left(\left(p-k/2\right)^2-m_{\rm dyn}^2\right)+2\left(4m_{\rm dyn}^2-k^2\right)+4k_\mu p^\mu}
{\left(\left(p-k/2\right)^2-m_{\rm dyn}^2\right)\left(\left(p+k/2\right)^2-m_{\rm dyn}^2\right)}\\
	&=8i\lambda_S\int\frac{d^4p}{\left(2\pi\right)^4}\frac{1}{p^2-m_{\rm dyn}^2}
+4i\lambda_S\int\frac{d^4p}{\left(2\pi\right)^4}\frac{\left(4m_{\rm dyn}^2-k^2\right)}{\left(p^2-m_{\rm dyn}^2\right)\left(\left(p+k\right)^2-m_{\rm dyn}^2\right)}\,\nonumber.
\end{align}
The resummation \eqref{eq: BS} then implies
\begin{align}
	\Gamma\left(k\right)=-\frac{\lambda_S}{2}\Bigg(1&-8i\lambda_S\int \frac{d^4p}{\left(2\pi\right)^4}\frac{1}{p^2-m_{\rm dyn}^2}\nonumber\\
	&-4i\lambda_S\int \frac{d^4p}{\left(2\pi\right)^4}\frac{\left(4m_{\rm dyn}^2-k^2\right)}{\left(p^2-m_{\rm dyn}^2\right)\left(\left(p+k\right)^2-m_{\rm dyn}^2\right)}\Bigg)^{-1}\,,
\end{align}
where the cancellation of the first two terms follows if one assumes the gap equation \eqref{eq: gap} to be satisfied for $m_{\rm dyn}\ne0$. 
Consequently there is no explicit quadratic divergence in the propagator
\begin{align}
	\Gamma(k)&=\frac{\lambda_S}{2}\left(4i\lambda_S\int \frac{d^4p}{\left(2\pi\right)^4}\frac{\left(4m_{\rm dyn}^2-k^2\right)}{\left(p^2-m_{\rm dyn}^2\right)\left(\left(p+k\right)^2-m_{\rm dyn}^2\right)}\right)^{-1}\nonumber\\
	&=-\left(\frac{\left(4m_{\rm dyn}^2-k^2\right)}{2\pi^2}\int_0^1 dx\ln\left(\frac{M_\Lambda^2}{m_{\rm dyn}^2-x\left(1-x\right)k^2}\right)+\dots\right)^{-1}\,,
	\label{Gammap}
\end{align}
where ellipsis indicates finite terms.
The wavefunction renormalisation for the gravitino bound state is then
\begin{align}
	Z&=\frac{1}{2\pi^2}\int_0^1 dx\ln\left(\frac{M_\Lambda^2}{m_{\rm dyn}^2-x\left(1-x\right)k^2}\right)\nonumber\\
	&=\frac{1}{2\pi^2}\ln\left(\frac{M_\Lambda^2}{m_{\rm dyn}^2}\right)+{\cal O}\left(k^2\right)
	\simeq-\frac{1}{2\pi^2}\ln\left(\omega^{2}\right)\,.	
	\label{eq: Z}
\end{align}

\begin{figure}[h!]
  \centering
      \includegraphics[width=0.55\textwidth]{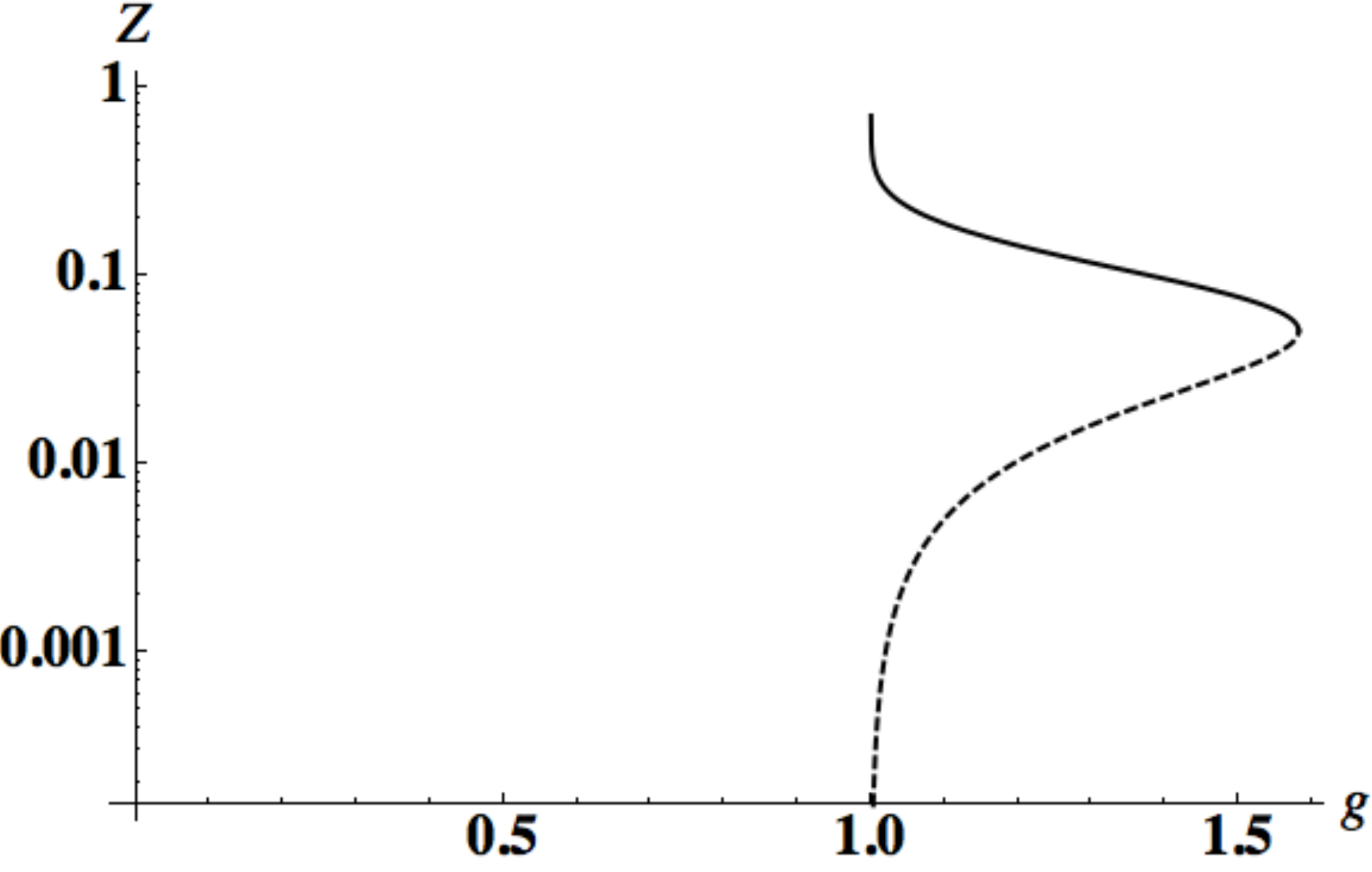}
    \caption{Principal (dashed) and lower branches of the wavefunction renormalisation \eqref{eq: Z}, where the multivalued structure is inherited from the dimensionless dynamical mass $\omega$ \eqref{musolution}.}
    \label{fig: Z}
\end{figure}

As is clear from Figure \ref{fig: Z}, only the lower branch is admissible if we wish to avoid a small $Z$.
This is of course the case if we wish to make use of the condensation mechanism to drive early universe inflation, by virtue of the factor of $1/Z$ carried by the slow roll parameters in \eqref{eq: modified slow roll}.

An interesting observation outside of the present context is that since $Z$ derives from the effect of resummed bubble graphs, we expect it to enjoy an $N$-fold enhancement in the instance of $N$ species of fermion propagating in the loops.
Indeed, this intuition is borne out in calculations of precisely this scenario \cite{Higashijima:2003et, Bardeen:1989ds}.
The suppression of slow roll parameters with a composite inflaton, and possible evasion of the $\eta$ problem, can then be engineered for sufficiently large $N$.

A subtle point also remains in that the naive mass prediction for the condensate, arising from the pole at $k^2=4m_{\rm dyn}^2$ in the propagator \eqref{Gammap}, would appear to suggest that since the condensate has exactly the mass of its constituents, it must therefore have vanishing binding energy.
This is not, of course, the hallmark of a robust phenomenon.

Where this intuition fails however is in that the condensate is not a non-relativistic bound state of the familiar type, and renormalisation group effects must be accounted for.
At the condensation scale we may imagine a dynamic equilibrium between the gravitino and condensate fields, where it is then sensible to have vanishing binding energy as the condensate forms on the edge of stability.
Below this scale however, the respective runnings of the scalar condensate and fermionic gravitino mass parameters should not match.
This renormalisation at a relevant infrared energy scale can then result in a positive binding energy \cite{Bardeen:1989ds}, as expected.

Having established in this chapter the possibility of dynamical mass generation for the gravitino, it should however be clear that something is missing from this picture.
As established previously, any gravitino mass is inextricably linked to a cosmological constant in supergravity, conspicuously absent from the flat-space considerations presented here.
This is furthermore compounded by the necessity of the gravitational degrees of freedom, and the as-of-yet unaddressed role they may play in this context
\footnote{As we will see, they are intrinsically linked to the stability of the condensate.}.

As such, this will be the central topic addressed in the following chapter. 
In so doing we will derive a curved-space effective potential for the scalar condensate, valid for maximally symmetric backgrounds.

%% file: Chapters/Condensation_II.tex
Having demonstrated the possibility of gravitino condensation via the flat-space approach of the previous chapter, we may now revisit the problem in a more complete fashion, taking into account both spacetime curvature and the gravitational degrees of freedom.
That this is a necessary step to concretely establish this phenomenon follows as a general consequence of the supersymmetric requirement of graviton modes in any supergravity theory, and the concomitant linkage between gravitino mass terms and the cosmological constant.

As stated, our approach is to leverage the four-fermion gravitino self-interaction terms which provide the spin-connection torsion necessitated by local supersymmetry. 
If the latter condense, the gravitino field would acquire a non-zero mass whilst leaving the graviton massless.
Local supersymmetry would then be broken dynamically, in the spirit of chiral symmetry breaking in the NJL model.

It has been conjectured that exactly such a mechanism could occur, with analyses based on the one-loop effective potential of a simple ${\mathcal N}=1$ supergravity model in Minkowski spacetime \cite{Jasinschi:1983wr,Jasinschi:1984cx}, with the choice of background being for computational simplicity and to permit unambiguous definition of the gravitino mass via the condensate field. 
Therein, the effective potential of the condensate field was seen to acquire a non-trivial minimum for some values of a momentum cutoff relative to the gravitational constant, thereby providing a Planck-scale dynamical gravitino mass.  

Although appealing in many respects, this approach was criticised in \cite{Buchbinder:1989gi,Odintsov:1988wz} as it ignores both the quantum fluctuations of the metric field and the possible role of spacetime curvature. 
Indeed, following the generic approach of \cite{Fradkin:1983mq} of calculating one-loop effective potentials in supergravity theories via expansion about (anti)de Sitter, rather than Minkowski, backgrounds, the authors of \cite{Buchbinder:1989gi} argue that integrating over metric fluctuations inevitably introduces imaginary terms into the effective potential, irrespective of the value of the background cosmological constant $\Lambda$, and for any non-trivial value of the gravitino condensate field.

Imaginary terms would of course indicate the instability of the non-trivial gravitino condensate vacuum, implying that there could be no possibility of breaking local ${\mathcal N}=1$ supersymmetry dynamically in this simple and universal manner. 
It should be noted of course that the `traditional' way of engineering local supersymmetry breaking in supergravity via firstly dynamically breaking global supersymmetry through, say, gaugino condensation, before communicating it to the gravitational sector, evades these arguments \cite{Ferrara:1982qs, Barbieri:1982eh, Nilles:1982ik}. 
We must apparently then conclude that the `traditional' approach is the only consistent way of dynamically breaking local supersymmetry in supergravity, carrying the price of necessarily coupling the theory to specific choices of matter fields and potential, and therefore losing a degree of universality.

It should be noted however that the arguments and analysis of \cite{Buchbinder:1989gi,Odintsov:1988wz} do not incorporate the role of the Goldstino, known to be necessary for the super-Higgs effect, as outlined in chapter 3, and the concomitant development of a gravitino mass.
It is therefore unclear whether these purported instabilities are truly pathological, or possibly gauge artefacts, or merely reflect the absence of the requisite degrees of freedom needed for the gravitino to become massive.

We shall therefore revisit the analysis and arguments of \cite{Buchbinder:1989gi,Odintsov:1988wz} from scratch in this chapter, with a view to fully incorporating the erstwhile absent super-Higgs effect. 
Given possible ambiguities that may arise from issues of gauge dependence \cite{Jackiw:1974cv}, in so doing we will pay particular attention to gauge fixing.
These steps are nonetheless important, in order to elucidate the possible gauge dependence of the final result.

As we will demonstrate in detail, the proper incorporation of the Goldstino in the general framework of \cite{Fradkin:1983mq} enables the dynamical breaking of local supersymmetry in supergravity prior to its coupling to arbitrary matter or gauge fields. 
More specifically, we find that an analysis of a one-loop effective potential which fully incorporates gravitational fluctuations about (anti)de Sitter backgrounds, reveals the existence of stable non-trivial vacua, contrary to the claims of \cite{Buchbinder:1989gi,Odintsov:1988wz}. 

The double-well shape of the resultant effective potential and its capability to vanish at non-trivial minima also indicates that the contribution to the cosmological constant arising from the super-Higgs effect can be responsible for the overall vanishing of the effective vacuum energy in this context.
This is of course consistent with the generic features of dynamical supersymmetry breaking outlined in chapter 3, according to which the vacuum energy of broken global supersymmetry is necessarily positive, whilst vacua with broken local supersymmetry can still be characterised by zero vacuum energy.

Interestingly however, this analysis will also reveal that whatever physical instabilities arise necessarily do so from spin-two fluctuations only.
In that sense the authors of \cite{Buchbinder:1989gi,Odintsov:1988wz} were correct, in that these effects were invisible to the fermion-only analysis of \cite{Jasinschi:1983wr, Jasinschi:1984cx}.
We will also note, thanks to careful gauge fixing analysis, the presence of instabilities arising inevitably in certain unfortunate choices of gauge.

The material presented in this chapter is based upon the article \cite{Alexandre:2013iva}.

\section{Preliminaries}

Our starting point is the $\mathcal{N}=1$ supergravity action from chapter 3
\begin{align}
	&S=\int d^4x\,e\left(\frac{1}{2\kappa^2}e^{a\mu}e^{b\nu}R_{\mu\nu ab}\left(e\right)-\frac{1}{2}\overline{\psi}_\mu\gamma^{\mu\nu\rho}D_\nu\psi_\rho+\mathcal{L}_{\rm torsion}\right)\,,\nonumber\\
	&\mathcal{L}_{\rm torsion}
	=-\frac{\kappa^2}{32}\left(\left(\overline{\psi}^\rho\gamma^\mu\psi^\nu\right)\left(\overline{\psi}_\rho\gamma_\mu\psi_\nu+2\overline{\psi}_\rho\gamma_\nu\psi_\mu\right)
	-4\left(\overline{\psi}_\rho\gamma_\alpha\psi^\alpha\right)\left(\overline{\psi}^\rho\gamma_\beta\psi^\beta\right)\right)\,.
	\label{eq: torsion lagrangian 2}
\end{align}
where the old-minimal auxiliary fields $S$, $P$ and $A_\mu$ have been set for simplicity to their on-shell values, and so play no role in the following.

The torsion terms are firstly simplified via the gauge choice $\lambda=\gamma_\mu\psi^\mu=0$, which we may understand from now on as having been imposed. 
As outlined in chapter 3 this gauge conveniently removes the Goldstino from the supergravity action, leaving only a negative contribution to the cosmological constant. 

Given the freedom explored in the previous chapter to rewrite $\mathcal{L}_{\rm torsion}$ as scalar, pseudoscalar and pseudovector terms squared, we may make use of the general parametrisation
\begin{align}
	\mathcal{L}_{\rm torsion}=
	\lambda_{\rm S}\left(\overline\psi_\mu\psi^\mu\right)^2
	+\lambda_{\rm PS}\left(\overline\psi_\mu\gamma^{5}\psi^\mu\right)^2 
	+\lambda_{\rm PV}\left(\overline\psi_\mu\gamma^{5}\gamma^\nu\psi^\mu\right)^2\,,
\end{align}
where other possibilities such as vector squared vanish due to the Majorana nature of the gravitino.
To elucidate dynamical gravitino mass generation we may linearise these terms via the equivalence
	\begin{align}	
		\frac{1}{2\kappa^2}R\left(e\right)+\lambda_{\rm S}\left(\overline\psi_\mu\psi^\mu\right)^2+\dots\sim
		\frac{1}{2\kappa^2}R\left(e\right) - \sigma^2 + 2\sqrt{\lambda_{\rm S}}\sigma\left(\overline{\psi}_\mu\, \psi^\mu \right)+\dots ,
	\end{align}
which follows from the subsequent Euler-Lagrange equation for the auxiliary scalar $\sigma$. 
Since $\sigma^2>0$, comparison with the Einstein-Hilbert Lagrangian $\left(R-2\Lambda\right)$ indicates that this term functions as a positive cosmological constant, corresponding to a de Sitter background in our conventions.

The pseudoscalar and pseudovector interaction terms may be neglected in this regard as their linearisation does not produce terms of canonical form for mass terms.
As demonstrated in the previous chapter, we may self-consistently assume attractive coupling in the scalar channel and repulsive coupling in the pseudoscalar and pseudovector channels, ensuring the possibility of scalar condensation without unwanted Lorentz-violating condensates also forming.

Accounting for the Goldstino-induced negative cosmological constant term $f^2$, arising in \eqref{va3}, the relevant effective Lagrangian is then	\begin{align}
		&\mathcal{L}_{\rm eff}=
		\frac{1}{2\kappa^2}R(e) + f^2 - \sigma^2 
		-\frac{1}{2}\overline\psi_\mu\gamma^{\mu\nu\rho} D_\lambda\psi_\rho
		+2\sqrt{\lambda_{\rm S}}\sigma \left(\overline\psi_\mu\, \psi^\mu\right)\,.
		\label{actionauxiliary}
	\end{align}
Following the normalisation of \cite{Fradkin:1983mq} for the gravitino mass and noting $\gamma^{\mu\nu}\equiv\frac{1}{2}\left[\gamma^\mu,\gamma^\nu\right]$,
	\begin{equation}
	\frac{1}{2}\overline\psi_\mu\gamma^{\mu\nu\rho} D_\lambda\psi_\rho
	+m \left(\overline\psi_\mu\gamma^{\mu\nu}\psi_\nu\right)\bigg|_{\gamma\cdot\psi=0}
	=\frac{1}{2}\overline\psi_\mu\gamma^{\mu\nu\rho} D_\lambda\psi_\rho
	-m \left(\overline\psi_\mu\psi^\mu\right)\,,
	\label{normmass}
	\end{equation}	
and we may conclude that if loop effects endow $\sigma$ with a non-zero vacuum expectation value $\langle \sigma \rangle \ne 0$,
then there must be a dynamically generated effective gravitino mass of 
	\begin{equation}\label{ginomass}
		m =2 \sqrt{\lambda_{\rm S}}\left\langle\sigma\right\rangle\,, 
	\end{equation}
thus breaking local supersymmetry. 

\subsection{Cosmological constant}

We may then define a tree-level cosmological constant
\begin{align}
		\Lambda_0  \equiv \kappa^2 \left(\sigma^2- f^2\right)\,
		\Rightarrow\, \mathcal{L}_{\rm eff}=\frac{1}{2\kappa^2}\left(R\left(e\right)-2\Lambda_0\right)+\dots\,,
		\label{eq: bare}
\end{align}
which is dressed by quantum corrections to provide the renormalised cosmological constant $\Lambda$.  
Although de Sitter space may not be a solution of the bare $\mathcal{N}=1$ supergravity equations of motion, it may in this sense be a solution of a quantum effective action, after the fluctuations of the metric and other fields are taken into account. 

Following \cite{Fradkin:1983mq}, we then assume for the purposes of our analysis that one may compute the one-loop effective potential about a de Sitter background with a positive renormalised cosmological constant, $\Lambda > 0$, the value of which will ultimately be determined
via minimisation of the one-loop effective potential. 
To preserve supersymmetry at tree-level, we will ultimately find that, as one may expect, the bare cosmological constant $\Lambda_0$ must be negative.
This corresponds to anti de Sitter space in our conventions.

As discussed in chapter 3, local supersymmetry mandates an apparent gravitino mass term in the presence of a non-zero cosmological constant \cite{Townsend:1977qa}. To ensure that any such gravitino mass generated is indeed physical, we will then take the limit $\Lambda \to 0$ to demonstrate the perseverance thereof in the flat space-time limit.

Since $\Lambda$ may equivalently be viewed as the overall energy density at the level of the one-loop effective potential $V_{\rm eff}$, a self-consistency condition of the limit $\Lambda \to 0$ is that $V_{\rm eff}\left(\left\langle\sigma\right\rangle\neq0\right)=0$, ensuring that $\Lambda$ is indeed vanishing at whatever non-trivial minima we find.
This then sets the value of $f^2$, which enters via the bare cosmological constant $\Lambda_0$ in equation \eqref{eq: bare}. 
 
\subsection{Calculation overview} 
 
Our algorithm is therefore to firstly compute the one-loop effective potential for the condensate mode $\sigma$ in de Sitter space, appropriately Euclideanised.
Then, we will solve the corresponding effective equations in the limit of vanishing renormalised cosmological constant $\Lambda$, providing a straightforward interpretation of any resultant signatures of gravitino mass. 

One may interpret this excursion through the, generally non-supersymmetric, de Sitter space purely in the spirit of Euclidean continuation, as is appropriate for path integrals, enhancing convergence as a consequence of the underlying compactness, and providing a physical theory in the limit $\Lambda\to 0$.
We may however note a tension here in that whilst we are considering a Majorana gravitino field, there are in fact no Majorana representations on Euclidean de Sitter space \cite{Pilch:1984aw}.
Since we are approaching this continuation largely as a technical necessity, we will however proceed with the understanding that a physical result will only be obtained upon the conclusion of these manipulations.

Computing a one-loop effective potential for a given theory in a non-trivial background requires a number of operations that we must take account of. 
Firstly, one must expand to compute fluctuations of the action to quadratic order about the classical background, via decompositions of 
the type $\tilde g_{\mu\nu}\to g_{\mu\nu}+h_{\mu\nu}$, where $g_{\mu\nu}$ is for our purposes the standard Euclidean $dS_4$ metric, for which
\begin{align}
	R_{\lambda\mu\nu\rho}=\frac{\Lambda}{3}\left(g_{\lambda\nu}g_{\mu\rho}-g_{\lambda\rho}g_{\mu\nu}\right)\,, \quad
	R_{\mu\nu}=\Lambda g_{\mu\nu}\,,\quad
	\int d^4x\,\sqrt{g}=\frac{24\pi^2}{\Lambda^2}\,.
\end{align}
Since all fermionic terms in \eqref{actionauxiliary} are already quadratic, and Lorentz invariance forbids any fermionic background terms, working to one-loop in this instance provides the advantage of decoupling the graviton and gravitino sectors.

Given that we are ultimately computing the one-loop effective potential for the auxiliary field $\sigma$ in order to assess $\left\langle\sigma\right\rangle$, it is sufficient to generally identify $\sigma$ with its vacuum expectation value. 
Indeed, as an auxiliary field it does not propagate at tree level, and so a kinetic term, which would be obtained via integration over other fields for some non-uniform $\sigma$ configuration, would be therefore purely one-loop. 
Hence the influence of the propagation of $\sigma$ on the effective potential, which is already of one-loop order, must be of at least two-loop order. 

We must also further decompose every field present, including ghosts, into those that are `natural' to our background geometry; more precisely, 
to fields that correspond to irreducible representations of the underlying SO(5) isometry group.
It is the spectra of these fields that can be reliably found via knowledge of the underlying representation theory, which then permits us to compute their contributions to effective potential.

\section{Bosonic sector \label{sec:quadr}}

As detailed in appendix A, varying the Ricci scalar to quadratic order in the fluctuation $h_{\mu\nu}\equiv\overline h_{\mu\nu}+g_{\mu\nu} h/4$, yields in the vierbein formalism \footnote{Where there are extra terms present relative to the metric formalism.}
	\begin{align}
		\frac{1}{4\kappa^2}\int d^4 x\sqrt{g}&\left(\frac{1}{2} \overline h_{\mu\nu}\left(-\nabla^2+X_1\right) \overline h^{\mu\nu}
		-\frac{1}{8}h\left(-\nabla^2-X_2\right)h
		-\left(\nabla^\mu  \overline h_{\mu\nu}-\frac{1}{4}\nabla_\nu h\right)^2\right)\,,\nonumber\\
		&X_1
		=\frac{13}{6}\Lambda-\frac{3}{2}\Lambda_0\,,\quad
		X_2
		=\frac{5}{2}\Lambda_0-\frac{1}{2}\Lambda\,.
	\end{align}

To now decompose into `irreducible fields' we apply a standard transverse-traceless decomposition to the fields and their functional measures
	\begin{align}
		&\label{eq2} V_\mu=V_\mu^\perp+\nabla_\mu\phi\,, \quad 
		\nabla^\mu V_\mu^\perp=0\,,\quad
		\mathcal{D}V=\mathcal{D}V^\perp\mathcal{D}\phi\sqrt{\det\Delta_0\left(0\right)}\\
		&\label{decomp2}
		\overline h_{\mu\nu}=\overline h_{\mu\nu}^\perp+\nabla_\mu\xi_\nu^\perp+\nabla_\nu\xi_\mu^\perp
		+\nabla_{\mu}\nabla_{\nu}\chi
		-\frac{1}{4}g_{\mu\nu}\nabla^2\chi\,,\quad
	 	g^{\mu\nu}\overline h_{\mu\nu}=0\,,\\
		&\nabla^\nu\overline h_{\mu\nu}^\perp=0\,,\quad
		\mathcal{D}\overline{h}
		=\mathcal{D}\overline{h}^\perp\mathcal{D}\xi^\perp\mathcal{D}\chi\sqrt{\det\Delta_1\left(-\Lambda\right)\Delta_0\left(-\frac{4}{3}\Lambda\right)\Delta_0\left(0\right)}\,,
	\end{align}
where we have also defined a class of integer-spin operators for constant $X$
\begin{align}
		\Delta_s\left(X\right)\;&\equiv\left(-\nabla^2+X\right)\,,
		\label{decompositions1}
	\end{align}
understood to variously act upon post-decomposition scalar, vector and tensor fields.
These are the operators whose spectra we shall ultimately compute. 

There are however extra zero-modes present for these decomposed operators, which must be correctly accounted for.
In the simple example
\begin{align}
	A_\mu\left(-\nabla^2+X\right)A^\mu
	=\delta^{\mu\nu}A_\mu^\perp\Delta_1\left(X\right)A_\nu^\perp+\phi\Delta_0\left(X-\Lambda\right)\Delta_0\left(0\right)\phi\,,
\end{align}
the operator $\left(-\nabla^2+X\right)$ has no non-trivial zero modes, however $\Delta_0\left(0\right)$  gives a zero eigenvalue acting on any constant field.
Upon taking account of these modes later we will however find that in the limit $\Lambda\to0$ their contributions are subleading.
Identifying these modes with the underlying isometries of de Sitter spacetime, their disappearance in this limit should of course be unsurprising.

To counteract the complication of introducing these extra fields, we will make use of the following identities, which are derived by straightforward substitution for an $S^4$ background 
	\begin{align}		
	\label{eq: vector identity}			  	
		&V^\mu\left(\left(-\nabla^2+X\right)\delta_{\mu\nu}+Y\nabla_{\mu}\nabla_{\nu}\right)V^\nu\\\nonumber
		&=\begin{pmatrix}
			V_\mu^\perp\\
			\phi\\
		\end{pmatrix}^T\cdot
		\begin{pmatrix}
			\delta^{\mu\nu}\Delta_1\left(X\right)\\
			\left(1-Y\right)\Delta_0\left(0\right)\Delta_0\left(\frac{X-\Lambda}{{1-Y}}\right)
		\end{pmatrix}_{\text{Diag.}}\cdot
		\begin{pmatrix}
			V_\nu^\perp\\
			\phi\\
		\end{pmatrix}\,,	
	\end{align}
\begin{align}
	&\overline{h}^{\mu\alpha}\left(\left(-\nabla^2+X\right)\delta_{\mu\nu}+Y\nabla_{\mu}\nabla_{\nu}\right)\overline{h}^{\nu}{}_\alpha\\\nonumber
	&=\begin{pmatrix}
		\overline{h}^{\perp\mu\alpha}\\
		\xi^{\mu\perp}\\
		\chi
	\end{pmatrix}^T\cdot
	\begin{pmatrix}
		\Delta_2\left(X\right)\\
		\left(2-Y\right)\Delta_1\left(-\Lambda \right)\Delta_1\left(\frac{3 Y \Lambda -10 \Lambda +6 X}{6-3 Y}\right)\\
		\frac{3}{16}\left(4-3Y\right)\Delta_0\left(0\right)\Delta_0\left(-\frac{4 \Lambda }{3}\right)
		\Delta_0\left(\frac{4 ((3 Y-8) \Lambda +3 X)}{12-9 Y}\right)
	\end{pmatrix}_{\text{Diag.}}\cdot
	\begin{pmatrix}
		\overline{h}^{\perp}_{\mu\alpha}\\
		\xi_\mu^\perp\\
		\chi
	\end{pmatrix}\,,
\end{align}
for some constants $X$ and $Y$.

\subsection{Gauge fixing}
In the bosonic sector there are two symmetries present; local Lorentz and infinitesimal coordinate transformations.
We may fix the former by setting to zero the antisymmetric part of the vierbein \cite{VanNieuwenhuizen:1981ae}, resulting in ghost fields which are non-propagating and can then be disregarded here.

A standard two-parameter covariant gauge fixing term \cite{Fradkin:1983mq} may be added to remove the coordinate gauge freedom 
	\begin{align}
		S_B^{(GF)}=-\frac{1}{4\kappa^2}\frac{1}{\alpha}\int d^4 x\sqrt{g}\left(\nabla^\mu  \overline h_{\mu\nu}-\frac{1+\beta}{4}\nabla_\nu h\right)^2\,,
		\label{bosonic gauge fixing}
	\end{align}
necessitating addition of the ghost action
	\begin{align}
		S_B^{(GH)}=\frac{1}{4\kappa^2}\frac{1}{\alpha}\int d^4 x\sqrt{g}\;
		\overline{C}^\mu\left(\left(-\nabla^2-\Lambda\right)\delta_{\mu\nu}+\frac{\beta-1}{2}\nabla_{\mu}\nabla_{\nu}\right)C^\nu~,
	\end{align}
for a complex anticommuting vector field $C$.

Making use of the first identity \eqref{eq: vector identity} and integrating yields the ghost partition function for the bosonic sector
	\begin{align}
		\mathcal{Z}^{(GH)}_B
		=\det\Delta_1\left(-\Lambda\right)\Delta_0\left(\frac{4 \Lambda }{\beta -3}\right)\,,
	\end{align}
where we have disregarded irrelevant prefactors.

\subsection{Physical gauge}\label{sec:physical gauge}

There exists however a secondary approach to the issue of fixing a gauge. 
This is achieved via so-called `physical' gauges; gauges representing an alternative to the conventional Faddeev-Popov method, inasmuch as they are based upon simply isolating gauge degrees of freedom essentially disregarding them, leaving only the `physical' degrees of freedom left \cite{Percacci:2013ii}. 

In practice this is achieved in the following way.
As is well-known, gauge fixing is required in order to render path integrals well defined.
Absent this, we naturally overcount field configurations which are related by gauge transformations and thus physically equivalent.
Gauge fixing conditions remedy this by specifying a hypersurface which intersects each orbit of the gauge group, with the inclusion of a Faddeev-Popov ghost determinant in the path integral measure cancelling the non-trivial curvature associated to non-Abelian gauge symmetries.
This is not however the only path by which we may proceed.

Re-examining the gauge fixing condition \eqref{bosonic gauge fixing} in the illuminative limit $\alpha\to0$, we may strongly impose \footnote{Which is to say, use at the level of the action, rather than solely imposing via Lagrange multiplication.} 
the condition
	\begin{align}
		\nabla_\mu \overline h^\mu{}_\nu-\frac{\beta+1}{4}\nabla_\nu h=0\,.
	\end{align}
For $\beta=0$, substituting the decomposition \eqref{decomp2} then implies that 
\begin{align}
	\nabla_\mu\nabla^\mu\xi_\nu+\nabla_\mu\nabla_\nu\xi^\mu=0\,,\quad
	\xi_\mu=\xi_\mu^{T}+\nabla_\mu\chi\,.
\end{align}

We may therefore conclude that $\xi_\mu$ is a Killing vector field, and as such, generates the underlying diffeomorphism symmetry present.
Strongly imposing that $\xi_\mu=0$ is therefore a `physical' gauge condition, in that we disregard the components of the graviton which correspond to general coordinate transformations.
Functionally integrating over these degrees of freedom then yields an infinite constant prefactor corresponding to the volume of the diffeomorphism group, which is unimportant for our purposes.

There do exist some additional complications to this approach regarding accounting for the zero-modes of $\xi_\mu$, which should not necessarily be disregarded even if $\xi_\mu=0$. 
As however in the previous instance of extra zero-modes arising from field decompositions, their contributions are subleading in the limit $\Lambda\to0$. 

As effective potentials such as these are often gauge dependent, the direct physical gauge approach may be viewed as more `natural' in this context.
As leveraged in a number of situations similar to these \cite{Percacci:2013ii, Gaberdiel:2010xv, Zhang:2012kya}, it can certainly offer significant computational simplification too.

These considerations may of course also be applied to the gauge fixing of the fermionic sector, however for the purposes at hand it suffices to utilise this procedure only for the bosonic fields.
As we will see, it is the behaviour of the bosonic sector that ultimately dictates the stability of the effective potential in this context.

\subsection{Bosonic partition function}

Incorporating all additional terms up until now, the quadratic gravitational action becomes
	\begin{align}
		S_B^{(2)}=\frac{1}{4\kappa^2}\int d^4 x\sqrt{g}\bigg(&\frac{1}{2} \overline h_{\mu\nu}\left(\left(-\nabla^2
		+X_1\right)\delta^{\mu}{}_\alpha+2\left(1+\frac{1}{\alpha}\right)\nabla^{\mu}\nabla_\alpha\right) \overline h^{\nu\alpha}\\\nonumber
		&-\frac{\left(3\alpha+\beta^2\right)}{16\alpha}h\left(-\nabla^2-\frac{2X_2\alpha}{\left(3\alpha+\beta^2\right)}\right)h
		-\frac{\alpha+\beta}{2\alpha}\overline{h}_{\mu\nu}\nabla^{\mu}\nabla^{\nu}h\bigg)\,,
		\label{gauge fixed action}
	\end{align}
which would be straightforward to functionally integrate, were it not for the final term. 

However, \eqref{decomp2} firstly implies that $\overline{h}^\perp_{\mu\nu}$ is conserved and so cannot mix with $h$ here.
Furthermore, on an Einstein background there is no $\left(\xi^\perp,h\right)$ mixing \cite{Lauscher:2001ya}.
Eliminating the final $\left(\chi,h\right)$ mixing could be achieved via a so-called `diagonal gauge' \cite{Percacci:2013ii}, however this would then be incompatible with a Landau-DeWitt gauge choice, known to correspond to the unique gauge-invariant one-loop effective potential in the case of pure Einstein gravity \cite{Fradkin:1983nw}.
As such, we will proceed instead without yet specifying a gauge. 

The scalar part of the action is schematically of the form
	\begin{align}
		\frac{1}{4\kappa^2}\int d^4 x\sqrt{g}\left(
		\begin{pmatrix}
			h && \chi
		\end{pmatrix}
		\cdot
		\begin{pmatrix}
			A_1 && B\\
			B && A_2
		\end{pmatrix}
		\cdot
		\begin{pmatrix}
			h \\ 
		\chi
		\end{pmatrix}\right)\,,
	\end{align}
with matrix elements	
	\begin{align}
		A_1&=-\frac{1}{16\alpha}\left(-\left(3\alpha+\beta^2\right)\nabla^2-2X_2\alpha\right)\,,\nonumber\\
		A_2&= -\frac{3 (\alpha +3)}{16 \alpha}\Delta_0\left(0\right)\Delta_0\left(-\frac{4}{3}\Lambda\right)\Delta_0\left(\frac{4 (\alpha -3) \Lambda -6 \alpha  X_1}{3 (\alpha +3)}\right)\,\nonumber\\
		B&=-\frac{3\left(\alpha+\beta\right)}{16\alpha}\Delta_0\left(0\right)\Delta_0\left(-\frac{4}{3}\Lambda\right)\,.
	\end{align}
Integration then provides an expression for the bosonic partition function $\mathcal{Z}^{(B)}$ as
	\begin{align}
		&\mathcal{Z}^{(GH)}_B
		\left(\frac{\det\Delta_0\left(-\frac{4}{3}\Lambda\right)\Delta_0\left(0\right)}
		{\det\Delta_2\left(X_1\right)\Delta_1\left(\alpha\left(\frac{2 }{3}\Lambda-X_1\right)-\Lambda\right)
		\left(A_1A_2-B^2\right)}\right)^{1/2}\,,\\
		&=\mathcal{Z}^{(GH)}_B
		\bigg(\det\Delta_2\left(X_1\right)
		\Delta_1\left(\alpha\left(\frac{2}{3}\Lambda-X_1\right)-\Lambda\right)\nonumber\\
		&\hspace{3cm}\times\Delta_0\left(\frac{A_3+ \sqrt{A_4+A_5}}{6\left(\beta-3\right)^2}\right)
		\Delta_0\left(\frac{A_3- \sqrt{A_4+A_5}}{6\left(\beta-3\right)^2}\right)\bigg)^{-1/2}\,,
		\nonumber
	\end{align}
	\begin{align}
		A_3&= 4\Lambda  \left(6 \alpha +\beta^2+6 \beta -9\right)
		-6 X_1 \left(3 \alpha +\beta^2\right)
		+6(\alpha +3) X_2\,,\nonumber\\
		A_4&=4 (2 \Lambda(6\alpha+\beta(\beta +6)-9)
		-3X_1 (3\alpha+\beta^2)
		+3(\alpha +3) X_2)^2\,,\nonumber\\
		A_5&=48 (\beta -3)^2 X_2 (3 \alpha X_1-2 (\alpha -3)\Lambda)\,,
	\end{align}
disregarding again an irrelevant prefactor.

A first important check is to reproduce known results from the literature.
Making the substitutions $\left\{X_1\to\frac{8}{3}\Lambda-2\Lambda_0,X_2\to2\Lambda_0,\beta\to1,\alpha\to0\right\}$, corresponding to the action of Einstein gravity in the metric formalism, in Landau-DeWitt gauge, we find 
	\begin{align}\label{grlimit}
		\mathcal{Z}^{(B)}&\to\left(\frac{\det\Delta_1\left(-\Lambda\right)\Delta_0\left(-2\Lambda\right)}
		{\det\Delta_2\left(\frac{8}{3}\Lambda-2\Lambda_0\right)
		\Delta_0\left(-2\Lambda_0\right)}\right)^{1/2}\,,
	\end{align}
which is precisely the partition function given in \cite{Fradkin:1983mq}.
In other gauges, equivalent results similarly follow. 

\section{Fermionic sector} 

We will follow largely the same approach on the fermionic side as that utilised in the bosonic sector.
However, rather than starting from the Euclidean action, we will utilise \eqref{actionauxiliary} and perform the analytic continuation at an opportune moment.

Given the absence of fermionic background terms, the action
	\begin{align}
		S_F^{\left(2\right)}
		=\int d^4 x\sqrt{-g}\left(-\frac{1}{2}\overline{\psi}_\mu\gamma^{\mu\nu\rho} D_\nu\psi_\rho
		+2\sqrt{\lambda_S}\sigma \left(\overline{\psi_\mu}\psi^\mu\right)
		\right)\,,
		\label{fermion action}
	\end{align}
is already quadratic in the fluctuations, so no further expansion is necessary.

From the standard decompositions for the fields and their functional measures
	\begin{align}
		\psi_\mu&=\phi_\mu+\frac{1}{4}\gamma_\mu\varphi\,, \quad 
		\gamma^\mu\phi_\mu=0\,,\quad
		\phi_\mu=\phi_\mu^\perp+\left(D_\mu-\frac{1}{4}\gamma_\mu\slashed{D}\right)\vartheta\,,
		 \label{eq: fermion decomposition 1} \\
		&D^\mu\phi_\mu^\perp=0\,,\quad
		\mathcal{D}\psi_\mu=\frac{\mathcal{D}\phi^\perp\mathcal{D}\varphi\mathcal{D}\vartheta}{\sqrt{\det\Delta_{1/2}\left(-\frac{4}{3}\Lambda\right)}}\,,
		\label{eq: fermion decomposition} 
	\end{align}
where $\varphi=0$ in our gauge choice $\gamma_\mu\psi^\mu=0$. 
We again define a class of constant-$X$ operators, for half-integer spins
	\begin{align}\label{decompositions}
		\Delta_{1/2}(X)\;\varphi
		&\equiv\left(-D^2+\Lambda+X\right)\varphi\,,\\
		\Delta_{3/2}(X)\;\phi_\mu^\perp
		&\equiv\left(-D^{2}+\frac{4}{3}\Lambda +X\right)\phi_\mu^\perp\,,
		\nonumber
	\end{align}
with explicit $\Lambda$ terms for coherence with the literature and future convenience.

Euclideanising via
	\begin{align}
		\gamma^0\to i\gamma^0_E~, \quad
		\gamma^j\to \gamma^j_E~, \quad
		e^0\to e^0_E~, \quad 
		e^j\to ie^j_E\,,
	\end{align}
then implies the transformation $\slashed{D}\to i\slashed{D}_E$, which, since Dirac operators `square' to the Laplacian, in the sense that
	\begin{align}
		\slashed{D}_E^2=-D^2+\frac{R}{4}\,,
	\end{align}
provides the useful transformation 
	\begin{align}\label{dirac square}
		-\slashed{D}^2\to-D^2+\frac{R}{4}\,.
	\end{align}
We may then simultaneously Euclideanise the theory and remove the $\slashed{D}$ operators.

Using \eqref{dirac square} as appropriate, we may streamline computations as before via a general $S^4$ identity given in \cite{Fradkin:1983mq} for some constant $X$
	\begin{align}
		&\frac{1}{2}\overline{\psi}_\mu\gamma^{\mu\nu\rho} D_\nu\psi_\rho
		-X \left(\overline{\psi}_\mu\psi^\mu\right)
		=\frac{1}{2}\overline{\phi}^\perp_\mu\left(\slashed{D}-X\right)\phi^{\perp\mu}\\
		&+\frac{3}{16}
		\begin{pmatrix}
			\overline\vartheta \\
			 \overline\varphi
		\end{pmatrix}^T\cdot
		\begin{pmatrix}
			\left(\slashed{D}+2X\right)\Delta_{1/2}\left(-\frac{4}{3}\Lambda\right) && -\Delta_{1/2}\left(-\frac{4}{3}\Lambda\right)\nonumber \\
			-\Delta_{1/2}\left(-\frac{4}{3}\Lambda\right) && -\left(\slashed{D}-2X\right)
		\end{pmatrix}\cdot 
		\begin{pmatrix}
			\vartheta \\ 
			\varphi
		\end{pmatrix}\,.
	\end{align}

\subsection{Gauge fixing}

Although a gauge condition has already been imposed in the fermionic sector, to implement it consistently with a view to preserving local supersymmetry we may firstly take a more general approach, before specialising specifically to $\gamma_\mu\psi^\mu=0$.

Fixing the local supersymmetry present in the gravitino sector mandates that we supplement \eqref{fermion action} with some gauge-fixing term.
In order to preserve on-shell supersymmetry, this term may be derived via the variation of the bosonic gauge fixing term \eqref{bosonic gauge fixing} under the supersymmetry transformation $\delta h_{\mu\nu}=\kappa\bar{\epsilon}\gamma_{(\nu}\psi_{\mu)}$, where $\epsilon$ is assumed to be Killing, therefore obeying the $S^4$ relation $D_\mu\epsilon=\frac{1}{2}\sqrt{-\frac{R}{12}}\gamma_\mu\epsilon$ \cite{Fujii:1985bg}.

As known from other circumstances \cite{Percacci:2013ii,Kojima:1993hb}, a strict proportionality between bosonic and fermionic gauge fixing terms is difficult to engineer due to $\phi^\perp_\mu$ terms in the variation of \eqref{bosonic gauge fixing}.
A compromise is then to find a proportionality in the following fashion.

Varying before subsequently taking the $\gamma$-trace, with a mind to eventually applying the constraint $\gamma_\mu\psi^\mu=\varphi=0$, we find
	\begin{align}\label{fermion gauge 1}
		\gamma^\mu\left(\nabla_{\nu}\delta \overline h^\nu{}_\mu+\frac{\beta+1}{4}\nabla_\mu \delta h\right)
		= \frac{3}{2}\kappa\bar{\epsilon}\left(D^2+\frac{R}{12}\right)\vartheta\,,
	\end{align}
which suggests, as a consequence of the relation
\begin{align}\label{fermion gauge 2}
		\left(\slashed{D}+\sqrt{-\frac{R}{3}}\right)F
		= \left(D^2+\frac{R}{12}\right)\vartheta\,,
	\end{align}
the following gauge-fixing term	 
	\begin{align}
		S_F^{(GF)}=\frac{1}{2}\int d^4 x\sqrt{-g}\;\overline{F} \left(\slashed{D}+\sqrt{-\frac{R}{3}}\right)F\,, \quad 
		F= \left(\slashed{D}-\sqrt{-\frac{R}{3}}\right)\vartheta\,.
		\label{fermion gauge 3}
	\end{align}

Varying $F$ about the classical background, where $S=P=A_\mu=0$, we can then find the corresponding ghost action via decomposition of $\delta\psi_\mu$ in the supersymmetry variation
\begin{align}
	\delta\psi_\mu=\frac{1}{\kappa}D_\mu\epsilon\,,
\end{align} 
to yield
	\begin{align}
		\left(D^2-\frac{R}{12}\right)\left(\delta\vartheta-\frac{\epsilon}{\kappa}\right)=0\,,
	\end{align} 
which results in the ghost action
	\begin{align}
			S_F^{(GH)}=\frac{1}{\kappa}\int d^4 x\sqrt{-g}\;\overline\eta\left(\slashed{D}-\sqrt{-\frac{R}{3}}\right)\eta\,,
	\end{align}
for some spin $1/2$ complex commuting field $\eta$.

Requiring on-shell gauge independence mandates that we also take account of so-called third, or Nielsen-Kallosh, ghosts \cite{Bastianelli:2006rx, Nielsen:1978mp, Kallosh:1978de}, which arise from the non-trivial $\left(\slashed{D}+\sqrt{-\frac{R}{3}}\right)$ operator present in the gauge fixing condition \eqref{fermion gauge 3}.
Exponentiating, these take the form
	\begin{align}
		S_F^{(NK)}=\int d^4 x\sqrt{-g}\;\left(\overline\omega \left(\slashed{D}+\sqrt{-\frac{R}{3}}\right)\omega
		+\overline\rho \left(\slashed{D}+\sqrt{-\frac{R}{3}}\right)\rho\right)~,
	\end{align}
for some commuting Dirac and anticommuting Majorana spinor fields $\rho$ and $\omega$, respectively.

Functional integration then gives an expression for the fermionic sector ghost partition function $\mathcal{Z}^{(GH)}_F$ as
	\begin{align}
		\left(\det\left(\slashed{D}-\sqrt{-\frac{R}{3}}\right)\right)^{-1}\left(\det\left(\slashed{D}+\sqrt{-\frac{R}{3}}\right)\right)^{-1/2}
		=\left(\det\Delta_{1/2}\left(-\frac{R}{3}\right)\right)^{-3/4}\,,
	\end{align}
where we are implicitly leveraging \eqref{dirac square} to provide the relation
	\begin{align}
		\det\left(\slashed{D}\pm X\right)
		=\left(\det\Delta_{1/2}\left(|X|^2\right)\right)^{1/2}\,,
	\end{align}
which is true modulo additional zero-modes incurred by the field decomposition, as in the bosonic case.

\subsection{Fermionic partition function}

Combining the various gravitino terms, noting that the spin 1/2 field $\varphi$ in the decomposition \eqref{eq: fermion decomposition 1} vanishes as a consequence of the gauge condition $\gamma_\mu\psi^\mu=0$, we find the quadratic gravitino terms
	\begin{align}
		-\frac{1}{2}\overline{\phi}^\perp_\mu\left(\slashed{D}-2\sqrt{\lambda_S}\sigma\right)\phi^{\perp\mu}
		-\frac{11}{16}\overline{\vartheta}\left(\slashed{D}+\frac{12}{11}\sqrt{\lambda_S}\sigma-\frac{8}{11}\sqrt{-\frac{4}{3}\Lambda}\right)\Delta_{1/2}\left(-\frac{4}{3}\Lambda\right)\vartheta\,,
	\end{align}
which integrate to give the total fermionic sector partition function, including Jacobian factors,
	\begin{align}
		\mathcal{Z}^{(F)}
   		&=\left(\frac{\det\Delta_{3/2}\left(4\lambda_S\sigma^2\right)\Delta_{1/2}
   		\left(\left|\frac{12}{11}\sqrt{\lambda_S}\sigma-\frac{8}{11}\sqrt{-\frac{4}{3}\Lambda}\right|^2\right)}
		{\left(\det\Delta_{1/2}\left(-\frac{4}{3}\Lambda\right)\right)^3}\right)^{1/4}\,,
	\end{align}
and we have again leveraged \eqref{dirac square} to equate
	\begin{align}
		\left(\det\left(\slashed{D}\pm X\right)_{\phi^\perp}\right)
		=\left(\det\Delta_{3/2}\left(|X|^2\right)\right)^{1/2}\,,
	\end{align}
which is true modulo additional zero-modes incurred by the field decomposition, as in the bosonic case.

\section{One-loop partition function \label{sec:potential}}

Having derived the fermionic and bosonic partition functions, we may now compute the one-loop effective potential via
	\begin{align}
		\Gamma=-\ln \left(\mathcal{Z}^{(B)} \mathcal{Z}^{(F)}\right)
		=\frac{1}{2}\ln\det\Delta_{2}\left(X_1\right)+\dots,
	\end{align}
making use of the functional determinant techniques detailed in appendix D, such as 
	\begin{align}\label{mudef}
		\ln\det\left(\frac{\Delta_s}{\mu^2}\right)=
		-\frac{1}{2}B_0 L^4-\frac{1}{2}B_2 L^2
		-B_4\left(\ln\left(\frac{L^2}{\mu^2}\right)
		-\gamma\right)
		+\zeta_s\left(0\right)\ln\left(\frac{\Lambda}{3\mu^2}\right)
		-\zeta_s'\left(0\right)\,,
	\end{align}
where $\gamma$ is the Euler-Mascheroni constant, we have the cut-off $\epsilon=\left(\mu^2/L^2\right)\to0$, and the extra zero modes arising from field decompositions have not yet been accounted for.	
It is important to note that as $\epsilon$ arises from \eqref{eq: Mellin} in appendix D, a proper time integral, $\epsilon\to0$ is the short-time and thus high energy limit.
Decreasing $\mu$ therefore corresponds to flowing from the infrared to the ultraviolet.

As we are investigating a simple model of supergravity, anticipating an embedding into a more ultraviolet-complete theory, we set aside for now the question of renormalisability implicitly encoded in the polynomial divergences above
\footnote{As an aside, $B_0$ is always zero for supersymmetric theories, even if the symmetry is broken, since it corresponds to a supertrace over the number degrees of freedom present.}. 

Focusing instead on the finite parts of the `decomposed' effective potential $V_{\rm eff}$ and the resultant effective equations, 
we may represent the former as
	\begin{align}\label{one-loop action}
		\Gamma=&S
		+\left(B_4-N\right)\ln\left(\frac{\Lambda}{3\mu^2}\right)
		-B_4'~, \quad 
		N=14-\frac{1}{2}\times 8=10\,,\quad
		V_\text{eff}=-\frac{\Lambda^2}{24\pi^2}\Gamma\,,
	\end{align}
where $N$ is the number of extra zero-modes incurred by our decompositions, as alluded to previously and first elucidated in \cite{Fradkin:1983mq}, $24\pi^2/\Lambda^2$ is the usual spacetime volume for an $S^4$ of radius $\sqrt{3/\Lambda}$, and 
	\begin{align}
		S=-\frac{1}{2\kappa^2}\int d^4x\;\sqrt{g}\left(R-2\Lambda_0\right)
		=-\frac{12\pi^2}{\Lambda^2\kappa^2}\left(R-2\Lambda_0\right)\,,	\quad
		R=4\Lambda
	\end{align}
	\begin{align}\nonumber\label{B_4}
		B_4=&\frac{1}{2}\zeta_2\left(0,X_1\right)
		-\frac{1}{4}\zeta_{3/2}\left(0,4\lambda_S\sigma^2\right)
		-\zeta_1\left(0,-\Lambda\right)+\frac{1}{2}\zeta_1\left(0,\alpha\left(\frac{2}{3}\Lambda-X_1\right)-\Lambda\right)\,,\\
		&-\frac{1}{4}\zeta_{1/2}\left(0,\left|\frac{12}{11}\sqrt{\lambda_S}\sigma-\frac{8}{11}\sqrt{-\frac{4}{3}\Lambda}\right|^2\right)
		+\frac{3}{4}\zeta_{1/2}\left(0,-\frac{4}{3}\Lambda\right)-\zeta_{0}\left(0,\frac{4\Lambda}{\beta-3}\right)\,,\nonumber\\
		&\quad+\frac{1}{2}\zeta_{0}\left(0,\frac{A_3+ \sqrt{A_4+A_5}}{6\left(\beta-3\right)^2}\right)
		+\frac{1}{2}\zeta_{0}\left(0,\frac{A_3- \sqrt{A_4+A_5}}{6\left(\beta-3\right)^2}\right)\,,
	\end{align}
	\begin{align}\label{B_4'}
		&B_4'=\frac{1}{2}\zeta_2'\left(0,X_1\right)+\dots
	\end{align}

We emphasise at this point the utility of the asymptotic forms for $\zeta_s$ and $\zeta'_s$
\begin{align}
	\zeta_s\left(0,X\right)\bigg|_{\Lambda\to0}\sim\frac{6s+3}{4}\frac{X^2}{\Lambda^2}\,,\quad
	\zeta'_s\left(0,X\right)\bigg|_{\Lambda\to0}\sim\frac{6s+3}{4}\frac{X^2}{\Lambda^2} \left(\frac{3}{2}-\ln\left(\frac{3X}{\Lambda}\right)\right)\,,
	\label{eq: zetas}
\end{align}
derived in the appendix.
These relations permit us to find an expression for the effective potential in terms of elementary functions, rather than the awkward integrals of special functions from which they originally derive in this instance, enabling a full investigation of $V_{\rm eff}$.

We should also address the presence of an $\mathcal{O}\left(\ln\left(\Lambda\right)\Lambda^{-2}\right)$ term in \eqref{eq: zetas}, which would naturally dominate over any classical contributions in the limit $\Lambda\to0$, apparently indicating the failure of our one-loop approach.
Thankfully, there is however a precise cancellation of any such terms at the level of the effective potential.
Specifically, for each $\zeta'_s$ in \eqref{one-loop action} there is a corresponding $\zeta_s$ with the same coefficient and opposite sign.
Since each $\zeta_s$ gains a factor of $\ln\left(\Lambda\right)$ in \eqref{one-loop action}, the potentially ruinous terms cancel exactly with  corresponding ones arising from \eqref{B_4'}, the cancellation also serving as a non-trivial check of our manipulations up to that point.

As indicated earlier, the contribution to $V_{\rm eff}$ of the $N$ extra zero-modes in \eqref{one-loop action} can also clearly be seen to vanish in the flat space limit.

\subsection{Imaginary terms} 
\label{sec: imaginary terms}

One may also be concerned by the $\ln\left(X\right)$ terms in the above, as $X$ may become negative.
Any resultant imaginary terms in the effective potential, if not artefacts of the one-loop formalism, would then indicate instability and therefore the impossibility of dynamical gravitino condensation in this manner.
This may be addressed in two distinct ways.

Given the freedom to vary $f$ we will firstly consider $\Lambda_0<0$, whereupon after varying $\sigma$ we self-consistently find non-trivial minima with $\sigma^2\leq f^2$.
This is sensible in the present context given the general incompatibility of supersymmetry with de Sitter vacua; if $\Lambda_0$ is positive we expect tree-level breaking and any subsequent gravitino condensation may be irrelevant.

Furthermore, it is notably suggested that quantisation of metric fluctuations about $dS_4$ generically leads to positive Planckian values for $\Lambda$ \cite{Fradkin:1983mq}.
As we are considering the renormalised cosmological constant $\Lambda=\Lambda_0+\mathcal{O}\left(\hbar^2\right)$, to arrive at $\Lambda\sim0$ it is then unsurprising that $\Lambda_0$ be negative to cancel out the positive energy density associated to quantisation.

Given that $X_1\to-3\Lambda_0/2$ for $\Lambda\to0$, inspection of \eqref{B_4} reveals that in that limit the only $\zeta_s'$ functions from \eqref{B_4'} which could be problematic in this sense are 
	\begin{align}\label{zetas2}
		\left\{\zeta'_1\left(0,\frac{3}{2}\alpha \Lambda_0\right),\quad
		\zeta'_{0}\left(0,\frac{A_3-\sqrt{A_4+A_5}}{6\left(\beta-3\right)^2}\right),\quad
		\zeta'_{0}\left(0,\frac{A_3+\sqrt{A_4+A_5}}{6\left(\beta-3\right)^2}\right)\right\}\,,
	\end{align}
which correspond to the fields $\xi_\mu^\perp$, $\chi$ and $h$.
In the gauge $\alpha\to0$, these become
	\begin{align}\label{zetas}
		\left\{\zeta'_1\left(0,0\right)\,,\quad
		\zeta'_{0}\left(0,\frac{3 \left(\beta ^2+5\right)\Lambda_0}{(\beta -3)^2}\right)\,,\quad
		\zeta'_{0}\left(0,0\right)\right\}\,,
	\end{align}
providing, via \eqref{B_4'}, an imaginary contribution to the effective potential which may be freely tuned via the gauge parameter $\beta$. 
This suggests of course that any such terms are non-physical.

Whilst we cannot tune any such terms exactly to zero, on account of the real-valued nature of $\beta$, they may in practice be removed via choices of $\beta$ for which
 	\begin{align}
		\text{Im}\left(\frac{24\pi^2}{\Lambda^2}V_{\text{eff}}\left(\left\langle\sigma\right\rangle\right)\right)=2n\pi\,, \quad
		n\in\mathbb{Z}\,,
	\end{align}
for some non-trivial minimum $\left\langle\sigma\right\rangle$ satisfying $\text{Re}\left(V_{\text{eff}}\left(\left\langle\sigma\right\rangle\right)\right)=0$.
As this is somewhat inelegant however, we will not make use of it in what follows. 

Instead we will make use of the so-called `physical' gauges outlined in \ref{sec:physical gauge}.
This amounts in the present context to disregarding the components of the graviton which we can identify with general coordinate transformations; the fields $\xi_\mu^\perp$ and $\chi$ in our notation.

It is the $`A_3+\sqrt{A_4+A_5}$' term which provides the only non-problematic element of \eqref{zetas2}, which by comparison with the pure Einstein gravity case via \eqref{grlimit} can also be seen to correspond to the trace of the graviton, $h$.
Interestingly, it is then precisely $\xi_\mu^\perp$ and $\chi$ which correspond to the problematic $\zeta'$ functions in \eqref{zetas2}. 
Any imaginary terms in this context must arise solely from what the physical gauge procedure identifies as gauge, rather than physical, degrees of freedom. 
In a physical gauge these $\zeta'$ functions would be absent, assuring the reality of the action for negative $\Lambda_0$.

\subsection{Effective potential} 
\label{sec: effective potential}

Although in principle we may tune $\beta$ to eliminate any imaginary terms in the effective potential, we will instead take the simpler approach of using a physical gauge.
Having derived the effective potential in generality, specialising to this gauge is straightforward; we set $\alpha=\beta\to0$ and excise the fields $\xi_\mu^\perp$ and $\chi$, along with the ghosts that were introduced to cancel out their degrees of freedom \footnote{Which regardless do not contribute for $\Lambda\to 0$.}.

Functional integration over these gauge degrees of freedom then results in an infinite constant prefactor arising from the volume of the diffeomorphism group, which is unimportant for our purposes and can be disregarded.
As also noted in \ref{sec:physical gauge} there are extra zero modes arising from the field decomposition which should generally be accounted for, but whose contributions are subleading in the limit $\Lambda\to0$ and so can also be neglected in the following.

In the limit $\Lambda \to 0$, $\alpha=\beta\to0$, we find
	\begin{align}\label{Veff2}
		&V_{\text{eff}}=
		-\frac{\Lambda_0}{\kappa^2}
		+\Lambda_0^2\left(-\frac{45}{256 \pi ^2}  \ln \left(-\frac{6 \Lambda_0}{\mu^2}\right)
		+\frac{135}{512 \pi ^2}
		+\frac{45  \ln \left(2\right)}{128 \pi ^2}\right)\\
		&+\lambda_S^2 \sigma^4\left(\frac{15289}{29282 \pi ^2}  \ln\left(\frac{\lambda_S \sigma ^2}{ \mu ^2}\right)
		-\frac{45867 }{58564 \pi^2}
		+\frac{ \ln \left(4\right)}{\pi ^2}
		+\frac{648 \ln \left(\frac{24}{11}\right)}{14641 \pi^2}
		-\frac{15289 \ln \left(2\right)}{14641 \pi ^2}\right)\,.
		\nonumber
	\end{align}

We firstly address the $\ln\left(-\Lambda_0\right)=\ln\left(\kappa^2\left(f^2-\sigma^2\right)\right)$ term in \eqref{Veff2}, which has the capability to destabilise the potential for $f^2<\sigma^2$.
As mentioned previously, given the general incompatibility of de Sitter space with supersymmetry, this should not be surprising. 
Given the intention to break local supersymmetry dynamically, tree-level breaking via a positive bare cosmological constant $\Lambda_0$ should render subsequent breaking via a dynamically generated $\left\langle\sigma\right\rangle$ an impossibility.

As such, for a given value of the renormalisation point $\mu$ we must find the value of $f$ ensuring self-consistent minima $\left\langle\sigma\right\rangle$ satisfying $\sigma^2<f^2$ to ensure $\Lambda_0<0$ and thus a real effective potential.
If this condition is violated, we expect an imaginary contribution
	\begin{align}
		\frac{45 i}{256\pi}\Lambda_0^2\,,
	\end{align}
which, sensibly, vanishes when $\Lambda_0=0$.

It is furthermore interesting to note that this instability arises precisely from the irreducible spin two parts of the spectrum; absent their contribution, we find \eqref{Veff2} again albeit lacking the $\Lambda_0^2\left(\dots\right)$ term. 
The resultant potential is therefore real for all $f$, $\sigma$. 

It is therefore true that the authors of \cite{Buchbinder:1989gi,Odintsov:1988wz} were correct to be critical of \cite{Jasinschi:1983wr, Jasinschi:1984cx} regarding the importance of gravitational degrees of freedom.
However, their finding of imaginary contributions for any non-trivial value of $\sigma$ can be traced back to the absence of the Goldstino in their formalism.
Without an extra degree of freedom to absorb the gravitino cannot of course become massive, reflected here in the implicit choice of $f=0$ in \cite{Buchbinder:1989gi,Odintsov:1988wz}, ensuring that their $\Lambda_0$ is necessarily positive.

Having derived the effective potential for the condensate mode we may now examine in greater detail the phenomenon of gravitino condensation, and the physical consequences thereof.
This will be the topic of the next chapter.

%% file: Chapters/Analysis.tex
Having derived both the one-loop effective potential for the scalar gravitino condensate $\sigma$, and the wavefunction renormalisation necessary for the canonical normalisation thereof, we now turn to quantitative analysis of the resulting canonically normalised potential.
In so doing, we will assess the suitability of this approach for both supersymmetry breaking, and early universe inflation.

Establishing various preliminary details in section \ref{sec: Preliminaries}, we will firstly delineate the role of the various parameters appearing in the potential.
This will demonstrate as a consequence the existence of a super-Higgs phase, indicating the development of a dynamical gravitino mass and the associated breaking of local supersymmetry.
As expected this phase exists in a low energy regime, which we may interpret to be the regime below the condensation scale.

Given the existence of such a phase, in section \ref{sec: Supersymmetry breaking} we will analyse the prospects and phenomenology of supersymmetry breaking in this manner.
We demonstrate that, as is expected from a gap-equation based approach, suitably low breaking scales can always be achieved at the expense of tuning the coupling $\lambda_S$ close to criticality, and explore some of the related contextual issues. 

From the tempting interpretation of a scalar condensate as the inflaton, the inflationary possibilities of the canonically normalised potential are then explored in section \ref{sec: Inflation}.
Whilst, as we demonstrate, phenomenologically viable inflation is possible, it is also shown that it cannot coexist with equally viable local supersymmetry breaking.
As such, we are seemingly faced with an exclusive choice between the two. 

An alternative to the obvious inflationary scenario is also assessed, where, in the spirit of the original Starobinsky model, the incorporation of one-loop effects leads to an effective $R^2$ term in the effective description below the condensation scale.
Transforming to the Einstein frame then yields a canonically normalised potential for the `scalaron' degree of freedom associated to this extra term, which is capable for certain parameter choices of providing inflation compatible with the Planck 2015 results.

The results and analysis presented in this chapter are largely based on \cite{Alexandre:2013iva, Alexandre:2013nqa, Alexandre:2014lla}, albeit with some modification owing to improved understanding of the Fierz ambiguity and the associated non-trivial wavefunction renormalisation.

\section{Preliminaries}
\label{sec: Preliminaries}

To recapitulate, as derived in chapter 5 we have the following one-loop effective potential for the condensate mode $\sigma$ in the flat-space limit
	\begin{align}
		&V_{\text{eff}}=
		\label{eq: veff}
		-\frac{\Lambda_0}{\kappa^2}
		+\Lambda_0^2\left(-\frac{45}{256 \pi ^2}  \ln \left(-\frac{6 \Lambda_0}{\mu^2}\right)
		+\frac{135}{512 \pi ^2}
		+\frac{45  \ln \left(2\right)}{128 \pi ^2}\right)\\
		&+\lambda_S^2 \sigma^4\left(\frac{15289}{29282 \pi ^2}  \ln\left(\frac{\lambda_S \sigma ^2}{ \mu ^2}\right)
		-\frac{45867 }{58564 \pi^2}
		+\frac{ \ln \left(4\right)}{\pi ^2}
		+\frac{648 \ln \left(\frac{24}{11}\right)}{14641 \pi^2}
		-\frac{15289 \ln \left(2\right)}{14641 \pi ^2}\right)\,,
		\nonumber
	\end{align}
where 
\begin{itemize}
	\item $\Lambda_0=\kappa^2\left(\sigma^2-f^2\right)$ is the tree-level cosmological constant, which we require to be negative, corresponding to anti-de Sitter space in our conventions, for stability of the potential.
	\item $f$ is the scale of global supersymmetry breaking within the chiral supermultiplet coupled to supergravity to provide the Goldstino.
	In principle there may also be other contributions to $f$ arising from whatever other vacuum expectation values are present.
	Regardless, we will generally set $f$ to the natural mass scale in the theory, the reduced Planck mass $M_P$.
	\item $\lambda_S$ is the four-fermion coupling into the scalar channel arising from the \\$\lambda_S\left(\overline\psi_\mu\psi^\mu\right)^2$ term in the supergravity Lagrangian, unfixed in value due to the Fierz ambiguity.
	\item $\mu$ is the renormalisation point introduced during zeta function regularisation, deriving from a proper time cutoff in appendix D.
	Since $\mu\to 0$ is the short time, and thus high energy limit, increasing $\mu$ corresponds to moving from the ultraviolet to the infrared.
\end{itemize}

\subsection{Canonical normalisation}

Given the normalisation inherited from
\begin{align}
	\frac{1}{2}\kappa^2Z\partial_\mu\sigma\partial^\mu\sigma = \frac{1}{2}\partial_\mu\tilde{\sigma}\partial^\mu\tilde{\sigma}\,,
\end{align} 
to arrive at the canonically normalised potential we must firstly rescale by the identification $\tilde{\sigma}\equiv\kappa\sqrt{Z}\sigma$, with wavefunction renormalisation derived in chapter 4
\begin{align}
	Z=-\frac{1}{2\pi^2}W\left(\frac{2\pi^2}{\lambda_SM_\Lambda^2}-1\right)\,,\quad
	\frac{2\pi^2}{M_\Lambda^2}<\lambda_S\leq \frac{2\pi^2}{M_\Lambda^2}\frac{1}{1-e^{-1}}\,,
	\label{eq: Z 2}
\end{align}
where $W$ is the Lambert W-function, also known as the product logarithm, and $M_\Lambda$ is a flat space cut off.

The use of explicit cutoff regularisation in deriving the wavefunction renormalisation, and zeta function regularisation in deriving the effective potential is a somewhat awkward necessity, given that each approach is somewhat incompatible with the methods of the other.
This may be seen on one hand in the incompatibility of an explicit cutoff with diffeomorphism symmetry, and hence unsuitability in a gravitational context, and on the other in the reliance of the gap-equation based derivation of $Z$ on explicit quadratic divergences, which are invisible in regularisation methods based on analytic continuation \cite{capper}.

Ultimately, accommodating both is unproblematic in that we will simply set both $M_\Lambda$ and $\mu$ to $\mathcal{O}\left(M_P\right)$.
As with the assumption in the gap equation analysis of near-criticality for the coupling $\lambda_S$, this is of questionable validity, but it is in one sense motivated by the unfortunate circumstance of having to simultaneously make use of two distinct and incompatible regularisations.
Since we are operating at the proof of concept level, we will proceed under the assumption that this will not prove pathological.

We may note that any shifts in $M_\Lambda$ can be compensated accordingly by a shift in $\lambda_S$, which would consequently affect the $\lambda_S^2\sigma^4\left(\dots\right)$ term in \eqref{eq: veff}.
Given that we will ultimately be most interested in the regime when $\lambda_S$ is very close to the critical value of $2\pi^2M_\Lambda$, the compensatory shifts required will be minimal, and the knock-on effect on $\lambda_S^2\sigma^4\left(\dots\right)$ will be negligibly small.
It is then feasible to consequently assume $M_\Lambda$ constant in the regime of interest, even when varying $\mu$.

A further concern may be found in the multivalued structure of $Z$, inherited from $W$.
Since both the principal and lower branches of $W$ are equally valid solutions of the gap equation for the dynamical gravitino mass, we will simply be agnostic about a choice of branch, and evaluate the physical consequences in both instances.

Since $m_{\rm dyn}$ and $\left\langle\overline\psi_\mu\psi^\mu\right\rangle$ do not explicitly appear in \eqref{eq: veff}, care should be exercised in their normalisation.
Assuming that the previous relations \eqref{ginomass} may be simply imported results in a dynamical gravitino mass which is independent of $Z$, which we know from section \ref{sec: wavefunction renormalisation} not to be the case.

Instead we may revisit the original four-fermion linearisation relation, where in terms of the canonically normalised condensate and a presumed-to-be canonically normalised gravitino field we identify
\begin{align}
	\lambda_{\rm S}\left(\overline\psi_\mu\psi^\mu\right)^2\sim
	- \frac{\tilde\sigma^2}{\kappa^2} + \frac{2\sqrt{\lambda_{\rm S}}}{\kappa}\tilde\sigma\left(\overline{\psi}_\mu\, \psi^\mu \right)\,,
\end{align}
with factors of $\kappa$ inserted for convenience.
This then results in the relations
\begin{align}
	m_{\rm dyn}=\frac{2\sqrt{\lambda_S}}{\kappa}\left\langle\tilde\sigma\right\rangle
	=2\sqrt{\lambda_S Z}\left\langle\sigma\right\rangle\,,\quad
	\left\langle\overline\psi_\mu\psi^\mu\right\rangle
	=\frac{1}{\kappa\sqrt{\lambda_S}}\left\langle\tilde\sigma\right\rangle
	=\sqrt{\frac{Z}{\lambda_S}}\left\langle\sigma\right\rangle\,,
	\label{eq: normalised mass}
\end{align}
which have the `expected' scaling, in that decreasing the physical vacuum expectation value $\left\langle\tilde\sigma\right\rangle$ results in an accordingly lighter gravitino.
The expected behaviour of the dynamical mass for a near-critical $\lambda_S$ as found in section \ref{sec: gap equations} is then also reproduced, by virtue of the presence of $Z$.

This is a slightly unconventional normalisation, in that we would normally expect inverse powers of $Z$ to appear in mass parameters, with the associated condition that $Z=1$ at some fiducial scale.
However, as argued in the related context of top quark condensation \cite{Bardeen:1989ds}, we should in fact normalise so that $Z\to0$ at the condensation scale.
This is of course a consequence of the obvious fact that, unlike a conventional field, the condensate field should vanish at this point, and be absent above it.

In that case we then expect the various induced parameters to accordingly go to zero at the condensation scale, as local supersymmetry is restored and in concordance with \eqref{eq: normalised mass}.
This would of course be incompatible with a conventional normalisation involving inverse powers of $Z$, which could cause them to instead diverge.

\subsection{Parametric dependence}

\begin{figure}[h!!]
		\centering
		\includegraphics[width=0.5\textwidth]{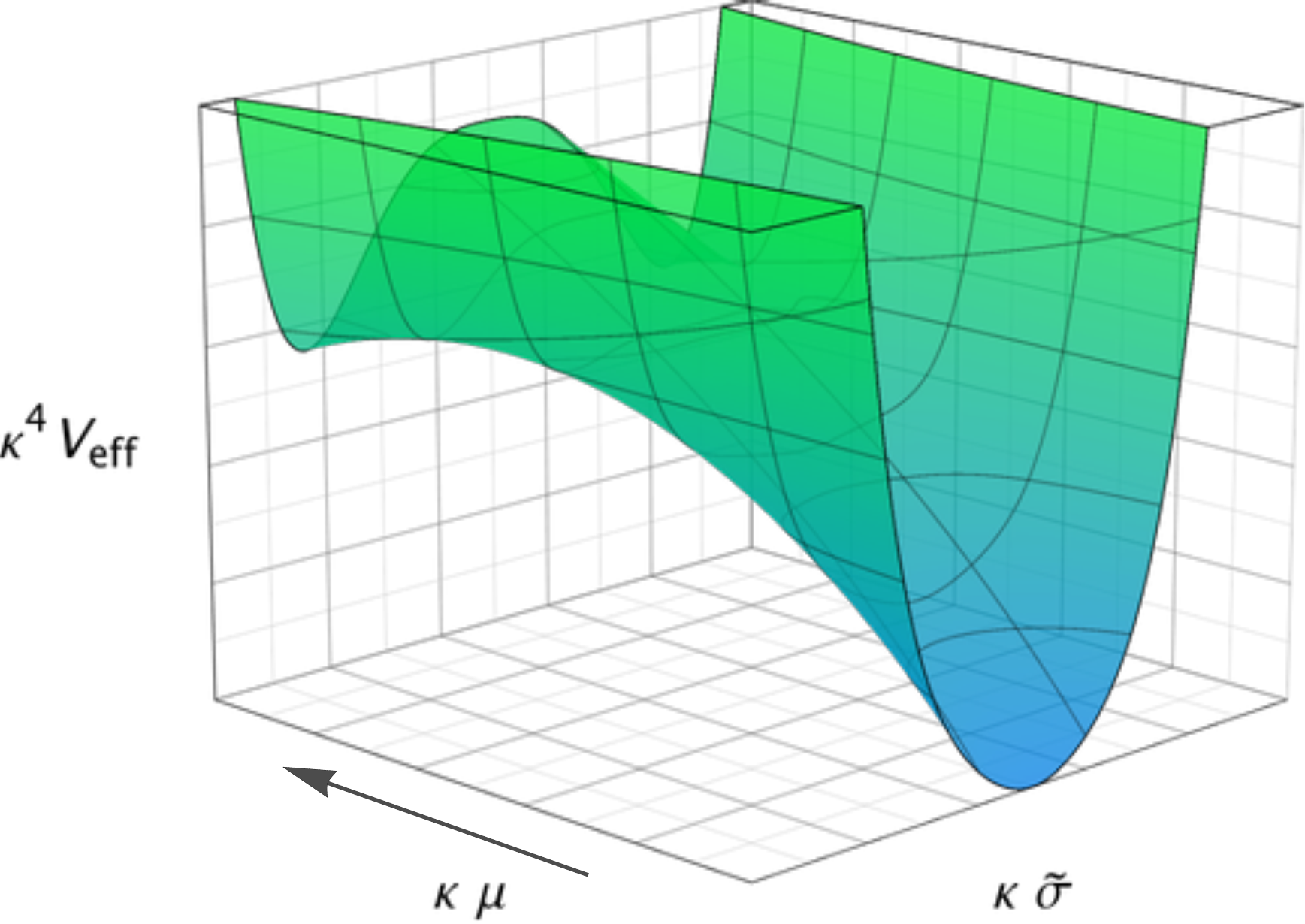}
		\caption{
		Scaling of the effective potential \eqref{eq: veff} under variations in $\mu$, whilst all other parameters are held fixed.
		As can be seen, the double-well shape associated to the super-Higgs effect emerges in the direction of increasing $\mu$.
		}
	\label{fig: mu scaling}	
\end{figure}

We may firstly assess the behaviour of \eqref{eq: veff} under variations in $\mu$.
As indicated via figure \ref{fig: mu scaling}, as one varies the renormalisation point $\mu$ from the ultraviolet to the infrared, the shape of the effective potential changes in such a way that the broken phase becomes available in the infrared, realising the super-Higgs effect. 
This is as expected from rather general features of dynamical mass in field theory, in that it is a low-energy phenomenon. 

\begin{figure}[h!!]
		\centering
		\includegraphics[width=0.5\textwidth]{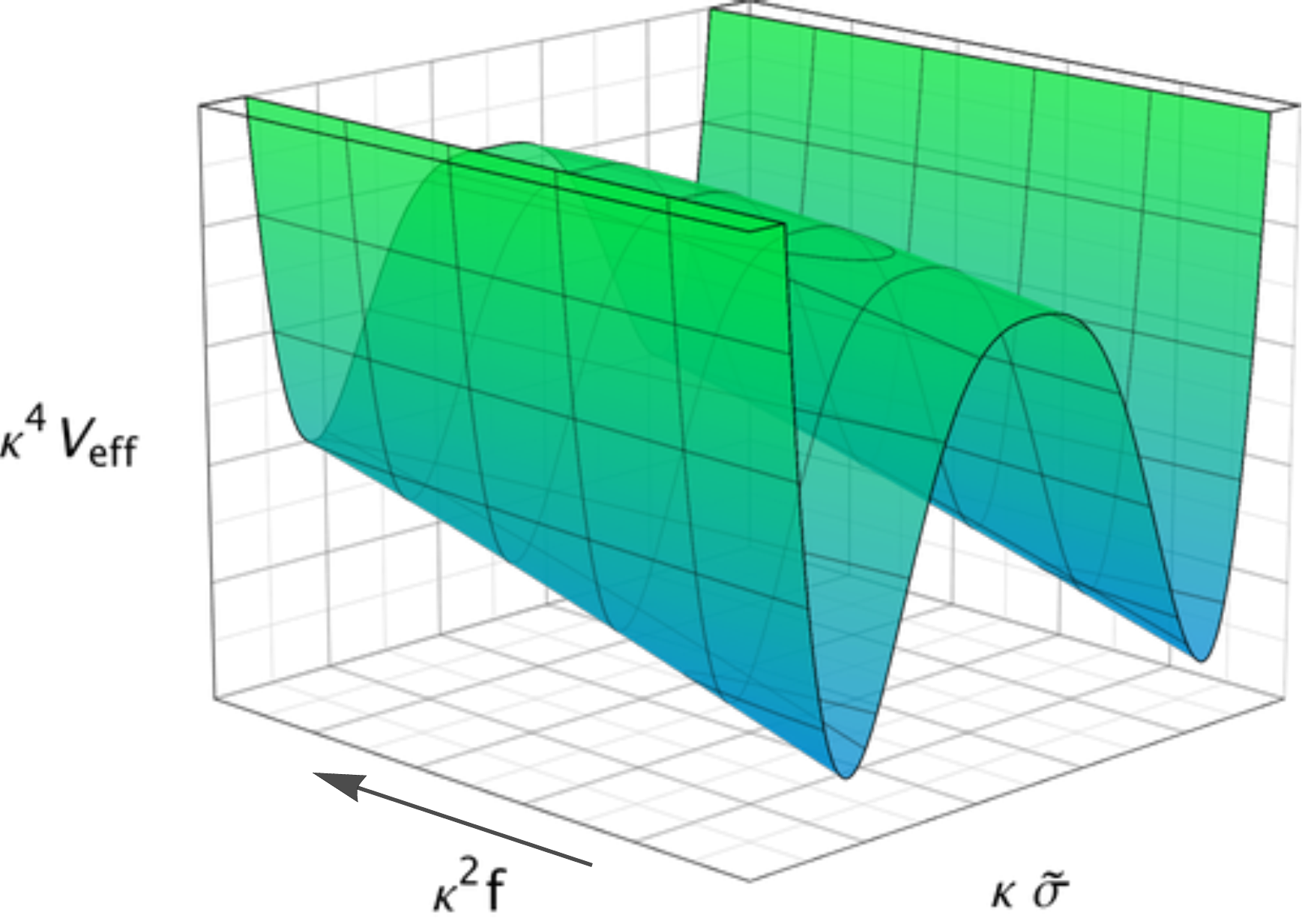}
		\caption{Scaling of the effective potential \eqref{eq: veff} under variations in $f$, whilst all other parameters are held fixed.
		Self-consistency of the flat space limit then fixes the value of $f$ via \eqref{eq: V=0}.}
	\label{fig: f scaling}
\end{figure}

As may be expected from the role of $f$ in the tree-level cosmological constant, scaling $f$ in this broken-symmetry phase corresponds to varying the overall energy density, shifting the potential vertically.
It is important to stress however that the value of $f$ is not strictly free. 
Rather, it is set as a self-consistency condition of the limit $\Lambda\to 0$ we have taken in deriving \eqref{eq: veff}.
Since any remnant energy density in \eqref{eq: veff} may be identified with a renormalised cosmological constant $\Lambda$, valid non-trivial minima in the broken-symmetry phase are those satisfying the condition
\begin{align}
	{V}_{\rm eff}\left(\left\langle\tilde{\sigma}\right\rangle\right)=0\,,
	\label{eq: V=0}
\end{align}
which, by virtue of figure \ref{fig: f scaling}, we may see to occur for a given minima only for a set value of $f$.

Of course, this should not be misconstrued with solution of the cosmological constant problem.
Given the intricacies associated with quantum field theory in curved space time, the limit $\Lambda\to 0$ is largely one of convenience.

Lastly, we may assess the effect of scaling $\lambda_S$ within the range prescribed in \eqref{eq: Z 2}.
As may be expected from the role of $\lambda_S$ in the canonically normalised potential, any variation in the broken symmetry phase largely amounts to shifting $\left\langle\tilde{\sigma}\right\rangle$.
An increasing $\lambda_S$ then corresponds to an increasing $\left\langle\tilde{\sigma}\right\rangle$ in one branch of $Z$, and a decreasing $\left\langle\tilde{\sigma}\right\rangle$ in the other.

\begin{figure}[h!!]
		\centering
		\includegraphics[width=0.5\textwidth]{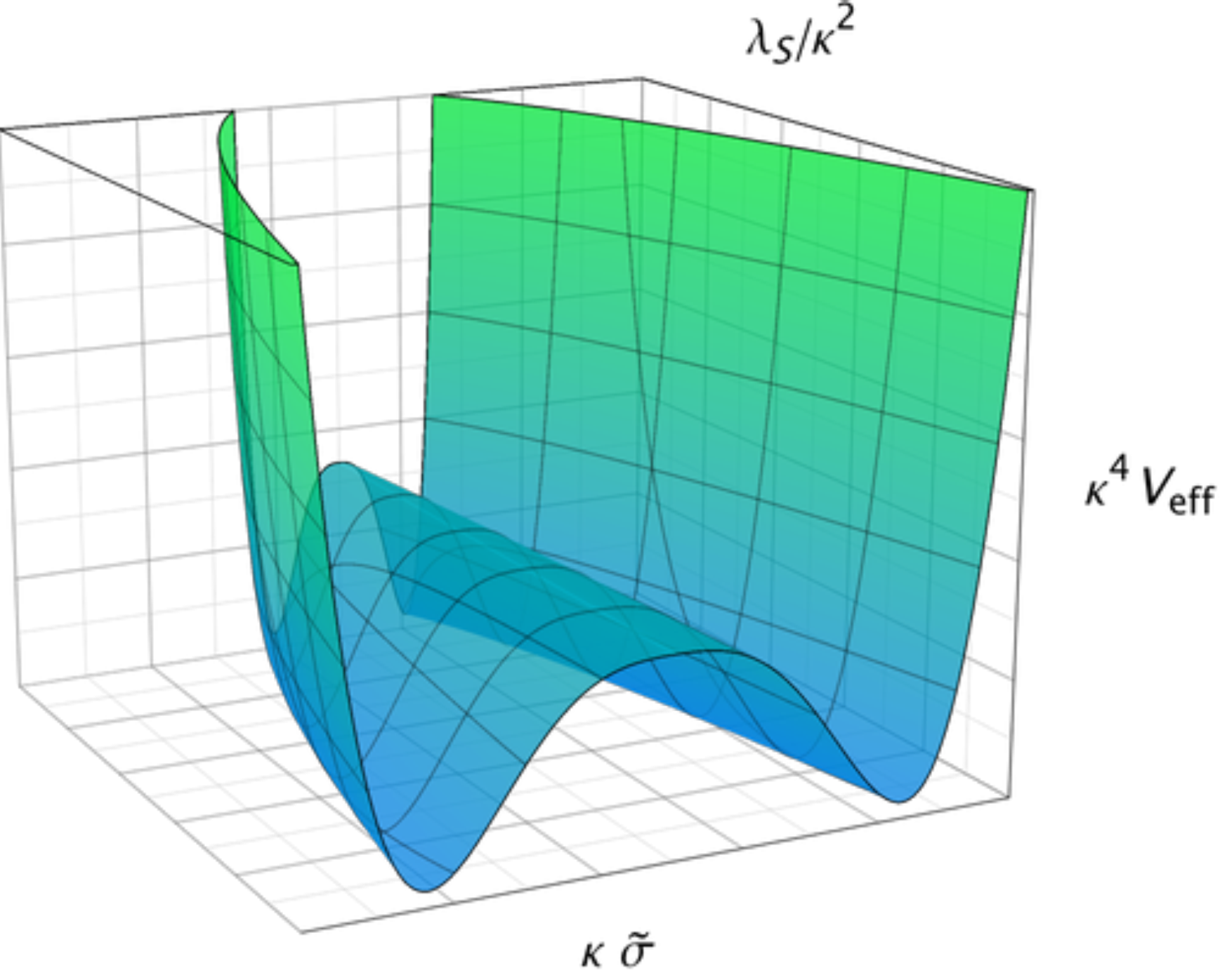}
		\caption{Scaling of the effective potential \eqref{eq: veff} under variations in $\lambda_S$, whilst all other parameters are held fixed.
		As can be seen, the super-Higgs profile of the potential is retained.
		Dependent on the branch choice of $Z$, $\left\langle\tilde{\sigma}\right\rangle$ is either an increasing or decreasing function of $\lambda_S$.}
	\label{fig: coupling scaling}	
\end{figure}

Given that the stability of the potential is inherited from the negativity of $\Lambda_0\equiv\kappa^2\left(\sigma^2-f^2\right)$, it is self-evident that \eqref{eq: veff} will be complex for some sufficiently large value of $\sigma$.
Whilst it is obviously important that the effective potential is real-valued in some suitable regime, we emphasise that any complexity beyond the self-consistent minima is unproblematic in the current context. 
Although it is convenient to think of varying $\tilde\sigma$, $\left\langle\tilde\sigma\right\rangle$ is a constant which is self-consistently determined by the minimisation procedure, and in that sense is non-dynamical. 

Furthermore, fluctuations about the minima may be characterised by a space-time dependent condensate $\tilde\sigma \left(x\right) = \left\langle\tilde\sigma\right\rangle + \hbar \tilde\sigma\left(x\right)$, with a mass of the order of the gravitino field, as expected from the parabolicity of the effective potential around the non-trivial minima. 
Quantum fluctuations are, as expected, suppressed and therefore not capable of destroying the stability of the broken phase minima. 

In totality, these considerations imply that in suitable regions of the parameter space, we expect non-trivial minima satisfying the condition \eqref{eq: V=0}, with a resultant gravitino mass defined in terms of the canonically normalised condensate.
Whilst the dynamical gravitino mass we find via this procedure is not strictly unique for a given value of $\lambda_S$, in that it can be modified by simultaneous tuning of $f$ and $\mu$, variation beyond an order of magnitude is generally not possible. 
	
\section{Supersymmetry breaking}
\label{sec: Supersymmetry breaking}

Having demonstrated the possibility of breaking local supersymmetry in this way, we now turn to assessing the resultant phenomenological suitability.
Given the input of a Planck-scale $f$ and near-critical coupling $\lambda_S$, the relevant order parameter is then the expectation value of the condensate $\left\langle\overline\psi_\mu\psi^\mu\right\rangle$, which we infer from 
\begin{align}
	\left\langle\tilde\sigma\right\rangle=\kappa\sqrt{\lambda_S}\left\langle\overline\psi_\mu\psi^\mu\right\rangle.
\end{align}

As this breaking is taking place in the gravitational sector, it is only natural to communicate it to the visible sector via gravitational effects.
More precisely, since the gravitino and standard model sectors are inevitably coupled via gravity, we expect incomplete cancellation of soft gravitational corrections below the condensation scale, suppressed by powers of the gravitational coupling $\kappa^{-1}=M_P$. 

Given that the observable consequences depend on the details of this mediation mechanism, and furthermore on the overall context of unification within which the standard model fields are incorporated, it suffices to demonstrate that the order parameter $\left\langle\overline\psi_\mu\psi^\mu\right\rangle$ may be made arbitrarily small. 
Phenomenologically viable supersymmetry breaking can then in principle always be achieved via this mechanism, at the cost of whatever tuning of $\lambda_S$ is required.

\subsection{$Z$ dependence}

\begin{figure}[h!!]
		\centering
		\includegraphics[width=0.5\textwidth]{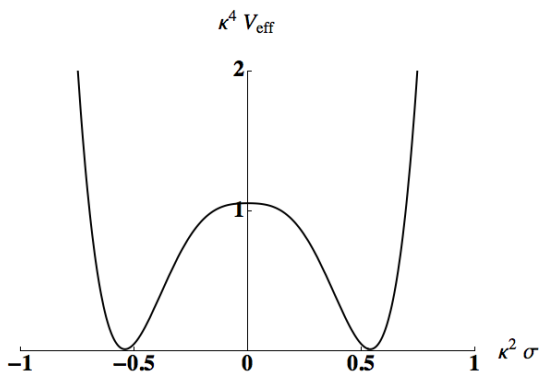}
		\caption{The effective potential \eqref{eq: veff} prior to rescaling by $Z$, for $\kappa^2f=1$, $\lambda_S/\kappa^2=\left(2\pi^2+0.1\right)$ and $\kappa\mu=2.7\,$.
		As can be seen, there is a vacuum expectation value $\kappa^2\left\langle\sigma\right\rangle\sim 0.55$ satisfying the condition $V_{\rm eff}\left(\left\langle\sigma\right\rangle\right)=0$. 
		In practice, the value of $\left\langle\sigma\right\rangle$ is somewhat characteristic in that it cannot be reduced more than an order of magnitude without encountering instabilities.}
	\label{fig: fiducial potential}	
\end{figure}

This may most straightforwardly be seen via the fiducial non-canonically normalised potential plotted in Figure \ref{fig: fiducial potential}, which has $f, \left\langle\sigma\right\rangle\sim\mathcal{O}\left(M_P^2\right)$, $\mu\sim \mathcal{O}\left(M_P\right)$ and $\lambda_S\sim 2\pi^2/M_\Lambda^2$.
Given the existence of the formal limit
\begin{align}
	W_0\left(\frac{2\pi^2}{\lambda_SM_\Lambda^2}-1\right)\bigg|_{\lambda_S\to \frac{2\pi^2}{M_\Lambda^2}}\to 0\,,
	\label{eq: W limit}
\end{align}
where the zero subscript denotes the principal branch, it is clear from \eqref{eq: Z 2} that an arbitrarily small condensation scale can always be reached for $\lambda_S\sim 2\pi^2/M_\Lambda^2$
\footnote{There is a subtlety here in that Figure \ref{fig: fiducial potential} is presented for a fixed value of the coupling, whilst we are discussing varying $\lambda_S$ to reduce the resultant $\left\langle\tilde{\sigma}\right\rangle$. 
However, as can be seen from Figure \ref{fig: coupling scaling} the super-Higgs profile is preserved under this variation, ensuring straightforward extrapolation.}.

As discussed previously, we must make a choice between the purportedly high-mass and low-mass branches of the $Z$.
What is however interesting is that the principal branch used in \eqref{eq: W limit} is in fact the former branch, and not the latter.
It is of course counterintuitive that the apparently high-mass branch of the wavefunction renormalisation ultimately gives a lower mass condensate, and vice-versa.

A straightforward explanation of this phenomenon may be found in noting that for a simple parabolic potential the higher mass branch would correspond to increased curvature about the origin.
In the present context of a double-well potential, increasing the curvature of the potential about the origin brings in turn the vacuum expectation values closer to the origin, counterintuitively lowering the associated gravitino mass scale.

\subsection{Soft terms}

As is familiar from the analogous case of gaugino condensation, in the simplest possible mediation scenario we expect purely on dimensional grounds soft supersymmetry-breaking terms in the resultant low energy theory such as
\begin{align}
	m_{\rm soft}\sim \frac{M_{\left\langle\tilde\sigma\right\rangle}^3}{M_P^2}\,,\quad
	\left\langle\overline\psi_\mu\psi^\mu\right\rangle\equiv M_{\left\langle\tilde\sigma\right\rangle}^3\,.
	\label{eq: mediation}
\end{align}
Requiring a TeV-scale $m_{\rm soft}$ then corresponds to $M_{\left\langle\tilde\sigma\right\rangle} \sim 10^{13}$ GeV.

Making use of the equation of motion for $\tilde\sigma$ we relate this via
\begin{align}	
	\left\langle\overline\psi_\mu\psi^\mu\right\rangle
	=\frac{1}{\kappa\sqrt{\lambda_S}}\left\langle\tilde{\sigma}\right\rangle
	=\sqrt{\frac{Z}{\lambda_S}}\left\langle\sigma\right\rangle\,,
\end{align}
where, by virtue of the $W_0$ series expansion for a near-critical $\lambda_S$,
\begin{align}
	Z=-\frac{1}{2\pi^2}W_0\left(\frac{2\pi^2}{\lambda_SM_\Lambda^2}-1\right)\bigg|_{\lambda_S\to 2\pi^2/M_\Lambda^2}
	=\frac{\lambda_SM_\Lambda^2-2\pi^2}{4\pi^4}+\mathcal{O}\left(\lambda_SM_\Lambda^2-2\pi^2\right)^2\,.
\end{align}
For $\left\langle\sigma\right\rangle\sim M_P^2 $ we then have 
\begin{align}
	m_{\rm soft}\sim\frac{\sqrt{M_\Lambda^2-\frac{2\pi^2}{\lambda_S}}}{2\pi^2}\,.
\end{align}
Assuming $M_\Lambda\sim M_P$ it is then straightforward to note that the fine-tuning problem of the lightness of $m_{\rm soft}$ has now been supplanted by the fine-tuning of $\left(\lambda_S-\frac{2\pi^2}{M_\Lambda^2}\right)$.

That this is the case should not be inherently surprising, given that the quadratic divergences necessitating this tuning were absorbed in the gap equation used to derive $Z$.
Assuming a non-trivial solution corresponding to a dynamical gravitino mass does not negate their underlying presence and physical effect, so their resurgence via another quantity is to be expected.
This phenomenon is identical in analogous circumstances, such as the top-quark condensation scenario explored in \cite{Bardeen:1989ds}.

We may also draw a relevant parallel here with early universe inflation, where the apparent fine-tuning of the initial conditions of the universe is supplanted with the issues of the flatness of the inflaton potential and initial conditions of the inflaton.
If, analogously, tuning $\lambda_S$ is more `fundamental' than tuning $m_{\rm soft}$, then we have made progress.

Indeed, as is central to the topic at hand, we may approach supergravity as an effective description of some more fundamental theory, in the spirit of the NJL model and the relation it bears to QCD.
As explored in section \ref{sec: chiral symmetry breaking} we may then hope that as in the NJL model, the underlying microphysics which sources the quadratic divergences in the effective description also provide the characteristic non-perturbative factors of e.g. $\exp\left(-\mathcal{O}\left(10\right)\right)$ which are invisible at the level of the effective theory.

Indeed, in the the gravitino field-strength condensation scenario \cite{Konishi:1989em, Mangano:1988kf} briefly reviewed in chapter 3, the small quantity equivalent to $\left(\lambda_SM_\Lambda^2-2\pi^2\right)$ required for successful phenomenology is achieved via a weighting factor arising from the action of the Eguchi-Hanson instanton
\begin{align}
	\exp\left(-16\pi^2X\right)\bigg|_{X\sim0.23}\sim 10^{-16}\,,
\end{align}
where $X$ is the unfixed prefactor of a topological density term in the action.

Whilst, as previously emphasised, the presence of four-fermion interactions in supergravity cannot be attributed to non-perturbative effects in the same straightforward manner as those arising in QCD, the physics which dictates the overall distribution into scalar, pseudoscalar and pseudovector channels, and therefore controlling the magnitude of $\lambda_S$, is unclear at this stage, and may be non-perturbative in origin.

\begin{figure}[h!!]
		\centering
		\includegraphics[width=0.5\textwidth]{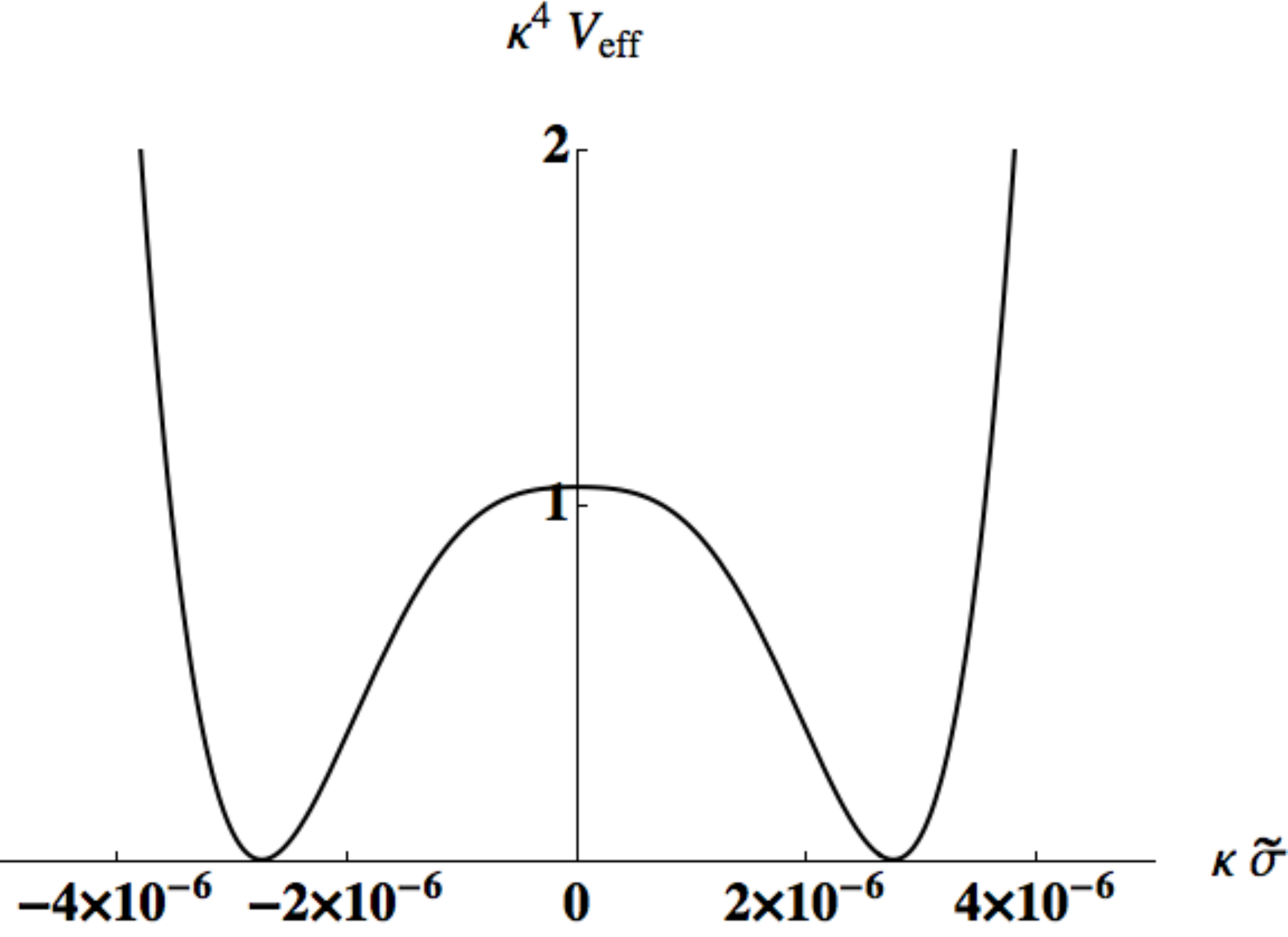}
		\caption{The canonically normalised effective potential for $\kappa^2f=1$, $\lambda_S/\kappa^2=\left(2\pi^2+10^{-8}\right)$, and $\kappa\mu=2.75\,$.
		As can be seen, there is a vacuum expectation value $\kappa\left\langle\tilde{\sigma}\right\rangle\sim 10^{-6}$ satisfying the condition ${V}_{\rm eff}\left(\left\langle\tilde{\sigma}\right\rangle\right)=0$.
		This corresponds to a dynamical gravitino mass $m_{\rm dyn}\sim 10^{13}$ GeV.}
		\label{fig:fiducial2}
\end{figure}

It should of course also be noted that other scenarios exist, which, by virtue of altering the mediation relation \eqref{eq: mediation} connecting $m_{\rm soft}$ to $\left\langle\tilde{\sigma}\right\rangle$, accordingly require far less tuning of $\lambda_S$ to achieve $m_{\rm soft}$ at an acceptable scale.
One such example is the split supersymmetry scenario, where $m_{\rm dyn}\lesssim 10^{13}\, \rm{GeV}$, rather than $\mathcal{O}\left(m_{\rm soft}\right)$ as in \eqref{eq: mediation}, can be accommodated with acceptably low soft supersymmetry-breaking terms \cite{ArkaniHamed:2004yi}.
A canonically normalised effective potential chosen to achieve a suitable $m_{\rm dyn}$ in this context is given in Figure \ref{fig:fiducial2}.
The palatability of the `tuning' associated with introducing and arranging new fields and couplings to accomplish such a goal is, of course, a matter of taste.

Given the reliance of these arguments on the prefactor provided by the wavefunction renormalisation, a particular concern may be that the calculation of $Z$ in chapter 4 is deficient in that it ignores gravitational degrees of freedom.
A further development would then be a fully curved-space derivation of $Z$, possibly following the approach of \cite{Candelas:1975du}.

This issue aside, in totality these considerations establish that gravitino condensation can break both local supersymmetry and generate a hierarchy from Planck scale inputs, from which suitably low-scale soft supersymmetry breaking terms can then arise.
Naturally, it should be stressed that this is not a complete scenario. 
We have given little thought to the incorporation of an observable sector and the mediation of supersymmetry breaking from the gravitino condensate to these fields.

There is however a useful aspect to this state of affairs, in that, given the generality of our approach and the universality of the gravity supermultiplet in $\mathcal{N}=1$ supergravity, this mechanism can be imported without modification into any $\mathcal{N}=1$ scenario and leveraged appropriately.
As we will discuss in the following, further utility may also be found in an application to early universe inflation.

\section{Inflation}
\label{sec: Inflation}

Given that we have derived a canonically normalised potential for the scalar mode $\tilde{\sigma}$, corresponding to a phase transition we expect to occur in the early universe, it is of course tempting to make use of this setting for the purposes of inflation.
This may be further motivated both by the fact that nature already makes use of fermionic condensation mechanisms in QCD and low-temperature superconductivity, and by the connections of supergravity to ultraviolet physics.

There are however a number of caveats that we may first of all raise.
\begin{itemize}
	\item Whilst fully incorporating gravitational degrees of freedom, and having been derived via excursion into de Sitter space, considerations of computational tractability limit the potential \eqref{eq: veff} to the flat space limit, rather than the de Sitter vacuum we know to characterise inflation.
	\item The wavefunction renormalisation required for canonical normalisation of the potential was computed in chapter 3 in a flat background, neglecting the conceivably important role of gravitational degrees of freedom.
	\item Furthermore, we expect the deviation between the Minkowskian and de Sitter wavefunction renormalisations to be maximal at the origin of the potential, when the non-zero vacuum energy may be interpreted as a cosmological constant. 
	This is precisely the region we are interested in for inflationary purposes.
\end{itemize}
These considerations aside, we may proceed nonetheless. 

\subsection{Slow roll}

As discussed in chapter 2, for successful inflation we require the slow roll parameters to be very small.
Since our potential is of the double-well type, the maximally flat and therefore optimal region for inflation is about the origin. 
Computing the slow roll parameters in the small $\tilde\sigma$ limit we find
\begin{align}	
	\eta_{\tilde\sigma} &=-\frac{8 \left(-45 \kappa^4f^2 \ln \left(\frac{6 \kappa^2f^2}{\mu ^2}\right)+45 \kappa^4f^2 (1+\ln (4))+128 \pi ^2\right)}{f^2\kappa^4 Z \left(-90 	\kappa^4f^2 \ln \left(\frac{6 \kappa^2f^2}{\mu^2}\right)+45 \kappa^4f^2 (3+\ln (16))+512 \pi ^2\right)}+\mathcal{O}\left(\tilde{\sigma}^2\right)\,,\nonumber\\
	\epsilon_{\tilde\sigma}&=\frac{32 \tilde{\sigma} ^2 \left(-45\kappa^4 f^2 \ln \left(\frac{6 \kappa^2f^2}{\mu ^2}\right)+45 \kappa^4f^2 (1+\ln (4))+128 \pi ^2\right)^2}{f^4 \kappa ^6 Z^2 \left(-90 \kappa^4f^2 \ln\left(\frac{6 \kappa^2f^2}{\mu ^2}\right)+45 \kappa^4f^2 (3+\ln (16))+512 \pi ^2\right)^2}+\mathcal{O}\left(\tilde{\sigma}^4\right)\,.
   	\label{eq: V slow roll}
	\end{align}
	
It is worthwhile noting that even given broken supersymmetry some higher order contributions which may ruin the flatness of the potential are absent, as they are proportional to the difference in number of bosonic and fermionic degrees of freedom present, and that others are furthermore suppressed in the small $\tilde\sigma$ limit.

Given the similarity between the bracketed terms in the numerator and the denominator, acceptable inflation about the origin can only be realised when $\kappa^4f^2 Z>>1$.
This is inevitably at odds with the $Z<<1$ branch we require for acceptable supersymmetry breaking. 
Our first result is then the apparent impossibility of simultaneously viable inflation and supersymmetry breaking via this mechanism.
That said, it may be the case that we have the former with viable supersymmetry breaking achieved via other means.

In hindsight this conclusion should not be too surprising. 
Noting that the height of the potential at the origin is a function of $f$ and $\mu$ only, we can see that if low $m_{\rm soft}$ corresponds to a small vacuum expectation value for the condensate, as is perfectly plausible, the associated potential will be steeply curved and thus unsuitable for inflation.
The converse follows similarly.

This is in seeming contradiction with a common phrasing of the $\eta$ problem, discussed in section \ref{sec: eta problem}, as the requirement that the inflaton be suitably light.
We may note however that the mass of the inflaton here is dynamically determined, rather than being simply encoded in the second derivative of the potential, and so there is in actuality no discrepancy.
Indeed, in this context the `mass' of the condensate at the origin of the potential is subtle, in that if supersymmetry is not yet broken we cannot ascribe a mass to it at all.
Nevertheless, we may simply assess the suitability of the potential via the slow roll parameters encoded in \eqref{eq: V slow roll}, sidestepping thorny issues of interpretation.

Given the form of \eqref{eq: V slow roll}, we may also note that to leading order in $\tilde{\sigma}$, 
\begin{align}
	\epsilon_{\tilde\sigma} = \frac{1}{2}\kappa^2\tilde{\sigma}^2\eta_{\tilde\sigma}^2\,,
\end{align}
implying that $|\epsilon_{\tilde\sigma}|<<|\eta_{\tilde\sigma}|$ for successful inflation about the origin. 
From the Planck best fit value \cite{Planck:2015xua} for the scalar spectral index
\begin{align}
	n_s=1-2\epsilon_{\tilde\sigma}-\eta_{\tilde\sigma}\simeq0.9603\pm0.0073\,,
\end{align}
we can then conclude that fifty to sixty e-folds prior to the end of inflation we require $|\eta_{\tilde\sigma}|\simeq0.04$.

\begin{figure}[h!!]
		\centering
		\includegraphics[width=0.8\textwidth]{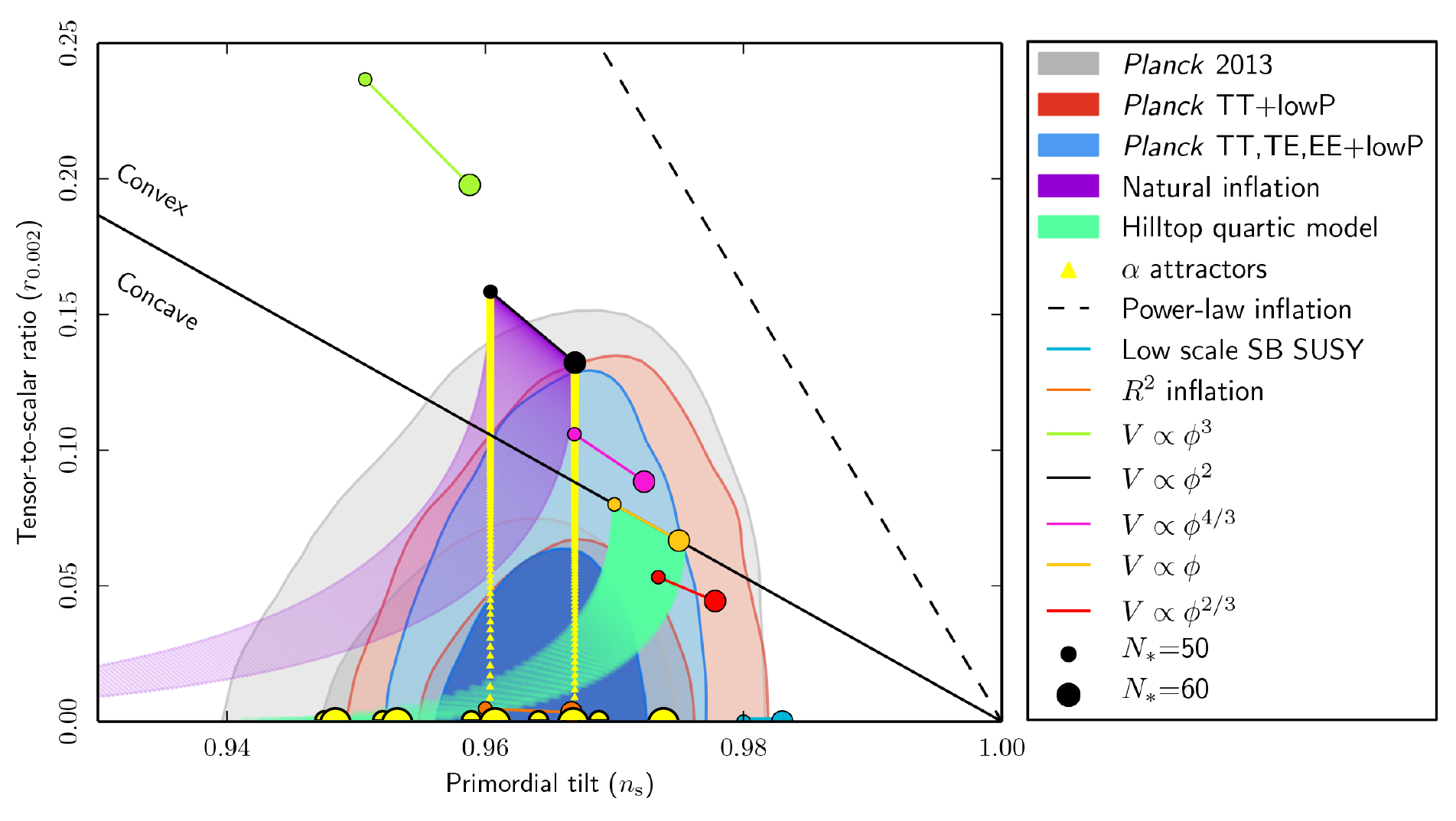}
		\caption{Planck 2015 68\% and 95\% marginalised confidence levels for $n_s$ and $r$ taken from \cite{Ade:2015lrj}, with the inflationary predictions of the canonically normalised potential \eqref{eq: veff} for various values of the coupling $\lambda_S$ overlaid in yellow.
		The leftmost points correspond to $\lambda_SM_\Lambda^2=2\pi^2+10^{-1}$ and the rightmost to $\lambda_SM_\Lambda^2=2\pi^2+10^{-9}$, with the remaining points interpolating in-between.
		For all points shown the tensor to scalar ratio is $\mathcal{O}\left(10^{-6}\right)$.}
	\label{fig: Planck 2015}	
\end{figure}

Since $|\epsilon_{\tilde\sigma}|<<0.04$, the tensor to scalar ratio $r\simeq16\epsilon_{\tilde\sigma}$ should be similarly negligible. 
We then expect the inflationary observables to lie along the line $r\sim0$, with a varying $n_s$ achieved via shifting the coupling $\lambda_S$.
This is borne out in Figure \ref{fig: Planck 2015}, where the tensor to scalar ratio and scalar spectral index are numerically found and overlaid onto the Planck 2015 inflationary constraints for $n_s$ and $r$.

It is important to emphasise that $f$ cannot be varied freely, in that it must be set for a given value of $\mu$ to ensure a potential with the appropriate super-Higgs profile, and satisfying the condition $V_{\rm eff}\left\langle\tilde\sigma\right\rangle=0$.
There is of course a continuum of pairs of values such that these conditions are met, however, given that appropriate inflationary phenomenology requires that $r\sim 0$, their effect on the $\left(n_s,r\right)$ plane is restricted to shifts in the $n_s$ axis, which are equivalently achieved by changing $Z$ via the coupling $\lambda_S$.

In the absence of a convenient rationale for the proximity of $\lambda_S$ to criticality, bearing in mind the overall NJL approach at hand and the related comments of the previous section, it then suffices to take some appropriate fiducial values for $f$ and $\mu$ as being sufficiently representative of the phenomenologically viable regime.
In practice, particularly convenient values of $\mu$ and $\sqrt{f}$ are $3.45$ and $0.82\times 10^{19}$ GeV respectively \footnote{That the chosen value of $\mu$ exceeds the Planck scale, $1.22\times10^{19}$ GeV, may cause concern.
However, given the provenance of this particular parameter as a proper-time regulator, implying that the ultraviolet limit is perhaps counterintuitively $\mu\to 0$, the physical significance of crossing this boundary is unclear. 
Equivalent results with a lower value of $\mu$ could however simply be achieved via tuning $\lambda_S$ closer to criticality.}.

As can be seen from Figure \ref{fig: Planck 2015}, inflation compatible with the Planck constraints is indeed possible. 
Whilst the smallness of the predicted tensor to scalar ratio is unfortunately beyond the sensitivity of future CMB experiments \cite{Baumann:2014nda}, this inflationary scenario would of course be falsified by the future observation of a non-zero tensor to scalar ratio.

Whilst this setting may ostensibly compare unfavourably to other inflationary scenarios, given the tuning required in $\lambda_S$, one should bear in mind that \eqref{eq: veff} is a one-loop, rather than tree-level, potential.
Given the considerations outlined in section \ref{sec: inflation in supergravity} regarding the influence of quantum effects on inflationary physics, caution should be exercised in drawing comparisons to the wide variety of tree-level inflaton potentials existing in the literature \cite{Martin:2013tda}. 

The end to inflation is achieved when the slow-roll conditions are violated, once the scalar $\tilde{\sigma}$ begins to move quickly on Hubble scales.
Ultimately the universe will then be reheated via oscillations of the condensate mode about the non-trivial minimum, with standard model particles generated via the decays of the condensate, the precise details of which are outwith the present analysis.

\subsection{Starobinsky inflation}

There can however exist an alternative to this `hilltop' type inflation, arising from the effective theory describing physics below the condensation scale.
This would occur subsequently, washing out whatever signatures were present of the previous mechanism.
As demonstrated in \cite{Alexandre:2013nqa} in the `canonical superconformal supergravity' context, integrating out the then-massive gravitino in a de Sitter background may generate the effective Lagrangian
\begin{align}
	\frac{1}{2\kappa^2}\int d^4x\,\sqrt{g} \left(R+\frac{R^2}{6M_{R^2}^2}\right)\,,
	\label{eq: R squared}
\end{align}
which is characteristic of the well-known Starobinsky model of inflation \cite{Starobinsky:1980te, Vilenkin:1983xp}. 
Via the transformations
\begin{align}
	\tilde{g}_{\mu\nu}=\left(1+\frac{\varphi}{3M_{R^2}^2}\right)g_{\mu\nu}\,,\quad
	\varphi'=\sqrt{\frac{3}{2}}M_P\ln\left(1+\frac{\varphi}{3M_{R^2}^2}\right)\,,
\end{align} 
this may be rendered  \cite{Whitt:1984pd} into the Einstein-frame Lagrangian
\begin{align}
	\frac{1}{2\kappa^2}\int d^4x\,\sqrt{\tilde{g}}\left(\tilde{R}+\partial_\mu\varphi'\partial^\mu\varphi'-\frac{3}{2}M_P^2M_{R^2}^2\left(1-e^{-\sqrt{\frac{2}{3}}\frac{\varphi'}{M_P}}\right)^2\right)\,,
	\label{eq: starobinsky}
\end{align}
with the field $\varphi'$ now serving as the inflaton.

Given $M_{R^2}\sim10^{-5}M_P$ \cite{Planck:2015xua}, which arises implicitly for certain parameter choices in the context of gravitino condensation, \eqref{eq: starobinsky} then gives the Planck-2015 compatible $R^2$ result shown in Figure \ref{fig: Planck 2015}.

We may emphasise that this procedure is indeed largely in the spirit of the original Starobinsky model, which is based upon self-consistent solution of the semiclassical Einstein equations incorporating one-loop contributions of quantum matter fields.
Therein, the expectation value of the resultant one-loop energy-momentum tensor is of the form
\begin{align}
	\left\langle T_{\mu\nu}\right\rangle\sim H_{\mu\nu}+\dots\,,
\end{align}
where $H_{\mu\nu}$ is obtained precisely by varying the $R^2$ term in \eqref{eq: R squared}.

Given the agreement with experimental constraints and overall simplicity of the model, exemplified by \eqref{eq: R squared} and the resultant inflationary relations
\begin{align}
	n_s\sim 1-\frac{2}{N}\,, \quad
	r\sim \frac{12}{N^2}\,,
\end{align}
it is unsurprising that there exists a large body of literature on Starobinsky inflation.
Of particular relevance to the present context are other implementations of this model in the wider context of supergravity \cite{Cecotti:1987sa,Ketov:2010qz,Ketov:2011rf,Ketov:2012yz,Farakos:2013cqa,Kallosh:2013lkr,Pallis:2013yda,Dalianis:2014aya,Ferrara:2014cca,Kounnas:2014gda,Kamada:2014gma,Ketov:2014hya,Basilakos:2014moa,Antoniadis:2014oya,Ketov:2014qha,Terada:2014uia,Diamandis:2014vxa,Lahanas:2015jwa,Garg:2015mra}, which are known to enjoy, amongst other benefits, a relaxation of the problem of initial conditions relative to the non-supersymmetric $R+R^2$ case \cite{Dalianis:2015fpa}, and connections to compactified string theories \cite{Ellis:2013xoa,Ellis:2013nxa}.

It should be noted however that the condensate-based derivation of \eqref{eq: R squared} is somewhat outside of the present `generic' supergravity context, in that it relies upon the specific setting of `canonical superconformal supergravity' \cite{Ferrara:2010in}. 
Therein the gravitino fields enjoy an enhanced four-fermion coupling, found in \cite{Alexandre:2013nqa} to be necessary for the resultant Starobinsky mass scale to satisfy the phenomenological requirement of $M_{R^2}\sim10^{-5}M_P$. 

Furthermore, the relevance of the Fierz ambiguity and non-trivial wavefunction renormalisation to the gravitino condensation scenario, as emphasised in chapter 4, was not appreciated and therefore taken into consideration in the analysis of \cite{Alexandre:2013nqa}.
As such it is unclear if these results carry over to the present context without modification, and it is topic of future research to revisit the scenario of Starobinsky inflation arising from massive-phase gravitino condensation with a view to fully incorporating these elements.

%% file: Chapters/Conclusions.tex
The central topic of this thesis is gravitino condensation, in the pursuit of a dynamical mechanism to break local supersymmetry.
In particular, our strategy has been to approach supergravity as an NJL-type effective description of some more fundamental microscopic theory, leveraging some of the intuition and tools arising from the study of chiral symmetry breaking via quark condensation in QCD.
We firstly summarise the various findings contained in this thesis, before concluding with some future prospects and questions raised.

\section{Summary}

Following the introductory and background material of the first three chapters, we firstly repurpose some tools from the study of chiral symmetry breaking  in chapter 4, in order to derive and solve the flat-space gap equation leading to a dynamical gravitino mass. 
The resultant theory is found in section \ref{sec: gap equations} to admit non-trivial solutions in a phase of sufficiently strong coupling, leading indirectly to issue of the ambiguity of the coupling into the scalar condensate channel.

This ambiguity is in part addressed in section \ref{sec: unwanted condensates}, where we demonstrate at least that for sufficiently strong scalar coupling it is possible to avoid the presence of pseudovector and pseudoscalar condensates in addition to the desired scalar solution.
This is of course important given the Lorentz-violating character of such objects.

Whilst the existence of these solutions is suggestive of the desired conclusion, it should be noted that this calculation does not incorporate the role of gravitational degrees of freedom, nor the presence of a cosmological constant.
Given the importance of these elements to the question of gravitino mass in the context of supergravity, discussed in section \ref{sec: cosmological constant}, it is clear that this result taken in isolation should be interpreted with caution.

This simplified approach is however useful to develop some of intuition for the full calculation in chapter 5, and furthermore necessary for the subsequent calculation in section \ref{sec: wavefunction renormalisation} of the bound state wavefunction renormalisation.
Therein we find a somewhat unusual multivalued result, based on the transcendental Lambert W function.

Building upon these foundations, we derive in chapter 5 the one-loop effective potential for the condensate mode in a de Sitter background, incorporating fully the previously neglected role of gravitational degrees of freedom and the cosmological constant. 
Leveraging this then allows us in section \ref{sec: imaginary terms} to revisit some of the claims in the literature, regarding the supposed pathological instability of any such gravitino condensate due to the effects of gravitational degrees of freedom \cite{Buchbinder:1989gi,Odintsov:1988wz}. 
Upon examining the resultant stability of the condensate, we demonstrate conclusively that, outside of certain unfortunate gauge choices, the stability of the condensate is directly linked to the sign of the tree-level cosmological constant.

This is, in hindsight, obvious, given the general incompatibility of unbroken supersymmetry with de Sitter vacua.
Crucially, it also illuminates the claims of the apparent impossibility of this approach to local supersymmetry breaking.
In so doing, we demonstrate that the supposed pathological instability in \cite{Buchbinder:1989gi,Odintsov:1988wz} is actually a consequence of the absence of Goldstino degrees of freedom in the formalism used, which are in turn necessary for their absorption by the gravitino via the super-Higgs effect.

Combining in chapter 6 the condensate wavefunction renormalisation from section \ref{sec: wavefunction renormalisation} and the one-loop effective potential from section \ref{sec: effective potential}, we arrive at the canonically normalised effective potential for the condensate mode.
Therein, we demonstrate the expected resurgence of the tuning associated to the lightness of the electroweak scale in the four-gravitino coupling into the scalar condensate channel, so that, given sufficient proximity of the coupling to a critical value, phenomenologically viable supersymmetry breaking may always be engineered.
We also illustrate the possibility of a suitable inflationary phase, characterised by a negligible tensor to scalar ratio. 
A future non-zero observation of this quantity would then succinctly rule out the inflationary aspect of this scenario. 
Notably, we also demonstrate that these circumstances cannot coexist in the basic supergravity setting, owing to their respective reliance upon opposing branches of the wavefunction renormalisation.

\section{Future prospects and outlook}

In closing, we may provide some directions for future research.
Primary amongst these is investigation into the microscopic origin of the scalar condensate coupling $\lambda_S$, with a view to understanding the proximity of this quantity to a certain critical value.
As explored in the text, the analogy between the present context and the breaking of chiral symmetry in QCD suggests a non-perturbative source, leading to relations of the type
\begin{align}
	\lambda_SM_\Lambda^2-2\pi^2\sim\exp\left(-\mathcal{O}\left(10\right)\right)\,,
\end{align}
which are invisible at the level of the effective description we operate within.
Recalling the arguments of sections \ref{sec: the nambu-jona-lasinio model} and \ref{sec: Gravitational instantons}, it seems plausible that the provenance of these effects may involve some aspect of string physics, or may possibly be gravitational in origin.

A related question pertains to the microphysics dictating the choice of branch for the condensate wavefunction renormalisation, which is at this stage similarly unclear, but nonetheless critical for the ultimate suitability of this mechanism for supersymmetry breaking or inflation.

It is furthermore interesting to speculate that gravitino condensation may be able to fulfil some useful role in circumstances where the analogous phenomenon of gaugino condensation is invoked.
One notable example where the latter mechanism is brought to bear is the issue of moduli stabilisation in string compactifications, a central problem in string phenomenology \cite{Denef:2008wq}.
It would of course be interesting to explore the incorporation of gravitino condensation into the toolbox of string phenomenology.

Going in the opposite direction there are also a number of `low-energy' considerations which may be explored.
Having established the possibility of suitably low-scale soft supersymmetry-breaking terms, given a sufficiently critical $\lambda_S$, it is natural to then consider the incorporation of this mechanism into more phenomenologically complete scenarios.
From there, the characteristic low energy signatures of this type of breaking may be explored.

Furthermore, it should be noted that we have only explored the inflationary aspects of this scenario to a limited extent.
Reheating and the end of inflation is for example a very rich subject upon which we have essentially not touched.
Another aspect of this is the need to reengineer the Starobinsky scenario outlined in the previous chapter into the generic supergravity context within which the remainder of the results of this thesis take place.

We resign the exploration of these various considerations, in whole or in part, to another day.

%% file: Chapters/Appendix_B.tex
\section*{Appendix A: Variation of Einstein-Hilbert action}

\addcontentsline{toc}{chapter}{Appendix A: Variation of Einstein-Hilbert action}

\setcounter{equation}{0}

\renewcommand{\theequation}{A.\arabic{equation}}

Our starting point is the metric-formalism variation of the Einstein Hilbert action to quadratic order in the traceless perturbation $\overline h_{\mu\nu}$
	\begin{align}
		\frac{1}{4\kappa^2}
		\int 
		\bigg(\frac{1}{2} \overline h_{\mu\nu}&\left(-\nabla^2
		+\frac{8}{3}\Lambda-2\Lambda_0\right) \overline h^{\mu\nu}
		-\frac{1}{8}h\left(-\nabla^2-2\Lambda_0\right)h
		-\left(\nabla^\mu  \overline h_{\mu\nu}-\frac{1}{4}\nabla_\nu h\right)^2\bigg)\,,
		\label{metric action}
	\end{align}
where the usual spacetime measure $d^4 x\sqrt{g}$ is implicit \cite{Fradkin:1983mq}.

We emphasise the implicit use of the metric formalism as, given the presence of fermions in the supergravity context, we must instead make use of the vierbein formalism. 
This then leads to extra terms which vanish on-shell in the quadratic action \cite{Percacci:2013ii}, which we should account for.
 
These contributions may be understood as arising from the first order variation of the gravitational action
	\begin{align}
		\left(G^{\mu\nu}+g^{\mu\nu}\Lambda_0\right)\delta g_{\mu\nu}
		=\left(G^{\mu\nu}+g^{\mu\nu}\Lambda_0\right)\delta e_{(\mu}{}^a\eta_{|ab|}e_{\nu)}{}^b\,,
	\end{align}
where $G^{\mu\nu}$ is the Einstein tensor.	
The second variation in the vierbein formalism then coincides with the result obtained from the metric formalism, along with the additional term
	\begin{align}
		&\frac{1}{4\kappa^2}\int \left(G^{\mu\nu}+g^{\mu\nu}\Lambda_0\right)\delta e_{(\mu}{}^a\eta_{|ab|}\delta e_{\nu)}{}^b
		=\frac{1}{16\kappa^2}\int \;g^{\mu\nu}\left(G+4\Lambda_0\right)\delta e_{(\mu}{}^a\delta e_{\nu) a}\nonumber\\
		&=\frac{1}{16\kappa^2}\int \;\left(\Lambda_0-\Lambda\right)h_{\mu\nu}^2
		=\frac{1}{4\kappa^2}\int \;\left(\frac{\Lambda_0-\Lambda}{2}\right)\left(\frac{1}{2}\overline{h}_{\mu\nu}^2+\frac{1}{8}h^2\right)\,,
	\end{align}
making use of $h_{\mu\nu}=2e_{(\mu}{}^a\delta e_{\nu) a}$ and $g^{\mu\nu}=e^{(\mu}{}_a\eta^{|ab|} e^{\nu)}{}_b$.

The action \eqref{metric action} then becomes in the vierbein formalism
	\begin{align}
		\frac{1}{4\kappa^2}\int &\left(\frac{1}{2} \overline h_{\mu\nu}\left(-\nabla^2+X_1\right) \overline h^{\mu\nu}
		-\frac{1}{8}h\left(-\nabla^2-X_2\right)h
		-\left(\nabla^\mu  \overline h_{\mu\nu}-\frac{1}{4}\nabla_\nu h\right)^2\right)\,,\nonumber\\
		&\qquad X_1
		=\frac{13}{6}\Lambda-\frac{3}{2}\Lambda_0\,,\quad
		X_2
		=\frac{5}{2}\Lambda_0-\frac{1}{2}\Lambda\,.
	\end{align}

%% file: Chapters/Appendix_C.tex
\section*{Appendix B: Gamma matrices}

\addcontentsline{toc}{chapter}{Appendix B: Gamma matrices}

In order to construct spinor representations of the Lorentz group we may make use of the Clifford algebra
\begin{align}
	\left\{\gamma^a,\gamma^b\right\}=2\eta^{ab}\,\mathbb{I}\,,
\end{align}
where the identity matrix carries spinor indices. 
Since this relation allows symmetric products of gamma matrices to be converted into the Minkowski metric, it is natural to expect that the Clifford algebra should have a basis in terms of antisymmetrised products of gamma matrices, such as
\begin{align}
	\gamma^{\mu\nu}=\frac{1}{2}\left(\gamma^\mu\gamma^\nu-\gamma^\nu\gamma^\mu\right)\,.
\end{align}
This basis should consist of the identity, the $D$ gamma matrices themselves, and all independent matrices that can be formed therefrom.

Since any antisymmetric combination of gamma matrices can be expressed
\begin{align}\label{gamma antisymmetrisation}
	\gamma^{\mu_1\mu_2\dots\mu_r}=\gamma^{\mu_1}\gamma^{\mu_2}\dots\gamma^{\mu_r}\,,\quad
	\mu_1\neq\mu_2\neq\dots\neq\mu_r\,
\end{align}
we expect independent products of gamma matrices to have at most rank $D$.

We can then construct a basis for the Clifford algebra via antisymmetrised products of gamma matrices, i.e. for $D=4$ we have the set 
	\begin{align}\label{clifford basis}
		\Gamma^A=\left\{\mathbb{I},\gamma^{\mu_1},\gamma^{\mu_1\mu_2},\gamma^{\mu_1\mu_2\mu_3},\gamma^{\mu_1\mu_2\mu_3\mu_4}\right\}, 
		\quad \gamma^{\mu_1\dots\mu_n}\equiv\gamma^{[\mu_1}\dots\gamma^{\mu_n]}\, ,
	\end{align}
where antisymmetrisation is always with unit weight.
For a rank $r$ combination, \eqref{gamma antisymmetrisation} indicates that there are ${D \choose r}$ independent matrices that can be formed. 
We therefore have $1+4+6+4+1=16$ independent matrices.
From this we then construct the (index-reversed) dual
	\begin{align}
		\Gamma_A=\left\{\mathbb{I},\gamma_{\mu_1},\gamma_{\mu_2\mu_1},\gamma_{\mu_3\mu_2\mu_1},\gamma_{\mu_4\mu_3\mu_2\mu_1}\right\}\, , 
	\end{align}
whose elements can only differ by a sign from those of \eqref{clifford basis} by virtue of \eqref{gamma antisymmetrisation}.

Given the cyclicity of the trace, and that the trace of an odd number of gamma matrices is zero, the identity matrix is the only element which can have a non-zero trace. 
The trace orthogonality condition $\Tr\left(\Gamma^A\Gamma_B\right)=4\delta^A{}_B$ must then always be satisfied, and \eqref{clifford basis} is an orthogonal basis for the Clifford algebra.

Since the highest rank element of the Clifford algebra often has a special role to play, we may define for $D=2m$
\begin{align}
	\gamma^*\equiv i^{m+1}\gamma_0\gamma_1\dots\gamma_{D-1}\,,
\end{align}
which, on account of \eqref{gamma antisymmetrisation}, gives the relation
\begin{align}
	\gamma_{\mu_1\dots\mu_D}=i^{m+1}\epsilon_{\mu_1\dots\mu_D}\gamma^*\,.
\end{align}
Amongst other useful relations, in $D=4$ this yields
\begin{align}
	\gamma_{\mu\nu\rho}=i\epsilon_{\mu\nu\rho\delta}\gamma^\delta\gamma^5\,,
\end{align}
which can be used to compactly express the Rarita-Schwinger equation as
\begin{align}
	\gamma^{\mu\nu\rho}\partial_\nu\psi_\rho=0\,.
\end{align}

Finally, we will require the Hermiticity properties of the gamma matrices
\begin{align}
	\gamma^\mu{}^\dagger=\gamma^0\gamma^\mu\gamma^0\,.
\end{align}

%% file: Chapters/Appendix_D.tex
\section*{Appendix C: Fierz identities}

\addcontentsline{toc}{chapter}{Appendix C: Fierz identities}

To assist in simplifying fermion bilinears, we may leverage some useful Fierz identities.
We may firstly write, suppressing Lorentz indices,
\begin{align}
	\left(\overline\lambda_1M\lambda_2\right)\left(\overline\lambda_3N\lambda_4\right)
	=\overline\lambda_{1\alpha}\lambda_2{}^\beta\overline\lambda_{3\gamma}\lambda_4{}^\delta M^\alpha{}_\beta N^\gamma{}_\delta\, ,
\end{align}
for some anticommuting spinors $\lambda_i$.

Any matrix $M$ may be expanded in the Clifford algebra basis \eqref{clifford basis}
\begin{align}	
	M=\sum_A m_A \Gamma^A\, ,
\end{align}
where the coefficients $m_A$ can be identified via
\begin{align}	
	\Tr\left(\Gamma_A M\right)
	=\Tr\left(\Gamma_A m_B \Gamma^B\right)
	=4 m_B \delta^B{}_A
	=4 m_A\, .
\end{align}

Identifying $M^\alpha{}_\beta N^\gamma{}_\delta$ as a matrix $P_\beta{}^\gamma\left(\alpha,\delta\right)$ for fixed $\left\{\alpha, \delta\right\}$, we can then write  
\begin{align}
	P_\beta{}^\gamma\left(\alpha,\delta\right)
	&=\frac{1}{4}\sum_A\Tr\left(p^\varphi{}_\epsilon\left(\Gamma_A\right)^\epsilon{}_\rho\right)\left(\Gamma^A\right)_\beta{}^\gamma
	=\frac{1}{4}\sum_Ap^\rho{}_\epsilon\left(\Gamma_A\right)^\epsilon{}_\rho\left(\Gamma^A\right)_\beta{}^\gamma\nonumber\\
	&=\frac{1}{4}\sum_A\left(M^\alpha{}_\epsilon N^\rho{}_\delta\left(\Gamma^A\right)^\epsilon{}_\rho\right)\left(\Gamma^A\right)_\beta{}^\gamma
	=\frac{1}{4}\sum_A\left(M\Gamma_A N\right)^\alpha{}_\delta\left(\Gamma^A\right)_\beta{}^\gamma \, .
\end{align}

This then yields the standard expansion of products of bilinears (noting a minus sign from anticommutativity of $\lambda_i$), where we reexpress $\left(\overline\lambda_1 M\lambda_2\right)\left(\overline\lambda_3 N\lambda_4\right)$ as
	\begin{align}
		-\frac{1}{4}\sum_A\left(\overline\lambda_1 M\Gamma_A N\lambda_4\right)\left(\overline\lambda_3\Gamma^A\lambda_2\right)
		=-\frac{1}{4}\sum_n\frac{1}{n!}\left(\overline\lambda_1M\gamma_{\mu_1\dots \mu_n}N\lambda_4\right)
		\left(\overline\lambda_3 \gamma^{\mu_n\dots \mu_1} \lambda_2\right)\, ,
		\label{eq: fierz expansion}
	\end{align}
with a factor of $1/n!$ is introduced to avoid overcounting of the same $\gamma_{\mu_1\dots \mu_n}$ matrix $n!$ times.

In the case at hand, significant simplifications are possible since we only have one spinor; the gravitino, which satisfies the Majorana condition
\begin{align}
	\overline{\psi}_\mu=-\psi_\mu^T C\,,\quad C^T=-C\,, \quad \gamma_\mu^T=-C\gamma_\mu C^{-1}\,.
	\label{eq: majorana condition}
\end{align}
Since the charge conjugation matrix $C$ then obeys the relations
\begin{align}
	\left(C\gamma^{\mu_1}\right)^T=\left(C\gamma^{\mu_1}\right)\,,\quad 
	\left(C\gamma^{\mu_1\mu_2}\right)^T=\left(C\gamma^{\mu_1\mu_2}\right)\,,
\end{align}
we may deduce that, as a consequence of the sign change arising from fermion exchange,
\begin{align}
	 \overline\lambda\gamma^{\mu_1}\lambda
	 =\overline\lambda\gamma^{\mu_1\mu_2}\lambda
	 =0\,,
	 \label{eq: vanishing bilinears}
\end{align}
and we need only consider expansion in a subset of our basis elements.
We note that the vanishing of these particular bilinears is a consequence of our sign convention in \eqref{eq: majorana condition}, and in other conventions $\overline\lambda\gamma^{\mu_1\mu_2\mu_3}\lambda=\overline\lambda\gamma^{\mu_1\mu_2\mu_3\mu_4}\lambda=0$ instead\cite{Freedman:2012zz}.

Furthermore, we may prove a useful quadrilinear identity
\begin{align}
	\left(\overline\lambda\lambda\right)^2
	=-\left(\overline\lambda\gamma^5\lambda\right)^2
	=\frac{1}{4}\left(\overline\lambda\gamma^5\gamma^\nu\lambda\right)^2\,,
	\label{eq: trilinear}
\end{align}	
which allows remaining basis elements, re-expressed via the useful identities 
\begin{align}
	\gamma_{\mu_1\mu_2\mu_3}=i\epsilon_{\mu_1\mu_2\mu_3\mu_4}\gamma^{\mu_4}\gamma^5, \quad
	\gamma_{\mu_1\mu_2\mu_3\mu_4}=-i\epsilon_{\mu_1\mu_2\mu_3\mu_4}\gamma^5\, ,
\end{align}
to be simplified further.

Making use of the projection matrices
\begin{align}
	P_{L/R}\lambda=\lambda_{L/R}\,,\quad
	\overline\lambda_{L/R}=\overline{\left(\lambda_{R/L}\right)}\,,
\end{align}
we may note that since a Majorana spinor $\lambda_{L/R}$ can have only two independent anticommuting components in $D=4$, any product of more than two $\lambda_{L}$ vanishes, so that
\begin{align}
	\left(\overline\lambda_{R}M\lambda_{L}\right)\lambda_{L}=0\,,
\end{align}
and likewise for $\lambda_{R}$.

Given that $\overline{\lambda}_R\lambda_R=\overline{\lambda}P_LP_R\lambda=0$, this then implies 
\begin{align}
	\left(\overline\lambda\lambda\right)\lambda_\alpha=
	\left(\overline\lambda_L\lambda_R\right)\lambda_{L\alpha}+\left(\overline\lambda_R\lambda_L\right)\lambda_{R\alpha}\,.
\end{align}
Similarly making use of $\overline{\lambda}_L\gamma^5\lambda_L=\overline{\lambda}_R\gamma^5\lambda_R=0$, we may express $\left(\overline\lambda\gamma^5\lambda\right)\left(\gamma^5\lambda\right)_\alpha$ as
\begin{align}
	\left(\overline\lambda_L\gamma^5\lambda_R\right)\left(\gamma^5\lambda_{L}\right)_\alpha
	+\left(\overline\lambda_R\gamma^5\lambda_L\right)\left(\gamma^5\lambda_{R}\right)_\alpha
	=-\left(\overline\lambda_L\lambda_R\right)\lambda_{L\alpha}-\left(\overline\lambda_R\lambda_L\right)\lambda_{R\alpha}\,,
\end{align}
where we have used $\gamma^5\lambda_{L/R}=\pm\lambda_{L/R}$.
Left multiplication with $\overline\lambda^\alpha$ then provides the first equality of \eqref{eq: trilinear}.

We may also note via \eqref{eq: vanishing bilinears} and \eqref{clifford basis} that there can be only three distinct non-vanishing bilinears for a single Majorana spinor.
This then implies that the Fierz expansion \eqref{eq: fierz expansion} may convert $\left(\overline\lambda\gamma^5\gamma^\nu\lambda\right)^2$ into a combination of $\left(\overline\lambda\gamma^5\lambda\right)^2$ and $\left(\overline\lambda\lambda\right)^2$ terms, which by the previous result may be further simplified into a term of just a single type.

Simple substitution into \eqref{eq: fierz expansion} for $M=\gamma^5\gamma^\nu$, $N=\gamma^5\gamma_\nu$ and $\lambda_1=\lambda_2=\lambda_3=\lambda_4$ then yields the final equality of \eqref{eq: trilinear},
in agreement, upon left multiplication, with an equivalent identity given in the appendix of \cite{VanNieuwenhuizen:1981ae}.

In the present case, noting the permutation of the first bilinear relative to \eqref{eq: torsion lagrangian 2},
\begin{align} 
	\mathcal{L}_{\rm torsion}=\frac{1}{32}\kappa^2\left(\left(\overline\psi^\nu\gamma^\mu\psi^\rho\right)\left(\overline\psi_\rho\gamma_\mu\psi_\nu+2\overline\psi_\rho\gamma_\nu\psi_\mu\right)\right)\,,
\end{align}
and we may expand $\left(\overline\psi^\nu\gamma^\mu\psi^\rho\right)\left(\overline\psi_\rho\gamma_\mu\psi_\nu\right)$, using $\gamma_\mu\psi^\mu=0$, into
\begin{align}\nonumber
	&-\left(\overline\psi^\nu\psi_\nu\right)\left(\overline\psi_\rho\psi^\rho\right)
	-\frac{1}{4}\left(\overline\psi^\nu\gamma^\mu\gamma^5\gamma^\alpha\gamma_\mu\psi_\nu\right)\left(\overline\psi_\rho\gamma^5\gamma_\alpha\psi^\rho\right)
	+\frac{1}{4}\left(\overline\psi^\nu\gamma^\mu\gamma^5\gamma_\mu\psi_\nu\right)\left(\overline\psi_\rho\gamma^5\psi^\rho\right)\,,\\
	&=-\left(\overline\psi^\nu\psi_\nu\right)\left(\overline\psi_\rho\psi^\rho\right)
	-\frac{1}{2}\left(\overline\psi^\nu\gamma^5\gamma^\alpha\psi_\nu\right)\left(\overline\psi_\rho\gamma^5\gamma_\alpha\psi^\rho\right)
	-\left(\overline\psi^\nu\gamma^5\psi_\nu\right)\left(\overline\psi_\rho\gamma^5\psi^\rho\right)\,,
\end{align}
and similarly, expand $\left(\overline\psi^\nu\gamma^\mu\psi^\rho\right)\left(\overline\psi_\rho\gamma_\nu\psi_\mu\right)$ into 	
\begin{align}
	&-\frac{1}{2}\left(\overline\psi^\nu\psi_\nu\right)\left(\overline\psi_\rho\psi^\rho\right)
	-\frac{1}{4}\left(\overline\psi^\nu\gamma^\mu\gamma^5\gamma^\alpha\gamma_\nu\psi_\mu\right)\left(\overline\psi_\rho\gamma^5\gamma_\alpha\psi^\rho\right)
	+\frac{1}{4}\left(\overline\psi^\nu\gamma^\mu\gamma^5\gamma_\nu\psi_\mu\right)\left(\overline\psi_\rho\gamma^5\psi^\rho\right)\,,\nonumber\\
	&=-\frac{1}{2}\left(\overline\psi^\nu\psi_\nu\right)\left(\overline\psi_\rho\psi^\rho\right)
	-\frac{1}{2}\left(\overline\psi^\nu\gamma^5\gamma^\alpha\psi_\nu\right)\left(\overline\psi_\rho\gamma^5\gamma_\alpha\psi^\rho\right)
	-\frac{1}{2}\left(\overline\psi^\nu\gamma^5\psi_\nu\right)\left(\overline\psi_\rho\gamma^5\psi^\rho\right)\, .
\end{align}

Simplifying via \eqref{eq: trilinear}, noting in particular that the first and last terms in each line then cancel, we may write 
\begin{align}
		\mathcal{L}_{\rm torsion}
		=-\frac{3}{16}\kappa^2\left(\overline\psi^\rho\psi_\rho\right)^2\,,
\end{align}
as made use in the text.

%% file: Chapters/Appendix_E.tex
\section*{Appendix D: Zeta function regularisation}

\addcontentsline{toc}{chapter}{Appendix D: Zeta function regularisation}

In this appendix we give details of some mathematical aspects of our approach towards construction of the one-loop effective potential. 
More precisely, we detail the use of the heat kernel in computing functional determinants, before specialising to the computation of the resultant zeta functions on $S^4$.
Finally, we demonstrate an asymptotic expansion which allows these zeta functions to be explicitly evaluated in the limit $\Lambda\to 0$.

\subsection*{The heat kernel}
Consider a second-order Laplace-type differential operator of the form $\Delta=-\Box+X$, for some constant $X$, defined on a smooth vector bundle over a compact, smooth $D$-dimensional Riemannian manifold without boundary. 
There are a countable number of eigenfunctions and corresponding eigenvalues of this operator, which may be spectrally decomposed into a complete orthonormal set of eigenfunctions $\phi_n$ with eigenvalues $\lambda_n$, of multiplicity $g_n$. 

The determinant of this operator may be expressed
	\begin{align}
		\prod_{n}^{\infty}\lambda_{n}^{g_n}\,,
	\end{align}
however as this obviously diverges, we shall instead define the zeta function
	\begin{align}\label{zeta}
		\zeta\left(z\right)\equiv\sum_{n}^{\infty}g_n\lambda_n^{-z}\,,
	\end{align}
convergent for $\Re\left(z\right)>2$, which can be extended in practice via analytic continuation to a meromorphic function of $z$ over the entire complex plane.
It is important to note that $\zeta\left(z\right)$ is regular at $z=0$, yielding the derivative
	\begin{align}
		\zeta'\left(0\right)=-\sum_n g_n\ln\left(\lambda_n\right)\,,
		\label{eq: zeta'(0)}
	\end{align}
		so that we may define $\det\left(\Delta\right)$ via
	\begin{align} 
		\exp\left(-\frac{d}{dz}\zeta\left(z\right)\bigg|_{z=0}\right)\,.
	\end{align}
Our task is then to compute the form of $\zeta'$ for a given operator 
\footnote{We should mention that there can exist a multiplicative anomaly for the functional determinant, in that $\det\left(AB\right)$ does not necessarily equal $\det\left(A\right)\det\left(B\right)$ \cite{Elizalde:1997nd}. 
In the absence of a convenient rationale for circumventing this issue, we will proceed as others do \cite{Beccaria:2015vaa} and assume that this does not invalidate our results.}.

A convenient way of encapsulating some of the behaviour of $\zeta$ is via the `trace over the heat kernel', defined thusly
	\begin{align}\label{a2}
		\Tr \;K\left(x,x,t,\Delta\right)\equiv\sum_i b_i\left(x,\Delta\right)t^{\left(i-D\right)/2}\,,
	\end{align}
valid for $t\to 0^{+}$, where $K$ satisfies the heat equation with boundary condition
	\begin{align}
		\frac{d}{dt}K\left(x,x',t,\Delta\right)+\Delta K\left(x,x',t,\Delta\right)=0\,, \quad
		K\left(x,x',0,\Delta\right)=\delta\left(x,x'\right)\,,
	\end{align}
and the $b_i\left(x,\Delta\right)$ are the heat kernel coefficients, which integrate to give spectral invariants of $\Delta$. 
Since the heat equation has the solution
	\begin{align} \label{a1}
		K\left(x,x',t,\Delta\right)=\sum_i\phi_i\left(x\right)\otimes\phi_i\left(x'\right)\exp\left(-t\lambda_i\right)\,,
	\end{align}
we can trace over \eqref{a1} and integrate to see that, since the $\phi_i$ form an orthonormal basis,
	\begin{align}
		&\sum_i\exp\left(-t\lambda_i\right)\left(\phi_i,\phi_i\right)\left(x\right)
		=\sum_i b_i\left(x,\Delta\right)t^{\left(i-D\right)/2}\,,\nonumber\\
		\label{a3} \sum_i\exp\left(-t\lambda_i\right)&
		=\sum_i  t^{\left(i-D\right)/2}\int\sqrt{g} \;b_i\left(x,\Delta\right)d^{D}x
		\equiv\sum_{i=0}  t^{\left(i-D\right)/2}B_i\left(\Delta\right)\,.
	\end{align}
Finally, we note that $\zeta\left(z\right)$ is related to $\sum_i\exp\left(-t\lambda_i\right)$ via the Mellin transform
	\begin{align}
		\zeta\left(z\right)
		=\frac{1}{\Gamma\left(z\right)}\int_{\epsilon\to 0^+}^{\infty}t^{z-1}\sum_i\exp\left(-t\lambda_i\right)dt
		=\frac{1}{\Gamma\left(z\right)}\int_{\epsilon\to 0^+}^{\infty}\sum_i t^{\left(i+2z-D-2\right)/2}B_i\left(\Delta\right)dt~,
		\label{eq: Mellin}
	\end{align} 
so that for $t\to 0^+$ we may expand the sum and extract information about $\zeta$ as necessary.
We also note that resummation of this series can yield contributions from the topological winding modes of the sphere \cite{Demmel:2014sga}, which are albeit unimportant in $\Lambda\to0$ limit we are ultimately interested in.

In addition to the finite piece $\Gamma_f\left(\Delta\right)$, we can then find for $D=4$
\begin{align}
		\ln\det\left(\frac{\Delta}{\mu^2}\right)=
		-\frac{1}{2}B_0 L^4-\frac{1}{2}B_2 L^2
		-B_4\left(\ln\left(\frac{L^2}{\mu^2}\right)
		-\gamma\right)
		+\Gamma_f\left(\Delta\right)\,,
	\end{align}
where we have a cut-off $\epsilon=\left(\mu^2/L^2\right)\to0$, and $\gamma$ is the Euler-Mascheroni constant. 
The mass dimensions of $\mu$ and $L$ are one, and it is important to note that as \eqref{eq: Mellin} can be thought of as a proper time integral, $\epsilon\to0$ is a short time and thus high energy cutoff.
Moving from the ultraviolet to the infrared therefore corresponds to the direction of increasing $|\mu|$.

Computing the form of the $b_i$ is straightforward in practice as their general forms are known \cite{Fradkin:1983mq}.
In $D=4$ we have
	\begin{align}
		B_p=\int d^4x\,\sqrt{g}\, \frac{\overline{b}_p}{16\pi^2}\,, \quad
		\overline{b}_0=\Tr\mathbb{I}\,,\quad
		\overline{b}_2=\Tr\left(\frac{1}{6}R-X\right)\,,\quad
		b_i=\left(4\pi\right)^2\overline b_i\,,
	\end{align}
where the trace is performed in the space of fields.	
As the heat kernel coefficients are straightforward to find for low $D$, our problem of evaluating functional determinants is now reduced to computing the form of $\Gamma_f\left(\Delta\right)$ for a given background. 

To connect the $\zeta$ function defined in \eqref{zeta} with the implicitly dimensionful eigenvalues in $\Delta\phi=\overline\lambda\phi$, we may  rescale via $\overline{\lambda}_n=\rho\lambda_n$ for $\rho\equiv\Lambda/3$, and write
\begin{align}
	\ln\det\left(\frac{\Delta}{\mu^2}\right)
	\equiv-\sum_n g_n\ln\left(\frac{\overline{\lambda}_n}{\mu^2}\right)
	=-\sum_n g_n\ln\left(\lambda_n\right)+\sum_n g_n\ln\left(\frac{\rho}{\mu^2}\right)\,.
\end{align}
Noting the relations \eqref{zeta} and \eqref{eq: zeta'(0)}, this then yields
\begin{align}
	\Gamma_f\left(\Delta\right)=\zeta\left(0\right)\ln\left(\frac{\rho}{\mu^2}\right)-\zeta'\left(0\right)\,,
\end{align}
in agreement with \cite{Hawking:1976ja}, further reducing our task to computation of just $\zeta\left(0\right)$ and $\zeta'\left(0\right)$.

\subsection*{Zeta functions}	
As $\zeta$ is defined by the eigenvalues and their degeneracies for a given operator, we must work in a framework where these quantities are known.
This is achieved in practice by specialising to `differentially constrained' and therefore irreducible operators; those corresponding to irreducible representations of the background isometry group.
For SO(5) these representations can be labelled by $(n,s)$, for the eigenvalue number $n$ and spin $s$, with corresponding quadratic Casmirs 
\begin{align}
		C_2\left(n,s\right)=n\left(n+3\right)+s(s+1)\,,
	\end{align}
and degeneracies then given by the dimension of the representation, 
	\begin{align}
		g(n,s)=\dim\left(n,s\right)&=\frac{1}{6}\left(2s+1\right)\left(n-s+1\right)\left(n+s+2\right)\left(2n+3\right)\,.
	\end{align}
In concordance with \cite{Fradkin:1983mq}, accounting for the spin and isospin contributions gives 
\begin{align}
	{\lambda}_n=C_2-s(s+2)+\overline X\,,\quad
	{\lambda}_n=C_2+\frac{3}{2}+\overline X\,, \quad
	X\equiv\rho\overline X\,,\quad
	\rho\equiv\Lambda/3\,,
\end{align}
in the whole and half-integer instances respectively, so that we find

\noindent {\bf Spin 0: $\left(n,0\right)$}
	\begin{align}
		{\lambda}_n=n^2+3n+\overline{X}, \quad
		g_n=\frac{1}{6}(n+1)(n+2)(2n+3)\,,
	\end{align}
{\bf Spin 1: $\left(n,1\right)$}	
	\begin{align}
		{\lambda}_n=n^2+3n-1+\overline{X}, \quad
		g_n=\frac{1}{2}(n+3)(2n+3)\,,
	\end{align}
{\bf Spin 2: $\left(n,2\right)$}
	\begin{align}
		{\lambda}_n=n^2+3n-2+\overline{X}, \quad
		g_n=\frac{5}{6}(n-1)(n+4)(2n+3)\,,
	\end{align}	
{\bf Spin 1/2: $\left(n\pm\frac{1}{2},\frac{1}{2}\right)$}	
Where both representations have the same spectra, for \\$\left(n-\frac{1}{2},\frac{1}{2}\right)$
	\begin{align}
		{\lambda}_n=(n+1)^2+\overline{X}, \quad
		g_n=\frac{2}{3}n(n+1)(n+2)\,,
	\end{align}
which is to be doubled.	

\noindent{\bf Spin 3/2: $\left(n\pm\frac{1}{2},\frac{3}{2}\right)$}	
Where both representations again have the same spectra, for $\left(n-\frac{1}{2},\frac{3}{2}\right)$
	\begin{align}
		{\lambda}_n=(n+1)^2+\overline{X}, \quad
		g_n=\frac{4}{3}(n-1)(n+1)(n+3)\,,
	\end{align}
which is to be doubled.	
We arrive at $\lambda_n$ in this final instance by incorporating both the spin 1 and spin 1/2 factors outlined above, so that $\lambda_n=C_2-3+3/2$.
Spinor representations may be incorporated into this framework via `squaring' the corresponding first-order operators to yield those of second-order.

With this in mind, our prior expression for $\zeta$ \eqref{zeta} can now be re-expressed more concretely as
	\begin{align}
		\zeta_s\left(z,\overline{X}\right)
		&=\sum_{n=0}^{\infty}g_n\lambda_n^{-z}
		=\frac{\left(2s+1\right)}{6}\sum_{n=0}^{\infty}\frac{\left(n-s+1\right)\left(n+s+2\right)\left(2n+3\right)}{\left(n\left(n+3\right)+2ns+\overline{X}\right)^{z}}\nonumber\\
		&=\frac{1}{3}\left(2s+1\right)F\left(z\,,\,2s+1\,,\,\left(s+\frac{1}{2}\right)^2\,,\,b_s\left(\overline{X}\right)\right)\,,
	\end{align}
where we have defined	
	\begin{align}
		&F\left(z,k,a,b\right)\equiv\sum_{v=\frac{1}{2}k+1}^{\infty}\frac{v(v^2-a)}{(v^2-b)^z}\,,\quad
		b_{1/2}\left(\overline{X}\right)=b_{3/2}\left(\overline{X}\right)\equiv-\overline{X}\,,\nonumber\\
		&b_0\left(\overline{X}\right)\equiv\frac{9}{4}-\overline{X}\,,\quad  
		b_1\left(\overline{X}\right)\equiv\frac{13}{4}-\overline{X}\,,\quad 
		b_2\left(\overline{X}\right)\equiv\frac{17}{4}-\overline{X}\,,
		\label{eq: F}
	\end{align}
where our sum starts from the minimal $v$ such that $g_n>0$, and therefore all possible negative and zero modes are included.

To enable analytical continuation we make a binomial expansion, following the appendix of \cite{Christensen:1979iy}, and insert $v=n+\frac{k}{2}+1$ to rewrite $F\left(z,k,a,b\right)$ as
\begin{align}
	\sum_{n=0}^{\infty}\sum_{r=0}^{\infty}\frac{\Gamma\left(r+z\right)}{\Gamma\left(z\right)\Gamma\left(r+1\right)}b^r \left(\frac{1}{\left(n+\frac{k}{2}+1\right)^{2r+2z-3}}-\frac{a}{\left(n+\frac{k}{2}+1\right)^{2r+2z-1}}\right)\,.
	\label{eq: F binomial}
\end{align}
Isolating the sum over $n$ yields generalised zeta functions
\begin{align}
	\zeta\left(z,s\right)\equiv\sum_{n=0}^\infty \frac{1}{\left(n+s\right)^{z}}\,,
\end{align} 
so that we then have
\begin{align}
	\sum_{r=0}^\infty \frac{\Gamma\left(r+z\right)}{\Gamma\left(z\right)\Gamma\left(r+1\right)}b^r\left(\zeta\left(2r+2z-3,\frac{k}{2}+1\right)- a \zeta\left(2r+2z-1,\frac{k}{2}+1\right)\right)\,.
\end{align}
By extracting the $r=0$ contribution, the first sum may be written
\begin{align}
	\zeta\left(2z-3,\frac{k}{2}+1\right)+
	\sum_{q=0}^\infty \frac{\Gamma\left(q+z+1\right)}{\Gamma\left(z\right)\Gamma\left(q+2\right)}b^{q+1}\zeta\left(2q+2z-1,\frac{k}{2}+1\right)\,,
\end{align}
for $q\equiv r-1$, so that by leveraging the relation $\Gamma\left(x+1\right)=x\,\Gamma\left(x\right)$, we may express $F\left(z,k,a,b\right)$ as 
\begin{align}
	\zeta\left(2z-3,\frac{k}{2}+1\right)
	+\sum_{r=0}^\infty \frac{\Gamma\left(r+z\right)}{\Gamma\left(z\right)\Gamma\left(r+1\right)} b^{r+1}\left(\frac{r+z}{r+1}- \frac{a}{b}\right) \zeta\left(2r+2z-1,\frac{k}{2}+1\right)\,.
	\label{eq: F zeta function}
\end{align}
Given that the first term may be evaluated via
\begin{align}
	\zeta\left(-n,\frac{k}{2}+1\right)
	=-\frac{B_{n+1}\left(\frac{k}{2}+1\right)}{n+1}\,,
\end{align}
where $B_n\left(x\right)$ is the $n$th Bernoulli polynomial, leading to
\begin{align}
	\zeta\left(-1,\frac{k}{2}+1\right)=-\frac{1}{12}-\frac{k^2}{8}-\frac{k}{4}\,,\quad
	\zeta\left(-3,\frac{k}{2}+1\right)=\frac{1}{120}-\frac{k^4}{64}-\frac{k^3}{16}-\frac{k^2}{16}\,,
	\label{eq: bernoulli polynomials}
\end{align}
our task is then to perform the sum over $r$.

It is firstly straightforward to note from the small $z$ expansions  
\begin{align}
	\Gamma\left(z\right)=\frac{1}{z}-\gamma+\mathcal{O}\left(z\right)\,,\quad
	\zeta\left(2z+1,X\right)=\frac{1}{2z}-\digamma\left(X\right)+\mathcal{O}\left(z\right)\,,
\end{align}
where $\digamma$ is the digamma function, that only the $r=\left\{0,1\right\}$ terms in \eqref{eq: F zeta function} contribute in the $z\to 0$ limit. 
These are, respectively,
\begin{align}
	\left\{-a\zeta\left(-1,\frac{k}{2}+1\right)\,,\quad
	\frac{b^2}{4}-\frac{ab}{2}\right\}\,.
\end{align}
In conjunction with \eqref{eq: bernoulli polynomials}, this then yields the result
	\begin{align}
		F\left(0,k,a,b\right)=\frac{1}{4}b\left(b-2a\right)+\frac{1}{24}a\left(3k^2+6k+2\right)-\frac{1}{64}k^2\left(k+2\right)^2+\frac{1}{120}\,,
	\end{align}
from which $\zeta_s\left(0\right)$ follows, suitably arranged into the form given in \cite{Fradkin:1983mq}.	

To compute $\zeta_s'\left(0\right)$ we may firstly note that
	\begin{align}
		\frac{d}{db}F'\left(0,k,a,b\right)=F\left(1,k,a,b\right)\,,
	\end{align}
so that it suffices to instead evaluate $F\left(1,k,a,b\right)$. 
Some care is required for the $z\to1$ limit for the $r=0$ term in the \eqref{eq: F zeta function} summation, so we firstly expand into
\begin{align}
	b\left(z- \frac{a}{b}\right) \zeta\left(2z-1,\frac{k}{2}+1\right)+
	\sum_{r=1}^\infty\ b^{r+1}\left(1- \frac{a}{b}\right) \zeta\left(2r+1,\frac{k}{2}+1\right)\,,
	\label{eq: F sum expanded}
\end{align}
where the $z\to1$ limit has been taken in the $r\geq1$ terms.
For the $r=0$ term, we have
\begin{align}
	b\left(z- \frac{a}{b}\right) \zeta\left(2z-1,\frac{k}{2}+1\right)
	\stackrel{z\to1}{=}\left(b-a\right)\left(\frac{1}{2\left(z-1\right)}-\digamma\left(\frac{k}{2}+1\right)\right)+\frac{b}{2}\,,
	\label{eq: F r=0 term}
\end{align}
where we have used the Taylor expansion
\begin{align}
	\zeta\left(2z-1,\frac{k}{2}+1\right)=\frac{1}{2\left(z-1\right)}-\digamma\left(\frac{k}{2}+1\right)+\mathcal{O}\left(z-1\right)\,.
\end{align}

By using the same expansion in reverse, we can then reincorporate \eqref{eq: F r=0 term} as an $r=0$ term in \eqref{eq: F sum expanded}, giving
\begin{align}
	F\left(1,k,a,b\right)=\zeta\left(-1,\frac{k}{2}+1\right)+\frac{b}{2}+\sum_{r=0}^\infty\ b^r\left(b- a\right) \zeta\left(2r+1,\frac{k}{2}+1\right)\,.
\end{align}
We may then leverage the identity
\begin{align}
	\sum_{r=0}^\infty b^r\zeta\left(2r+1,X\right)=-\frac{1}{2}\digamma\left(X\pm\sqrt{b}\right)\,,\quad
	f\left(x\pm y\right)\equiv f\left(x+ y\right)+f\left(x- y\right)\,,
\end{align}
which follows \cite{Allen:1983dg} from substitution of the $z=1$ Taylor series for the digamma function 
\begin{align}
	\digamma\left(X+b\right)=\sum_{r=0}^\infty\left(-1\right)^{r+1}b^r\zeta\left(r+1,X\right)\,.
\end{align}
In totality, this then implies that
	\begin{align}
		F\left(1,k,a,b\right)=\frac{1}{2}b-\frac{1}{12}
		-\frac{1}{8}k\left(k+2\right)
		-\frac{1}{2}\left(b-a\right)\digamma\left(\frac{k}{2}+1\pm\sqrt{b}\right)\,,
	\end{align}
so that, upon integrating both sides,
	\begin{align}\label{poly}
		F'\left(0,k,a,b\right)=\frac{b^2}{4}-\frac{b}{12}-\frac{b k\left(k+2\right)}{8}
		-\frac{1}{2}\int_0^b\left(y-a\right)\digamma\left(\frac{k}{2}+1\pm\sqrt{y}\right) dy+C\,,
	\end{align}
in agreement with \cite{Fradkin:1983mq,Allen:1983dg}, where $C$ is a real constant of integration
	\begin{align}
		C=&F'\left(0,k,a,0\right)=2\zeta_R'\left(-3,\frac{1}{2}k+1\right)-2a\zeta_R'\left(-1,\frac{1}{2}k+1\right)\,,\nonumber\\
		&\zeta_R\left(z,q\right)=\sum_{n=q}^{\infty}=\frac{1}{n^z}\,,\quad
		\zeta_R'=\frac{d\zeta_R\left(z,q\right)}{dz}\,.
	\end{align}

For large $b$ we can explicitly evaluate the integral above by shifting the measure via $y\to y^2$ and inserting the asymptotic expansion 
	\begin{align}
		\digamma\left(X\right)
		=\ln\left(X\right)-\frac{1}{2X}-\sum_{n=1}^{\infty}\frac{B_{2n}}{2nX^{2n}}
		=\ln\left(X\right)-\frac{1}{2X}-\frac{1}{12X^2}+\dots,
	\end{align}
where $B_n$ is the $n$th Bernoulli number. 
Integrating term by term, only the leading order contributes, yielding
	\begin{align}
		F'\left(0,k,a,b\right)
		\simeq\frac{1}{4}b^2-\int_0^{\sqrt{b}} y^3\log\left(\frac{k}{2}+1\pm y\right) dy
		\simeq\frac{b^2}{8} \left(3-2\log \left(-b\right)\right)\,.
	\end{align}

Combing these elements, we thus find that for small $\Lambda$
	\begin{align}
		\zeta_s\left(0,X\right)\simeq\frac{6s+3}{4}\frac{X^2}{\Lambda^2}\,,\quad
		\zeta'_s\left(0,X\right)\simeq\frac{6s+3}{4}\frac{X^2}{\Lambda^2} \left(\frac{3}{2}-\ln\left(\frac{3X}{\Lambda}\right)\right)\,.
	\end{align}
It is interesting to note that in the limit $\Lambda\to0$, the presence or absence of imaginary terms in the effective potential is a straightforward consequence of the sign of $X$, the argument of the functional determinant being evaluated.

%% file: Chapters/Acknowledgements.tex
\addcontentsline{toc}{chapter}{Acknowledgements}

This thesis would not have been possible without the expertise, supervision and collaboration of Jean Alexandre and Nikolaos Mavromatos.
I am furthermore indebted to the numerous former and present members of the Theoretical Particle Physics and Cosmology group of the King's College London Department of Physics, for providing a stimulating, enriching and enjoyable environment within which to work.
Special thanks are due to Spyros Sypsas for invaluable proofreading services, John Ellis and Malcolm Fairbairn for equally useful guidance and advice, Julia Kilpatrick for essential administrative support, and James Brister for collaboration on \cite{Alexandre:2012dm}.
The efforts of my examiners Richard Szabo and Xavier Calmet are also recognised for their helpful suggestions on improving the text, and the financial support of the King's College London Graduate School and Department of Physics is furthermore acknowledged.
Finally, I wish to recognise the continual support and encouragement of my parents John and Lesley.